\documentclass[
superscriptaddress,
twocolumn,
preprintnumbers,
nofootinbib,
 amsmath,amssymb,
 aps,prd,
]{revtex4-2}

\usepackage{graphicx}
\usepackage{dcolumn}
\usepackage{bm}
\usepackage[colorlinks,linkcolor=blue,citecolor=blue,urlcolor=blue ]{hyperref}
\usepackage{amsmath,amssymb,natbib,latexsym}
\usepackage[T1]{fontenc}
\usepackage[utf8]{inputenc}

\usepackage{xcolor}
\usepackage{xspace}
\newcommand{\sample}{{\rm {Y6\,BAO\,sample}}\xspace}
\newcommand{\mocks}{\textsc{ICE-COLA} mocks}
\newcommand\cosmolike{\textsc{CosmoLike}\xspace}
\newcommand{\namaster}{\textsc{NaMASTER}\xspace}
\newcommand{\der}{\mathrm{d}}
\newcommand{\healpix}{\textsc{HEALPix}\xspace}

\newcommand{\biasalpha}{\bar{\Delta} \langle  \alpha \rangle }

\newcommand{\zph}{z_{\rm ph}\xspace}
\newcommand{\zmc}{z_{\rm nn}\xspace}
\newcommand{\zmean}{\textsc{DNF\_Z}\xspace}
\newcommand{\zeff}{z_{\rm eff}\xspace}

\newcommand{\mocklike}{\textit{Mock-like}\xspace}
\newcommand{\datalike}{\textit{Data-like}\xspace}
\newcommand{\datalikemice}{\textit{Data-like-mice}\xspace}
\newcommand{\mice}{\textsc{MICE}\xspace}
\newcommand{\planck}{\textsc{Planck}\xspace}
\newcommand{\gold}{\textsc{Y6 gold}\xspace}

\newcommand{\aap}{A\&A}
\newcommand{\apjs}{ApJS}
\newcommand{\aj}{A.J.}
\newcommand{\mnras}{MNRAS}

\newcommand{\jcap}{JCAP}
\newcommand{\apjl}{ApJ}

\newcommand{\gray}[1]{\textcolor{gray}{#1}}

\newcommand{\Omegam}{\Omega_{{\rm m}}}
\newcommand{\Omegab}{\Omega_{{\rm b}}}
\newcommand{\OmegaL}{\Omega_{{\Lambda}}}

\usepackage[normalem]{ulem}

\usepackage{array}
\newcolumntype{Z}{>{\setbox0=\hbox\bgroup}c<{\egroup}@{}}

\newcommand\T{\rule{0pt}{2.6ex}}       
\newcommand\B{\rule[-1.2ex]{0pt}{0pt}} 

\usepackage{booktabs}

\newcommand{\appendixcite}[1]{\hyperref[#1]{\textcolor{blue}{Appendix \ref*{#1}}}}

\begin{document}

\preprint{DES-2023-0793}
\preprint{FERMILAB-PUB-24-0027-PPD}

\title[DES Y6 BAO]{Dark Energy Survey: A 2.1\%  measurement of the angular Baryonic Acoustic Oscillation scale at redshift $z_{\rm eff}=0.85$ from the final dataset}


\author{T.~M.~C.~Abbott}
\affiliation{Cerro Tololo Inter-American Observatory, NSF's National Optical-Infrared Astronomy Research Laboratory, Casilla 603, La Serena, Chile}
\author{M.~Adamow}
\affiliation{Center for Astrophysical Surveys, National Center for Supercomputing Applications, 1205 West Clark St., Urbana, IL 61801, USA}
\author{M.~Aguena}
\affiliation{Laborat\'orio Interinstitucional de e-Astronomia - LIneA, Rua Gal. Jos\'e Cristino 77, Rio de Janeiro, RJ - 20921-400, Brazil}
\author{S.~Allam}
\affiliation{Fermi National Accelerator Laboratory, P. O. Box 500, Batavia, IL 60510, USA}
\author{O.~Alves}
\affiliation{Department of Physics, University of Michigan, Ann Arbor, MI 48109, USA}
\author{A.~Amon}
\affiliation{Institute of Astronomy, University of Cambridge, Madingley Road, Cambridge CB3 0HA, UK}
\affiliation{Kavli Institute for Cosmology, University of Cambridge, Madingley Road, Cambridge CB3 0HA, UK}
\affiliation{Department of Astrophysical Sciences, Princeton University, Peyton Hall, Princeton, NJ 08544, USA}
\author{F.~Andrade-Oliveira}
\affiliation{Department of Physics, University of Michigan, Ann Arbor, MI 48109, USA}
\author{J.~Asorey}
\affiliation{Departamento de Física Teórica and Instituto de Física de Partículas y del Cosmos (IPARCOS-UCM), Universidad Complutense de Madrid, 28040 Madrid, Spain}
\author{S.~Avila}
\affiliation{Institut de F\'{\i}sica d'Altes Energies (IFAE), The Barcelona Institute of Science and Technology, Campus UAB, 08193 Bellaterra (Barcelona) Spain}
\author{D.~Bacon}
\affiliation{Institute of Cosmology and Gravitation, University of Portsmouth, Portsmouth, PO1 3FX, UK}
% ...

\collaboration{DES Collaboration}\email{des-publication-queries@fnal.gov}

\date{\today}

\begin{abstract}
We present the angular diameter distance measurement obtained with the Baryonic Acoustic Oscillation feature from galaxy clustering in the completed Dark Energy Survey, consisting of six years (Y6) of observations.  We use the Y6 BAO galaxy sample, optimized for BAO science in the redshift range $0.6<z<1.2$, with an effective redshift at $\zeff=0.85$ and split into six tomographic bins. The sample has nearly 16 million galaxies over 4,273 square degrees.
Our consensus measurement constrains the ratio of the angular distance to sound horizon scale to $D_M(\zeff)/r_d = 19.51 \pm 0.41$ (at 68.3\% confidence interval), resulting from comparing the BAO position in our data to that predicted by \planck $\Lambda$CDM via the BAO shift parameter $\alpha=(D_M/r_d)/(D_M/r_d)_{\planck}$.
To achieve this, the BAO shift is measured with three different methods, Angular Correlation Function (ACF), Angular Power Spectrum (APS), and Projected Correlation Function (PCF) 
obtaining
$\alpha= 0.952\pm0.023$, $0.962\pm0.022$, and $0.955\pm0.020$, respectively, which we combine to $\alpha= 0.957\pm0.020$, including systematic errors.
When compared with the $\Lambda$CDM model that best fits Planck data, this measurement is found to be 4.3\% and $2.1\sigma$ below the angular BAO scale predicted. To date, it represents the most precise angular BAO measurement
at $z>0.75$ from any survey and the most precise measurement at any redshift from photometric surveys.
The analysis was performed blinded to the BAO position
and it is shown to be robust against analysis choices, data removal,  redshift calibrations and  observational systematics. 
\end{abstract}

\maketitle


\section{\label{sec:intro}
Introduction
}

The Dark Energy Survey\footnote{\url{https://www.darkenergysurvey.org/}} (DES) is a stage-III photometric galaxy survey designed to constrain the properties of dark energy and other cosmological parameters from multiple probes \cite{DEtaskforce,DES05,Flaugher,DES_all_probes_Y1}. 
DES has performed state-of-the-art analyses of Weak gravitational Lensing (WL) by measuring and correlating the shape of more than 100 million galaxies \cite{y3-shapecatalog,y3-cosmicshear1,y3-cosmicshear2}. 
DES has also excelled in using Galaxy Clustering (GC) as a cosmological probe, either on its own or in combination with WL and other probes \cite{y3-galaxyclustering,y3-2x2ptaltlensresults,y3-2x2ptredmagic,y3-3x2ptkp}.  
These probes (WL, GC) have also been combined with galaxy cluster counts detected on DES \cite{desy1-clustercosmology,y1-cluster6x2pt} and with external CMB data \cite{y1-6x2pt,y3-6x2pt}. 
The DES Supernova program has also broken new grounds in constraining cosmology from $\sim1500$ type Ia supernovae \cite{SNkey,SNcosmo}. In addition to that, the large data-set and catalogues produced by the Dark Energy Survey, represent a unique source for other cosmological and astronomical analyses \cite{DESoverview,DESDR1,DESDR2,y3-gold,Y6_gold}.

The measurement of galaxy clustering (GC) within DES has traditionally been split into two main probes. On the one hand, we have the GC of the so-called {\it lens} samples that have been used mainly in combination with WL and other probes to constrain the amplitude of mass fluctuations
$\sigma_8$, the matter density
$\Omegam$ and other $\Lambda$CDM parameters as well as extensions of $\Lambda$CDM, such as the equation of state of dark energy, $w$. On the other hand, we have the measurement of the position of the Baryonic Acoustic Oscillation (BAO) peak in the clustering of galaxies from a different sample of galaxies, optimized for this science case. The BAO peak position can be used as a standard ruler to constrain the angular diameter distance to redshift relation and, in turn, constrain the expansion history of the Universe. In this work, we present the measurement of the BAO peak position from the final DES dataset, which includes 6 years (2013-2019) of observations. For the remainder, this data set will be referred to as Year 6 or Y6.  

The Baryonic Acoustic Oscillation (BAO) feature originated in early times when the Universe was in the form of a plasma in which photons and the baryonic matter were in continuous interaction. Thanks to this interaction, sound waves propagate in this plasma up to the drag epoch, when photons and the baryonic matter cease to interact. This leaves a preferred scale in the distribution of matter in the Universe, corresponding approximately to the sound horizon at decoupling, denoted by $r_d$. This scale can be measured as an excess of signal (a {\it peak}) in the two-point correlation function in different tracers of the matter distribution. The scale of this peak, $r_d$, remains fixed in comoving coordinates after recombination and, thus, can be used as a standard ruler to constrain the relation between redshift and the comoving angular diameter distance ($D_M(z)$) \cite{Peebles1970,SZ1970,Bond1984,Bond1987} \footnote{Technically, angular BAO constraints are only sensitive to the ratio $D_M(z)/r_d$. But since we can determine $r_d$ with great accuracy from CMB constraints and well understood physics, in practice the information we recover from late time BAO can be interpreted in terms of constraining the angular diameter distance, $D_M(z)$.}. This relation can be used to constrain the expansion history of the Universe and, hence, the nature of dark energy. Remarkably, the redshift range explored here corresponds to an epoch when the Universe expansion was about to transition from deceleration to acceleration, according to the standard model, hence being an excellent test for dark energy near this transition.

This acoustic peak was first seen in the CMB anisotropies with BOOMERanG and  MAXIMA experiments at the turn of the century \cite{Boomerang_BAO_peak,Maxima_BAO_Peak}. 
Half a decade later, the BAO peak was first measured in the distribution of galaxies by both 
the Sloan Digital Sky Survey (SDSS) \citep{2005ApJ...633..560E} and the 2-degree Field Galaxy Redshift Survey (2dFGRS) \citep{2001MNRAS.327.1297P,2005MNRAS.362..505C}.
Since then, a series of galaxy spectroscopic surveys have been designed to measure BAO at different redshifts. In particular, it is worth highlighting the 6-degree Field Galaxy Survey (6dFGS) \citep{2011MNRAS.416.3017B}, the WiggleZ dark energy survey \citep{2011MNRAS.415.2892B,2011MNRAS.418.1707B,2014MNRAS.441.3524K,2017MNRAS.464.4807H} and the [extended] Baryonic Oscillations Spectroscopic Survey ([e]BOSS), part of the SDSS series \cite{2015MNRAS.449..835R,2017MNRAS.470.2617A,2018MNRAS.473.4773A,ly-alpha_BOSS,ly-alpha_BOSS2,Ly-alpha_cross, 2015A&A...574A..59D,2017A&A...603A..12B,ly-alpha_eBOSS,eBOSS_final}. The last release by eBOSS/SDSS 
\cite{eBOSS_final}, represents the
state-of-the-art in spectroscopic measurements of BAO and the 
closure of stage-III spectroscopic surveys. A new generation of spectroscopic surveys (stage-IV), with the Dark Energy Spectroscopic Instrument \cite{DESI} and the European Space Agency mission Euclid \cite{Euclid} as the prime examples, recently started collecting data and have among their main design goals to measure the BAO peak with higher precision and at higher redshifts. 

In this context, the Dark Energy Survey, as a photometric survey designed simultaneously for multiple cosmological probes, can not measure the redshift of galaxies with high precision. Instead, we use the photometric redshift, $\zph$, based on the fluxes measured in five bands. This makes the measurement of distances between galaxies more challenging, losing part of the information and degrading the signal-to-noise ratio of the BAO signal. On the other hand, DES is able to detect a large number of galaxies and have a photometric redshift estimate for all of them (of the order of 100s of millions vs. 2 million spectra measured by SDSS in 20 years). Among those galaxies, we can select a sub-sample for which the redshift can be estimated at a $\sim3\%$ precision, giving us the opportunity to detect the angular component of the BAO with a precision competitive with stage-III spectroscopic surveys.  

In order to achieve competitive BAO measurements, the Dark Energy Survey collaboration has dedicated remarkable efforts in 
all the successive data batches (Year 1 or Y1, Year 3 or Y3 and now Y6) to this key analysis in parallel to other Galaxy Clustering projects. 
On the galaxy sample selection side, this work builds on the selection optimized in \cite{DESY1baosample} that we now re-optimized in our companion paper \cite{Mena23}. This selection is remarkably different to the ones applied to spectroscopic surveys \cite{BOSS_target,eBOSS_survey}, resulting in a much larger number of galaxies (16 million in Y6 DES BAO vs. <1 million in BOSS/eBOSS individual samples). The validation of these galaxy samples and the techniques for the correction of systematics build on \cite{sv_lss_sys,y1_lss_sys_kp, y3-galaxyclustering,y3-baosample}, which in turn build  on previous works \cite{ross11,Ho12,Leistedt13,Weaverdyck_2021}. In parallel, a large part of the tests performed to validate our analysis rely on having the order of 2000 simulations, with the techniques developed in \cite{halogen,DESY1baomocks,ice-cola,y3-baomocks}, similar to what it is standard in spectroscopic surveys \cite{Manera13,Zhao21}, but with the challenges of having a much larger number of galaxies and including the modeling of redshift errors. The techniques to obtain robust BAO measurements from the angular correlation function (ACF) and angular power spectrum (APS) were developed in \cite{Chan:2018gtc} and \cite{Camacho:2018mel}, respectively. Combining/comparing analyses from configuration and Fourier space is also a common practice in spectroscopic BAO analyses \cite[e.g.][]{Bautista21,GilMarin20,raichoor21,deMattia}. Here, we add a third way to analyse the data based on the projected correlation function (PCF), which builds upon the techniques developed in \cite{Ross:2017,PCF_method_Y3,PCF_Y3_BAO}.

All of the previous work led to a 4\% measurement of the angular diameter distance of the BAO peak in DES Y1 \cite{Y1_BAO} (at $z_{\rm eff}\sim0.81$) and a 2.7\% in DES Y3 (at $z_{\rm eff}\sim0.83$) \cite{y3-baokp}. 

The latter measurement already represented the tightest constraint from a photometric survey and the tightest constraint from any survey at an effective redshift $0.8<\zeff<1.4$. 
Another photometric BAO measurement at similar redshift is given by \cite{BAO_DECals}, with a $6.5\%$ precision at $\zeff=0.85$, and other BAO photometric measurements include \cite{2007MNRAS.378..852P,2009ApJ...692..265E,2010MNRAS.401.2477H, 2011MNRAS.411..277S,2011MNRAS.417.2577C,2012ApJ...761...13S,2012MNRAS.419.1689C}. 
When comparing to spectroscopic angular BAO measurement, at a similar redshift we find the eBOSS ELG with a $\sim 5\%$ precision at $\zeff=0.85$, weaker constraints that the DES Y3 BAO results. However, more precise angular BAO measurements are reported at higher and lower redshifts by BOSS (1.5\% at $\zeff=0.38$, 1.3\% at $\zeff=0.51$) and eBOSS ($1.9\%$ at $\zeff=0.70$,  $2.6\%$ at $\zeff=1.48$ and $2.9\%$  at $\zeff = 2.33$) \cite{2017MNRAS.470.2617A,2020MNRAS.500..736B,2020MNRAS.498.2492G,2020MNRAS.499.5527T,2021MNRAS.501.5616D,Tamone,deMattia,2020ApJ...901..153D}. 
The $2.1\%$ measurement we report in this paper is currently the tightest angular BAO measurement at an effective redshift larger than $\zeff=0.75$. 

In this work, we use the complete DES data set, Y6, to constrain the angular BAO. We follow a similar methodology to the Year 3 analysis, with three main changes. First, we re-optimize the sample in our companion paper \cite{Mena23} and extend it from $0.6<\zph<1.1$ to $0.6<\zph<1.2$, giving us an effective redshift of $z_{\rm eff}=0.851$. Second, we reinforce the redshift validation, considering several independent calibrations and quantifying its possible impact on the BAO measurement. Third, we provide BAO measurements from three types of 2-point clustering statistics: angular correlation function (ACF), Angular Power Spectrum (APS) and Projected Correlation Function (PCF). Our reported consensus result stems from the statistical combination of those three measurements.  

Finally, in the scientific community of cosmology, there is a growing awareness of the danger of confirmation biases affecting results in science. In order to mitigate this, many collaborations have built a series of protocols to {\it blind} the results of the analyses until these are {\it finalized}, with different criteria imposed on how to blind the data and when they are considered finalized. DES has built a strong policy in this direction and it is one pillar of the way we perform and present our analysis in this paper. The BAO analysis presented here and in previous DES data batches are likely the ones with the strongest blinding policies to this date. 

This paper is organized as follows. In \autoref{sec:data}, we describe the Y6 DES data and the BAO sub-sample, together with its mask, observational systematic treatment and redshift characterisation. In \autoref{sec:sims}, we describe the mock catalogues that are used to validate and optimize our analysis.
In \autoref{sec:methods}, we describe the methodology used to extract the BAO information. In \autoref{sec:validation}, we validate our analysis in three aspects: robustness against redshift distributions (\autoref{sec:redshift_val}), robustness of the modeling of individual estimators (ACF, APS, and PCF) against the mock catalogues (\autoref{sec:mocktests}) and robustness of our combined measurement (\autoref{sec:comb_mocks}). In \autoref{sec:blindtest}, we present a battery of tests performed on the data, while blinded, prior to decide whether it was ready to unblind and publish. 
In \autoref{sec:results}, we present the unblinded results on the BAO measurement and a series of robustness tests. Finally, we conclude in \autoref{sec:conclusions}.

\section{\label{sec:data} The Dark Energy Survey data}

\subsection{DES Y6 Gold catalogue}

The Dark Energy Survey data used in this analysis is
obtained from a subset of the wide-area imaging performed by
the survey in its five photometric bands, spanning a period
of approximately 6 years from 2013 to 2019, encompassing
the entire run of the project (DES Y6). In particular, we use the 
detections in the coadd catalogues, the details of
which are described in \cite{DESDR2}. This data set spans
the full 5,000 square degrees of the survey, reaching a depth
in the $[grizY]$ bands of $[24.7,24.4,23.8,23.1,21.7]$ for point sources, at a signal-to-noise ratio of 10.

The coadd catalogues are further enhanced into the \gold catalogue \cite{Y6_gold}, to include improvements in object photometry, star-galaxy
separation, quality flags, additional masking, and the
creation of ad-hoc survey property maps to be used to
mitigate clustering systematic effects. This catalogue is the
basis for the BAO sample, described in the following section.
Note that the core \gold catalogue has the same number of detections as the public DR2 data, but with additional columns and a flag identifying the object as part of the official footprint to be used in the cosmology analyses.

\subsection{The BAO-optimized sample}
\label{sec:sample}

The Y6 BAO sample is a subsample of the \gold catalogue described above and is fully described in the companion paper \citet{Mena23}. 
The procedure to build and characterize this sample builds up from those used in the Y1 and Y3 BAO samples, described in \cite{DESY1baosample} and \cite{y3-baosample}, respectively. 
The first criterion is to select galaxies above redshift $z\sim0.6$, where DES BAO measurements can be competitive. For that, in \cite{DESY1baosample}, we argued that a red selection as follows would already be a good starting point:
\begin{equation}
    1.7< (i-z) + 2(r-i).
    \label{eq:redselection}
\end{equation}
Additionally, red galaxies are expected to have better redshift estimates and higher galaxy bias, both improving the expected BAO signal. 
The Y6 data has an increased depth, resulting in better photometry and redshift estimations than the Y3 and Y1 catalogues. For this reason, we extended our redshift range of study to 
\begin{equation}
    0.6<\zph<1.2,
    \label{eq:redshiftrange}
\end{equation} 
whereas we studied $0.6<\zph<1.0$ in Y1 and $0.6<\zph<1.1$ in Y3. $\zph$ is the photometric redshift estimate and is given by the variable \zmean of the Directional Neighboring Fitting photo-$z$ code (DNF,  \cite{DNF}, more details in \autoref{sec:redshifts}). 
For most of the analysis, we will be splitting this sample into 6 tomographic bins given by $\Delta \zph=0.1$ that we will label as bins 1 to 6 in increasing order with redshift.

One of the main challenges in galaxy clustering with photometric samples is the estimation and validation of the redshift distribution $n(z)$. In Y3, an important step of the redshift validation was based on direct calibration with the VIPERS sample \cite{vipers}, which is complete up to 
\begin{equation}
    i<22.5.
    \label{eq:fluxcut}
\end{equation}
In order to ensure high quality in our validation pipeline, we include this selection in our sample definition. 
This selection did not need to be imposed in Y3, where the criterion was naturally met by the selection. 
By selecting bright galaxies we additionally expect this sample to be less affected by imaging systematics and also to have better estimates of the redshifts and their uncertainties. 

This leads us to our fourth main selection criterion: a redshift-dependent magnitude limit in the band $i$: $i<a+b\ \zph$. We have to choose a balance in our sample selection between having more galaxies or having lower redshift uncertainties. This idea was already implemented in Y1 \cite{DESY1baosample}, where $a$ and $b$ were optimized for the BAO distance measurement using a Fisher forecast based on sample properties such as number density and redshift distributions. The same selection was used in Y3 \cite{y3-baosample}, with the $a$ and $b$ values optimized from Y1. In Y6, however, having much deeper photometry, we expected the optimal sample to change. For that reason, in \cite{Mena23}, we have re-optimized the sample selection for $a$ and $b$ (after imposing \textcolor{blue}{Equations} \ref{eq:redselection}, \ref{eq:redshiftrange}, and \ref{eq:fluxcut}), finding our best BAO forecast for 
\begin{equation}
    i<19.64+2.894\ z_{\rm ph}.
    \label{eq:maglim}
\end{equation}
All the details of this optimization can be found in \cite{Mena23}.

Our Y6 BAO sample definition is given by the selections imposed by \textcolor{blue}{Equations} \ref{eq:redselection}, \ref{eq:redshiftrange}, \ref{eq:fluxcut} \& \ref{eq:maglim}. 
Additionally, as part of our quality cuts, we also apply a bright magnitude cut at $17.5<i$ to remove bright contaminant objects such as binary stars, as done in Y3. Stellar contamination is mitigated with the galaxy and star classifier \textsc{EXTENDED\_CLASS\_MASH\_SOF} from the Y6 GOLD catalogue.

The resulting catalogue, over the area described below (\autoref{sec:mask}), comprises a total of 15,937,556 objects, more than twice the Y3 BAO sample. In Y6, we additionally used unWISE infrared photometry \cite{unWISE} to estimate the residual stellar contamination in our sample, finding a stellar fraction of $f_{{\rm star}}=0.023,0.027,0.033,0.023,0.008,0.007$ for the redshift bins 1 to 6, respectively. The method to estimate this is briefly described in \cite{Mena23} and will be presented in detail in \cite{Y6_LSS_sys}.

\subsection{Angular mask}
\label{sec:mask}

\begin{figure}
    \includegraphics[width=84mm]{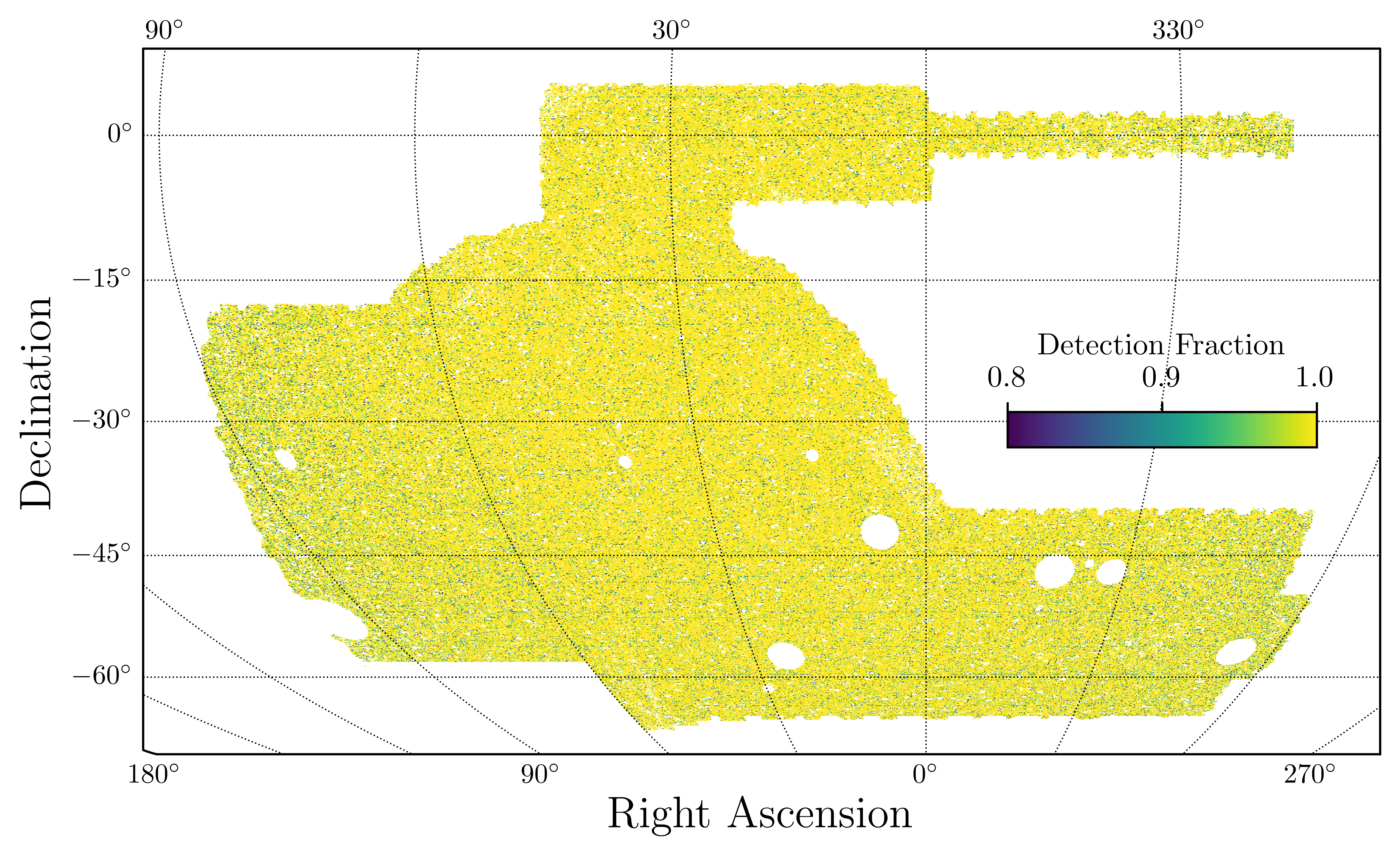}  
    \caption{Angular mask for the DES Y6 BAO analysis. The value plotted for each pixel represents its detection fraction. The total area of the mask, computed weighting by the detection fraction, is 4,273.42 deg${}^2$. 
    }
    \label{fig:footprint}
\end{figure}

The \sample is distributed over a footprint of 4273.42 deg$^2$, as shown in \autoref{fig:footprint}, defined at \healpix resolution of $N_{\rm side}$ = 4096 with a pixel area of $\sim 0.74$ arcmin$^2$. The final area results from applying several quality cuts: we impose pixels to have been observed at least twice in bands \emph{griz} and to have a detection fraction higher than 0.8. The detection fraction quantifies the fraction of the area of a pixel at resolution 4096 that is not masked by foregrounds, which is studied originally at higher resolution (details in \cite{Y6_mask}). We also exclude pixels that do not reach the depth of our sample: $i_{\rm lim} = 22.5$ at 10$ \sigma$.
We also veto pixels affected by astrophysical foregrounds such us bright stars, globular clusters or large nearby galaxies (including the Large Magellanic Cloud), see \cite{Y6_gold}. More details on the masking construction are given in \cite{Mena23}.  

Finally, we also mask outliers on the maps that trace galactic cirrus and image artifacts, amounting to $\sim 1.85\%$ of the area. Further details are given in the companion paper \cite{Mena23} and a full study of the effect of masking outliers of survey property maps is deferred to \cite{Y6_mask}.

\subsection{LSS systematics weights}
\label{sec:systematics}

Observing conditions as well as (galactic) foregrounds affect the fraction of galaxies that we are be able to detect in our sample. This will result in a detection fraction with a pattern in the observed sky that can lead to spurious galaxy clustering, if unaccounted for. In order to characterize this pattern, we use a series of survey property maps summarizing both the observing conditions and foregrounds.

We mitigate the impact of observational systematics by applying correcting weights to the galaxy sample with the   \emph{Iterative Systematics Decontamination} (ISD) method, used in other DES Galaxy Clustering analyses \cite{sv_lss_sys,y1_lss_sys_kp,y3-baosample,y3-galaxyclustering,Y6_LSS_sys}.
This method assumes a linear dependence between the observed galaxy number density and the survey property contamination template maps. A linear regression between the survey property and the number of galaxies is performed and its $\chi^2$ compared to that of a null correlation. The resulting $\Delta \chi^2$ is compared against 1000 lognormal mock catalogues, taking as reference the percentile 68 of their $\Delta \chi^2$ distribution, $\Delta \chi^2_{68}$.
Then, we consider a correlation of a given survey property map significant if $\Delta \chi^2 > T_{1D}\Delta \chi^2_{68}$, where 
$T_{1D}$ represents a threshold that is a free parameter of the ISD method. In Y3-BAO, we chose $T_{1D}=4$ (equivalent to a $2\sigma$ significance), a milder requirement than that used in the GC+WL analyses: $T_{1D}=2$. In Y6 BAO we lie on the more stringent side with a threshold $T_{1D}=2$.
Details on the decontamination methodology, the survey property maps used as contamination templates and the weight validation can be found in \cite{Mena23}. 

At the moment of construction of the \mocks\ described in \autoref{sec:sims}, the ISD weights were not finalized. In order to have a first estimate of the amplitude of clustering to fit the mocks, we used a preliminary version of the weights, based on the modified Elastic Net approach (\textsc{enet}), described in \cite{enet}.

Finally, we remark that the effect of these systematic weights has a relatively more important effect on large scales, requiring a thorough validation for studies such as the combination of GC with WL (the so-called 3$\times$2pt analysis), Primordial Non-Gaussianties, etc. 
However, this contamination typically has a very smooth pattern in clustering, not affecting the location of the BAO peak. 
Indeed, at early stages, we checked with lognormal mocks that (an early version of) the weights described here were not having any effect on the recovered BAO (last two entries of \autoref{tab:ACF_mocks}). 
Once the main pre-unblinding tests were passed (\autoref{sec:blindtest}) but prior to unblinding, we checked that when the systematic weights are ignored, the BAO position in Y6 moved only by 0.21\%, 0.04\% and 0.32\% for ACF, APS and PCF, respectively. This is below $0.2\sigma$ for all three estimators. We consider this error as a very upper limit of the possible residual effects from observational systematic and conclude that any remaining uncertainty on the weights should have a negligible impact on the BAO measured position.

\subsection{Photometric redshifts}
\label{sec:redshifts}

\begin{figure*}
\includegraphics[width=160mm]{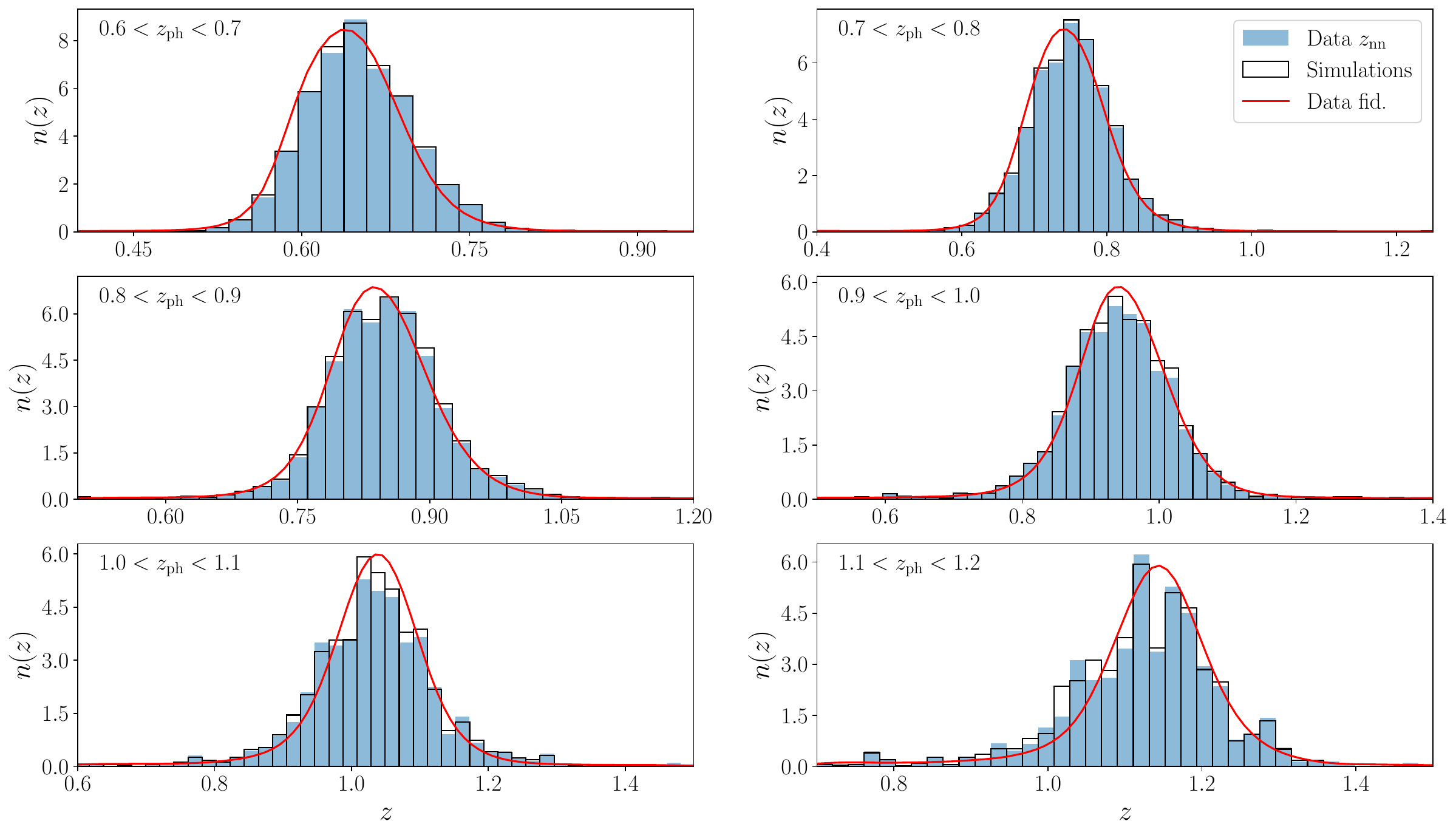}  
\caption{Redshift distributions of the 6 tomographic bins, split by $\zph$. In red, we show the fiducial $n(z)$ assumed for the Y6 BAO sample, see how these are constructed in \autoref{sec:redshifts}.  The blue histograms show the $n(\zmc)$ obtained from the DNF nearest neighbor estimation. $\zmc$ is the assumed input distribution for constructing the ICE-COLA simulations, whose final $n(z)$ distribution is shown as an empty histogram (black outline). }
\label{fig:nz}
\end{figure*}

As explained in \autoref{sec:sample}, we split our sample in tomographic redshift bins using the redshift estimation \zmean from DNF, which we describe below. But this estimation of redshift has a non-trivial uncertainty associated to it. This implies that the distribution of redshifts estimated from \zmean , $n(\zph)$, will not correspond to the true underlying distribution, $n(z)$, that will be more spread. In this section we study different ways to characterize that underlying distribution.

We consider the following methods:

\begin{itemize}
    \item Directional Neighborhood Fitting ({\bf DNF}, \cite{DNF}). This method computes the photometric redshift of each galaxy by comparing its colors and magnitudes to those of the training sample. For that, DNF uses a nearest neighbor algorithm with a directional metric that accounts simultaneously for magnitude and color. 
    It provides several outputs: 
    \begin{itemize}
        \item {{\textsc DNF\_Z}} \footnote{In previous papers and databases this was named {\textsc Z\_MEAN} inherited from other methods in which the main estimate of the redshift was the mean of a PDF.}, hereafter  \boldsymbol{{$\zph$}}, is computed from a regression on magnitude space. The regression is fitted from a set of neighbors from a reference sample of galaxies whose spectroscopic redshifts are known. That is the main estimate provided by the algorithm.
        \item {{\textsc DNF\_ZN}} \footnote{In previous papers and databases this was named {\textsc Z\_MC} inherited from other methods in which a secondary estimate of the redshift  was Monte-Carlo sampled from the PDF.} hereafter \boldsymbol{$z_{\rm nn}$} is the redshift of the nearest neighbor. The $\zmc$ stacking provides a good estimation for the redshift distribution $n(z)$ whenever the training sample is complete.
        \item {\bf PDF} provides the photometric redshift distribution for each galaxy computed from the residuals of the fit.
    \end{itemize}
    For the galaxies whose spectrum is in the redshift calibration sample, we do exclude this information in order to compute the summary statistics above ($\zph$, $\zmc$). 
    
    \item {\bf VIPERS} spectroscopic direct calibration. The VIPERS spectroscopic sample is complete for $i<22.5$ and $z>0.6$ \cite{vipers}. Since the DES footprint contains all of the area of VIPERS, we can construct a matched catalogue in the overlapping area (16.3 deg$^2$) and measure directly the $n(z)$ from a histogram of the spectroscopic redshifts. The limitation from this method comes from the effect of the sampling variance in this limited area when trying to extrapolate to the entire $>4,000$ deg$^2$ footprint.
    
    \item Clustering Redshift ({\bf WZ}) 
    Clustering redshift is a measurement where a sample of galaxies with unknown redshifts (in our case, the photometrically measured BAO galaxies) is angularly correlated with a sample of galaxies where the redshifts are known (a spectroscopic sample). Due to the clustering of galaxies, galaxies that are at the same redshift will tend to have a strong angular correlation compared to chance. Thus, computing the angular correlations of the BAO sample and spectroscopic samples at many thin redshift bins can give us a measure of the redshift distribution of the BAO sample. 

For our clustering redshift measurements, we utilize the final Baryon Oscillation Spectroscopic Survey (BOSS) LOWZ and CMASS galaxy samples \cite{reid16} and the final eBOSS, ELG \cite{raichoor21}, LRG and QSO samples \cite{ross20}. This is the same set of spectroscopic galaxies used for clustering redshifts in \cite{cawthon22}. These samples overlap approximately 15\% of the DES Y6 footprint. The methodology for the clustering redshifts measurements used here is nearly identical to that of \cite{cawthon22}, including choices of scales used, methods of uncertainty estimation and galaxy bias correction. 

\end{itemize}

We remark that the three different methods are largely independent of one another. A thorough description and comparison of these calibrations and combinations is performed in \cite{Mena23}, together with the description of the method used as the final choice for fiducial $n(z)$. 

\subsubsection*{Fiducial redshift distribution.} 

Our fiducial choice of $n(z)$ calibration combines the DNF information coming from the PDF with either WZ or VIPERS. We take the DNF PDF as the shape of our $n(z)$ to profit from its smoothness, although we know that this curve tends to overestimate the spread. On the other hand, we consider that for redshift bins 1-4 ($z_{\rm ph}<1$), WZ provides the most robust estimation of the mean and width of the distribution. Hence, we use the Shift-and-Stretch technique (see \cite{cawthon22,porredon22,Mena23}): we displace the PDF $n(z)$ distribution and widen/narrow it until it best fits a target $n(z)$ distribution, which in this case is the WZ. 
For $z_{\rm ph}>1$ there are not enough spectroscopic galaxies for precise enough WZ measurements and we trust better VIPERS direct calibration to estimate the mean and width of the distribution. Hence, for the bins 5-6 we use the PDF DNF \textit{shifted} and \textit{stretched} with VIPERS as target. 

The fiducial $n(z)$ is shown in red in \autoref{fig:nz}, compared to the data DNF $\zmc$ distribution (blue histogram) and to the simulations $n(z)$ (empty histogram), which is constructed to match the data $z_{\rm nn}$ distribution. More details about the simulations are found in \autoref{sec:sims}. The other $n(z)$ distributions mentioned in this section are shown in the companion paper \cite{Mena23}.

\subsubsection*{Calibration for PCF}
\label{sec:photoz_PCF_calibrate}

Whereas for two of our analyses (ACF, APS) we only use angular information, for the PCF method, we make use of the radial position of galaxies. In the methodology developed in \cite{PCF_method_Y3}, we model the 3D clustering as a weighted sum of the angular clustering in thin redshift bins. As part of this modeling, we require that we have the $n(z)$ distribution in thin $\zph$ bins. 

While redshift bins of equal width were considered in \cite{PCF_method_Y3}, here we increase the bin width with redshift because the photo-$z$ quality deteriorates substantially at high $z$, especially at $z\gtrsim 1$. The bin widths are set in geometric sequence with a ratio of 1.078 so that there are 22 bins in the range $0.6<\zph<1.2$. We follow exactly the same methodology as above. The first 17 bins (up to $\zph=1.02$) are calibrated with WZ and the remaining ones by VIPERS. We refer the readers to an appendix of \cite{Mena23} for more details.

\section{Simulations}
\label{sec:sims}

In order to create the galaxy mock catalogues (from now, \textit{mocks}) for the validation of the BAO analysis, we follow a practically identical approach as in Y3, but now calibrated on the Y6 sample. Hence, we describe the methodology briefly here and refer the reader to \cite{y3-baomocks} for more information. Part of the methodology to construct these mocks builds upon the methodology developed for Y1 \cite{DESY1baomocks}.

We created a set of 1952 mock catalogues, closely reproducing several crucial data attributes, including the observational volume, galaxy abundance, true and photometric redshift distribution, and clustering as a function of redshift.

To achieve this, we employed the ICE-COLA code \cite{ice-cola}, conducting 488 fast quasi-N-body simulations. These simulations utilize second-order Lagrangian Perturbation Theory (2LPT) in conjunction with a Particle-Mesh (PM) gravity solver. Our ICE-COLA algorithm extends the capabilities of the COLA method \cite{cola}, enabling on-the-fly generation of light-cone halo catalogues and weak lensing maps. 

Each simulation involving $2048^3$ particles enclosed within a box of 1536 Mpc$h^{-1}$ by side. In order to enhance our statistical power while keeping computational resources manageable, we replicated this volume 64 times using the periodic boundary conditions, effectively creating a full-sky light-cone extending up to redshift $z=1.43$. In these simulations, the ICE-COLA universe has the same cosmology as the benchmark MICE \textit{Grand Challenge} simulation \cite{MICE1,MICE2} (used for validation): $\Omegam=0.25$, $\OmegaL=0.75$, $\Omegab=0.044$, $n_s=0.95$, $\sigma_8=0.8$ and $h=0.7$.

Generating the galaxy mocks entailed populating halos based on a hybrid Halo Occupation Distribution and Halo Abundance Matching model, using two free parameters per tomographic bin. 
We calibrated these parameters through automatic likelihood minimization to match the clustering of the data. 
For that we use 3 points of the angular correlation at $0.5<\theta<1.0$ deg, while the rest of the correlation function was kept blinded.
Additionally, we derived photometric redshifts for the mock galaxies by applying a mapping between the true redshift and the observed redshift $\zph$. 
This mapping is constructed from the 2D histogram $N(\zph$, $\zmc)$ of the data with DNF, and assuming that it is a good representation of the $N(\zph$, $z_{\rm true})$. 
This choice is different to Y3, where we used $N(\zph$, $z_{\rm vipers})$ to characterize the redshift distribution. However, in Y6, we found that this characterisation is noise dominated in the higher redshift bins. 

Finally, we applied four non-overlapping Y6 footprint masks on each full-sky halo catalogue to multiply the number of galaxy mocks by four, allowing us to validate our analysis down to an increased accuracy.

\begin{figure*}
\includegraphics[width=\linewidth]{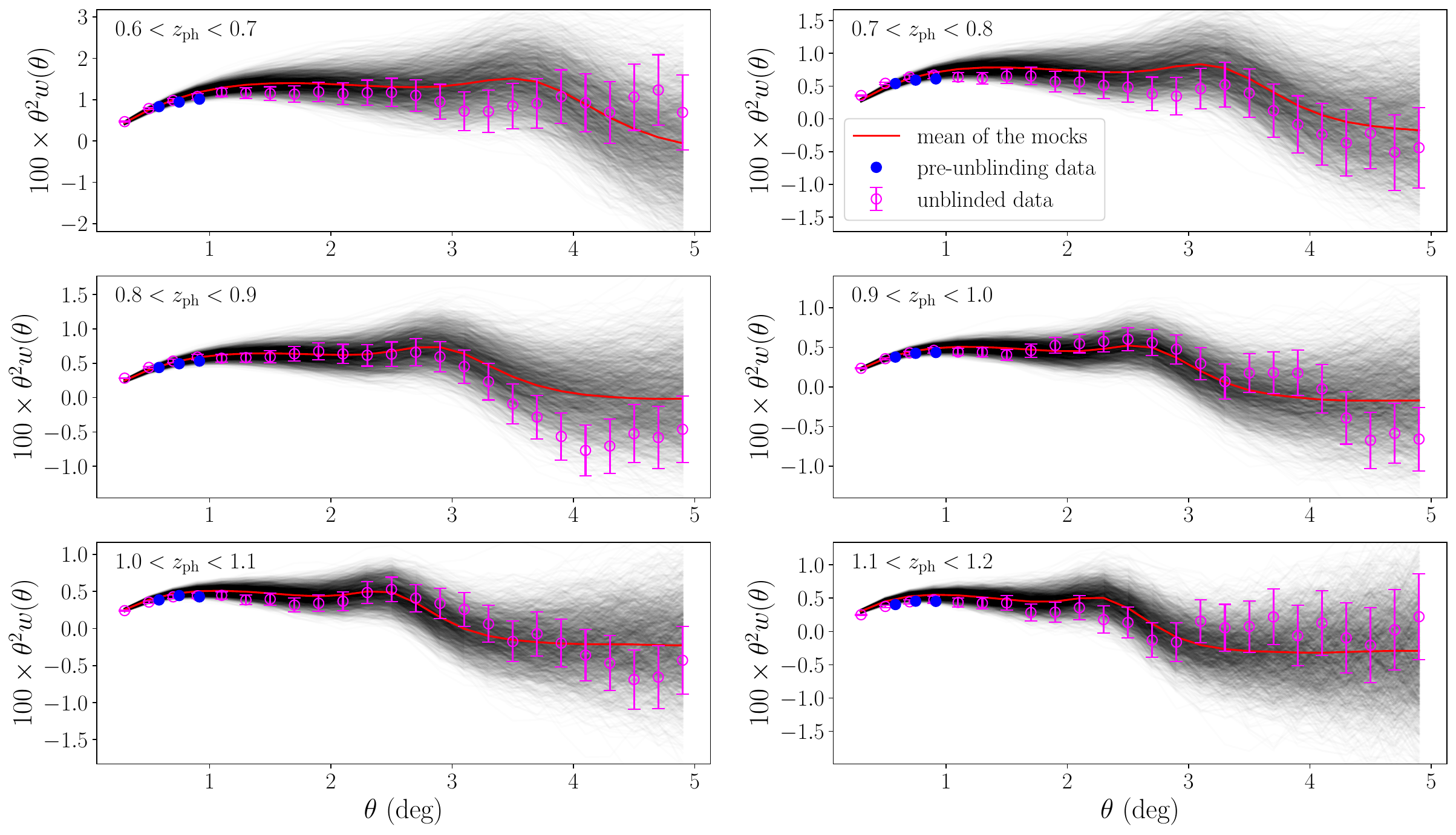}  
\caption{Angular correlation function of the mocks compared to the data. In red we show the mean of the ACF of all the mocks, whereas in black we show all the individual ACF of the 1952 \mocks. On filled circles we mark the 3 data points used for calibration prior to unblinding, whereas on empty circles we can see full unblinded ACF, with the fiducial error bars by \cosmolike (see \autoref{sec:covariance}).  
}
\label{fig:mocks_clustering}
\end{figure*}

As we already showed in \autoref{fig:nz}, the agreement between the $n(z)$ distribution of the mocks and the data $\zmc$ is excellent up to some noise. This is expected by the way we constructed the redshift errors on the mocks from the $N(\zph,\zmc)$ distributions. 
On the other hand, in \autoref{fig:mocks_clustering}, we show the galaxy clustering comparison of data versus mocks, finding  a good level of visual agreement. When comparing the galaxy biases (shown in \autoref{sec:setup} and mathematically introduced in \autoref{eq:pkmu}) some bins show some level of disagreement, partially due to the limitation in the number of scales and partially because of using a limited number of mocks for the calibration for the sake of reducing computing resources (see \cite{y3-baomocks} for details). Part of this disagreement may also come from using slightly different scales for the bias measurements and because the cosmology of the mocks will likely not correspond to the underlying one in the data. 
Additionally, when comparing the ACF of mocks and data, we find $\chi^2/d.o.f.=125/107$ for $\theta\in[0.5,5]$deg and $\chi^2/d.o.f.=89/95$ for $\theta\in[1,5]$deg, which indicates a good agreement, especially at large scales. Given this good agreement at the scales used for fitting the BAO, 
we do not expect that having a different best fit bias will affect the usage of the mocks for the purposes described below.

These simulations will have a crucial role to make different analysis choices and to validate the analysis pipeline. Generally, they will not be used for the covariance estimation, because we showed in \cite{y3-baomocks} that the replication of boxes explained above leads to significant spurious correlations between parts of the data vector. Our baseline covariance will be computed from theory using \cosmolike, see \autoref{sec:covariance}.

\section{Analysis methodology}
\label{sec:methods}

The methodology for the analysis we follow is very similar to that in Y3 \cite{y3-baokp}, with the main exception that we now include the projected correlation function, $\xi_{p}(s_\perp)$. 

\subsection{Analysis setups}
\label{sec:setup}

We consider three different main analysis setups for our analysis and validation, varying the cosmology, $n(z)$ and galaxy bias, depending on the particular needs of a particular analysis. 
We have one setup more oriented to test our methodology on the mocks (\textit{\mocklike}), our fiducial setup for the data assuming Planck cosmology (\textit{Data-like}) and a variation of it with the cosmology of the mocks (\textit{Data-like-mice}), all described below. 

\begin{itemize}

\item \textit{Mock-like}. The mocks are based on MICE Cosmology and we will be assuming it in this setup: $\Omegam=0.25$, $\OmegaL=0.75$, $\Omegab=0.044$, $n_s=0.95$, $\sigma_8=0.8$, $h=0.7$ and $M_\nu=0$ eV. The redshift distribution assumed is that of the mocks (empty histograms in \autoref{fig:nz}), which is based on $\zmc$.
Finally, we use the galaxy bias measured on the mock catalogues using the ACF in $1.5< \theta <5$ deg: $b~=~[1.663, 1.547, 1.633, 1.793, 2.038, 2.446]$ for the six bins, respectively.

\item \textit{Data-like}. This is the default setup for the data analyses. We assume Planck cosmology ($\Omegam=0.31$, $\OmegaL=0.69$, $\Omegab=0.0481$, $n_s=0.97$, $\sigma_8=0.8$, $h=0.676$ and $M_\nu=0.06 e$V \cite{Planck}), the fiducial $n(z)$ (combination of DNF PDF with either WZ or VIPERS, see \autoref{sec:redshifts}) and the galaxy bias measured on the data at angles $0.5< \theta <2$ deg: $b~=~[1.801, 1.805, 1.813, 1.957, 2.113, 2.413]$. 

This measurement of the bias was produced just before we started running the pre-unblinding tests, once the data validation was considered finalized, a much later stage than when the mocks were constructed. 

\item \textit{Data-like-mice}. This auxiliary setup is used to check how results change when assuming MICE cosmology. For that, we still assume the fiducial $n(z)$ and we refit the bias on the data, obtaining $b=[1.650, 1.640, 1.640, 1.752, 1.873, 2.108]$.  For comparison to the bias obtained in the mocks (\mocklike), the error on these biases (which will be larger than for \mocklike, since we here we are only using the scales $0.5< \theta <2$ deg) are $\sigma_b=[0.042, 0.044, 0.046, 0.050, 0.067, 0.102]$.

\end{itemize}

Some particular tests will require hybrid auxiliary setups that we will specify, but the majority of the analyses are run using one of the three above, especially the first two.

\subsection{Clustering measurements}
\label{sec:measurements}

\subsubsection{Random catalogue}

The starting point to measure all clustering statistics is the creation of a random catalogue with 20 times as many objects as our sample. The random catalogue is created by sampling the mask described in \autoref{sec:mask} with a \textsc{healpix nside} of 4096. We down-sample the pixels according to their fraction of coverage, which we remind the reader is always larger than 80\%. 

In \autoref{sec:stellar_contamination} below, we explain how we correct the clustering measurements from additive stellar contamination, quantified by $f_{\rm star}$. The method proposed there is equivalent to assigning all the objects in the random catalogue a weight of $1/(1-f_{\rm star})$.

\subsubsection{Angular correlation function: $w(\theta)$}

Once we have the random sample, the 2-point angular correlation function is estimated using the Landy-Szalay estimator \cite{LS1993}
\begin{equation}
w(\theta) = \frac{DD(\theta)-2DR(\theta)+RR(\theta)}{RR(\theta)},
\end{equation}
where $DD$, $DR$ and $RR$ are the normalized counts of data-data, data-random and random-random pairs, respectively, separated by 
$\theta\pm\Delta\theta/2$, with $\Delta\theta$ being the bin size. 
We start by computing the ACF with a bin size of $\Delta\theta=$0.05 degrees, which is the minimum bin that we consider, but the pair counts can be later combined in broader bins. Eventually, after testing different bin sizes in \autoref{sec:mocktests}, our default binning is set to $\Delta\theta=0.20$ deg.
We can see in \autoref{fig:mocks_clustering} that the BAO feature is located at $\sim 3$ deg and has a width of around $1$ deg. Hence, any of these configurations is able to resolve it. We will be considering a maximum separation of $5$ deg.

Before unblinding, we compared the ACF measurements with two different codes: TreeCorr \cite{2004MNRAS.352..338J} and CUTE \cite{cute}. The $\chi^2$ between the two measurements (with the full covariance) is found to be 0.05 and its root mean square relative error is $\sqrt{\frac{1}{N}\sum \left( \Delta w / \sigma\right)^2}=0.006$. With this excellent agreement and with the more detailed comparison performed in Y3, we consider the data vector to be validated.

\subsubsection{Angular power spectrum: $C_\ell$}

To estimate the clustering signal of galaxies in 
harmonic space, we use the Pseudo-$C_\ell$ (PCL) estimator \citep{2002ApJ...567....2H}.
In particular, we use the the \namaster\footnote{\url{https://github.com/LSSTDESC/NaMaster}} implementation \citep{2019MNRAS.484.4127A}.
We commence by constructing tomographic galaxy overdensity maps using the \healpix pixelization scheme at a resolution parameter of ${\rm NSIDE} = 1024$.
This corresponds to a mean pixel size of $\sim 0.06$ degrees, at least one order of magnitude below the expected angular separation of the BAO signal.
The equal-area pixelization facilitates the computation of galaxy overdensity maps as follows:
\begin{equation}
\label{eq:overdensity-map}
    \delta_p = \frac{N_p \sum_{p'} w_{p'}}{w_p \sum_{p'} N_{p'}} - 1
\end{equation}
where $N_p = \sum_{i\in p} v_i$ gives the weighted number of galaxies at a given pixel $p$, with $v_i$ representing the weight associated with the $i$-th galaxy as given by the systematics weights, \autoref{sec:systematics}.
Whereas $w_p$ gives effective fraction of the area covered by the survey at pixel $p$, as given by our mask, \autoref{sec:mask}.

The inherent discreteness of galaxy number counts introduces a shot-noise contribution to the auto-correlation galaxy clustering spectra, also known as noise-bias. 
We assume this noise to be Poissonian, and estimate it analytically following \citep{2019MNRAS.484.4127A,2020JCAP...03..044N,2021JCAP...10..030G}.
Subsequently, we subtract this estimated noise-bias from our power spectrum estimates.
Any deviations from the Poissonian approximation are expected in the form of an additive constant and are anticipated to be captured by broad-band terms in our template, having minimal impact on the BAO feature detection.

We bin the angular power spectrum estimates into bandpowers, assuming uniform weighting for all modes within each band.
Employing a piecewise-linear binning scheme, we construct contiguous bandpowers with varying bin widths of $\Delta\ell = 10$, $20$ and $30$, ranging from $\ell_{\rm min}=10$ up to $\ell=2048$. 
This binning strategy ensures adequate signal-to-noise ratios across the bandpowers while maintaining flexibility for scale cuts, see \autoref{tab:APS_mocks} for different analysis choices on the mocks.

After testing on the mocks, we adopted as fiducial choices $\ell_{\rm min}=10$, $\Delta\ell=20$ and an $\ell_{\rm max}$ scale-cut approximately corresponding to a $k_{\rm max}=0.211\, {\rm Mpc}^{-1}$ under the Limber relation, $k_{\rm max} = \ell_{\rm max} / r(\bar{z})$, evaluated at the mean redshift of each tomographic bin and the fiducial cosmology of the mocks.
We have verified that changing the cosmology to the Planck one does not introduce significant changes on our scale-cuts.
This $\ell$-binning allows us to resolve approximately five BAO cycles on each redshift bin (\autoref{fig:APS_data}). 
The resulting $\ell_{\rm max}$ values for each redshift bin are 510, 570, 630, 710, 730 and 770.
Finally, when constructing the likelihood, we consistently bin the theory predictions into the same bandpowers of the measurements following \cite{2019MNRAS.484.4127A}.

\subsubsection{Projected correlation function: $\xi_{p}(s_\perp)$}
\label{subsec:projectedclustering}

The Projected Correlation Function (PCF) method starts by computing the full 3D correlation function in terms of the \textit{observed} (in $\zph$-space) comoving distance between any pair of galaxies (or randoms or galaxy-random) along and across the line of sight: $s_{\parallel}$, $s_\perp$. 
For that, we transform $\zph$ to comoving distances using a fiducial cosmology (see \autoref{sec:setup} for the two cosmologies considered). 
Once we have that, we compute the anisotropic 3D correlation function in a similar way to the ACF, with the Landy-Szalay estimator: 
\begin{equation}
    \xi(s_{\perp},s_\parallel) = \frac{DD(s_{\perp},s_\parallel)-2\cdot DR(s_{\perp},s_\parallel) +RR(s_{\perp},s_\parallel)}{RR(s_{\perp},s_\parallel)} \, .
\end{equation}

We use a binning of $\Delta s_{\perp} = 1$ Mpc/$h$ and $\Delta s_{\parallel} = 1$ Mpc/$h$ (again, we recombine the pair-counts at a later step to obtain broader bins in $s_\perp$: $\Delta s_\perp=5 h^{-1}$Mpc). We compute these correlations both with \textsc{CUTE} and with \textsc{pycorr}\footnote{\url{https://github.com/cosmodesi/pycorr}}, finding good agreement between the two but the latter to be considerably faster and adopting it for our analysis. 

Once we have the 3D clustering, we integrate over the line of sight to obtain the PCF:
\begin{equation}
\xi_p(s_{\perp})= \frac{ \int_0^{1} W(\mu) \xi\big(s_\perp(s,\mu),s_\parallel(s,\mu)\big)\ d\mu }  { \int_0^{1} W(\mu) d \mu }  ,
\label{eq:3D}
\end{equation}
where $\mu$ is the orientation with respect to the line of sight ($\mu =$cos$\theta$, with tan$\theta$=$s_\parallel/s_\perp$) and $W(\mu)$ a weighting function that can be optimized.
Here, we follow an approach that is different to that of Y1 \cite{Y1_BAO} and Y3 \cite{y3-baokp} key papers, which were based on the methodology proposed in \cite{Ross:2017}, with a method to obtain the $W(\mu)$ from Fisher information. 
On follow-up analyses of Y3, we developed and applied a new version of the method that was able to account for non-Gaussian distribution of the redshift errors \cite{PCF_method_Y3,PCF_Y3_BAO}, unlike previous analyses. To increase the signal-to-noise and stability of the analyses, we apply a cut-off Gaussian weighting \cite{PCF_Y3_BAO}: 
\begin{equation}
\label{eq:cutoffGaussian_stacking}
W(\mu)=W_{\rm G}( \mu; \sigma_{\mu},\mu_{\rm max})  =
    \begin{cases}
      \exp \big({ - \frac{ \mu^2 }{ 2 \sigma_\mu^2 }   } \big)   &  \textrm{ if  $ \mu < \mu_{\rm max }  $} , \\
        0 &   \text{otherwise , } \\
    \end{cases}
\end{equation}
with $\mu_{\rm max } = 0.8$  and $\sigma_\mu=0.3$.

One of the advantages of PCF is that the BAO is always seen at the same position in  $s_\perp$ for different redshifts, if the assumed cosmology is roughly correct, contrary to ACF or APS. This allows us to consider all the tomographic bins together without losing too much information, i.e. the data compression is close to optimal. This was the approach taken in \cite{Y1_BAO,y3-baokp,PCF_Y3_BAO}. Here, we will keep this approach for visualisation purposes (in order to see one line with all the BAO SNR on it: left panel of \autoref{fig:PCF_data}), but not for the default BAO analysis. During the validation of the method with Y6 mocks, we found slightly tighter constraints on the BAO when considering the $N_z=6$ tomographic bins for the clustering measurements. This is expected given for $N_z = 6 $ we essentially combine data at the level of likelihood rather than data vector \cite{PCF_Y3_BAO}.   This also eases the comparison with the ACF and APS when we study isolating/removing one specific bin or similar tests. 

\subsubsection{Correcting for additive stellar contamination}
\label{sec:stellar_contamination}

As mentioned in \autoref{sec:sample}, we have quantified that between $f_{\rm star}=$0.7\% and 3.3\%, depending on the tomographic bin, of our objects are actually stars.  This has a multiplicative effect that is corrected with the systematic weights described in \autoref{sec:systematics} as other foregrounds or observational condition maps. However, stars have also an additive contribution to the observed number density of galaxies due to contamination. 
To first order, these stars can be considered un-clustered objects that contribute both to RR and DD equally, diluting all 2-point functions by a factor $(1-f_{\rm star})^2$. 
For this reason, we correct our measurements of $w(\theta)$, $C_\ell$ and $\xi_p(r_\perp)$ with a factor $(1-f_{\rm star})^{-2}$ more details in our companion paper \cite{Mena23} and in \cite{Krolewski:2019yrv,LSST:2019wqx}. 

This correction reaches up to a $6.5\%$ level on the clustering amplitude, although
we do not expect this correction to affect the measurement of the BAO that has a parameter $B$ absorbing the amplitude of the clustering (see \autoref{eq:template_all_parameters} below). 
Nevertheless, we include this correction in all our measurements from the data. 

\subsection{BAO template}
\label{sec:template}

Our approach to measure the BAO distance is based on a template fitting method.
In order to generate the BAO template for our observables, we first need to generate a reliable model for the 3D power spectrum ($P(k)$), from which the projected/angular clustering can be computed. For that, we follow the same methodology as in Y3, summarized below. 

We start from the linear power spectrum $P_{\rm lin}(k)$ generated by {\sc Camb} \citep{2000ApJ...538..473L}. The main modification to this model comes from the inclusion of the BAO peak broadening due to non-linearities \cite{CrocceScoccimarro_2008,PadmanabhanWhite_2009}. We model this by splitting the power spectrum into a {\it no-wiggle} ($P_{nw}$) and a {\it wiggle} ($P_{\rm lin}-P_{nw}$) component and smoothing the {\it wiggle} component anisotropically via $\Sigma$:
\begin{equation}
P(k,\mu) = (b+\mu^2 f)^2\left[ (P_{\rm lin}-P_{\rm nw})e^{-k^2\Sigma^2}+P_{\rm nw}\right],
\label{eq:pkmu}
\end{equation}
where we have also included the effect coming from galaxy bias ($b$) and redshift space distortions ($\mu^2 f$) \citep{1987MNRAS.227....1K}, with the latter one proportional to the growth rate $f$. 

We model the non-wiggle component using a 1D Gaussian smoothing in log-space following appendix A of \cite{2016JCAP...03..057V}. We also follow the infrared resummation model \cite{2015JCAP...02..013S,2016JCAP...07..028B} to compute the damping scale $\Sigma(\mu)$ \cite{2020JCAP...05..042I,2018JCAP...07..053I}, with respect to  the line of sight (see details in \cite{y3-baokp}).

Once we have a $P(k,\mu)$, we can decompose it into multipoles, $P_\ell(k)$, perform a Hankel transform to obtain the configuration space multipoles, $\xi_\ell (s)$, and then reconstruct the anisotropic redshift-space correlation function $\xi(s,\mu)$. 
From there, the angular correlation function is obtained by projecting 3D clustering into the angle subtended by two galaxies in the celestial sphere $\theta$. For that, we weight  $\xi(s,\mu)$ by the redshift distribution $n(z)$ (normalized to integrate to 1) of each tomographic redshift bin in a double integral:
\begin{equation}
\label{eq:w_template_schematic}
w(\theta) = \int dz_1\int dz_2 n(z_1) n(z_2)\xi \big( s(z_1,z_2,\theta),\mu(z_1,z_2,\theta) \big).
\end{equation}

We then compute the $C_\ell$ template by evaluating  $w(\theta)$ in 300 logarithmic spaced points from $0.001\deg$ to $179.5 \deg$ and transforming it to the harmonic space: 
\begin{equation}
\label{eq:Cl_template_Ltransform}
C_\ell = 2 \pi \int_{-1}^{1} d(\cos \theta) \,  w(\theta)  L_\ell (\cos \theta)
\end{equation}
where $L_\ell$ is the Legrendre polynomial of order $\ell$.  

Our modeling of the PCF starts by computing the general 
auto and cross-correlations ACF $w_{ij}$  using \autoref{eq:w_template_schematic} in thin photo-$z$ bins, whose calibration has been described in \autoref{sec:photoz_PCF_calibrate}. Then, we map the general ACF to PCF by
\begin{equation}
\label{eq:3D_model}
    \xi_p(s, \mu) = \frac{\sum_{ijk} f_{ijk} w_{ij}(\theta_k;z_i, z_j)}{\sum_{ijk} f_{ijk} } \, ,
\end{equation}
where $f_{ijk}$ denotes the weight accounting for the number of the cross bin pairs in  $w_{ij}(\theta_k)$ falling into the $s$ and $\mu$ bins. As for the data measurements, we project $\xi_{p}(s, \mu) $ to the transverse direction using weight \autoref{eq:cutoffGaussian_stacking} to get $\xi_{p}(s_\perp)$. We refer the reader to \cite{PCF_method_Y3} for more details. 

This way, the three templates corresponding to ACF, APS and PCF are all derived consistently.

Finally, our model ($M$) will contain the BAO template component ($T$) described above for $w(\theta)$, $C_\ell$, or $\xi_p(s_\perp)$, an amplitude rescaling factor $B$ and a smooth component $A(x)$:
\begin{equation}
\label{eq:template_all_parameters}
M(x) = B T_{\rm BAO, \alpha}(x^\prime) + A(x). 
\end{equation}

The term $A(x)$ is introduced to absorb smooth (not a sharp feature) components that may come from remaining theoretical or observational systematic errors in the clustering. We will model it as a sum of power laws and we will study in \autoref{sec:mocktests} what option gives the best behaviour when fitting the BAO  on the mock catalogues. 

In the case of the ACF, we have $x=\theta$ and $T$ corresponding to $w$ as given by \autoref{eq:w_template_schematic}. The rescaled coordinate $x^\prime$ is $\alpha \theta$, where $\alpha $ is the BAO shift parameter containing the cosmological information from the fit, and the function $A$ is modeled as
\begin{equation}
\label{eq:broadband_terms_w}
A(\theta)  = \sum_i \frac{ a_i }{ \theta^i }.
\end{equation}

For the APS, $T$ is the $C_\ell$ obtained from \autoref{eq:Cl_template_Ltransform},  $x=\ell$, $x^\prime= \ell / \alpha $, and  $A$ is 
\begin{equation}
\label{eq:broadband_terms_Cl}
A(\ell)  = \sum_i  a_i  \ell^i . 
\end{equation}

Finally, for the PCF,  $T(x)$ denotes $\xi_{\rm p}( s_\perp )$ from \autoref{eq:3D_model}, $x'=\alpha s_\perp$ and $A$ is
\begin{equation}
    A(s_{\perp})=\sum_i  \frac{ a_i}{s_{\perp}^{\ i}} \, . 
    \label{eq:broadband_terms_xi}
\end{equation}

\subsection{Covariance matrix}
\label{sec:covariance}

\subsubsection*{Covariance for ACF and APS}

Following our approach for the BAO analysis from Y3 data \cite{y3-baokp}, our fiducial covariance matrices are estimated analytically, using the \cosmolike code for ACF and APS \citep{cosmolike,cosmolike2020,cosmolike_curvedsky}. The covariance of the angular correlation function $w(\theta)$ at angles $\theta$ and $\theta'$ is related to the covariance of the angular power spectrum by
\begin{equation}
\begin{aligned}
&\mathrm{Cov}(w(\theta),\, w(\theta')) = \\
 &\sum_{\ell, \, \ell'}\dfrac{(2\ell +1)(2\ell'+1)}{(4\pi)^2}\overline{P_{\ell}}(\theta)\overline{P_{\ell'}}(\theta')  \mathrm{Cov}(C_{\ell}, C_{\ell'}) ,
\end{aligned}
\end{equation}
where $\overline{P_{\ell}}(\theta)$ are the Legendre polynomials averaged over each angular bin $[\theta_{\min}, \, \theta_{\max}]$ and are defined by
\begin{equation}
\overline{P_{\ell}} = \dfrac{\int_{x_{\min}}^{x_{\max}} \der x \, P_{\ell}(x) }{x_{\max} - x_{\min}} = \dfrac{\left[ P_{\ell+1}(x) - P_{\ell - 1}(x) \right]_{x_{\min}}^{x_{\max} }}{(2\ell+1)(x_{\max} - x_{\min}) },
\end{equation}
with $x= \cos \theta$ and $x_{\{\min, \, \max\}} = \cos \theta_{\{\min, \max\}}$ (see e.g. \cite{Crocce2011} for more details).

We have tested that including non-Gaussian contributions to the covariance estimation, such as the trispectrum and the super-sample covariance terms, does not impact our results. Given that, the Gaussian covariance of the angular power spectrum in a given tomographic bin is given by \cite{Crocce2011,cosmolike}
\begin{equation}
\mathrm{Cov}(C_{\ell}, C_{\ell'}) =  \dfrac{2\delta_{\ell \ell'}}{f_{\rm sky}(2\ell + 1)}\left(C_{\ell'} + \frac{1}{n_g}\right)^2,
\label{eq:cov-c_ell}
\end{equation}
where $\delta$ is the Kronecker delta function, $n_g$ is the number density of galaxies per steradian, and $f_{\rm sky}$ is the observed sky fraction,  which is used to account for partial-sky surveys. However, we go beyond the $f_{\rm sky}$ approximation by taking into account how the exact survey geometry suppresses the number of pairs of positions in each angular bin $\Delta \theta$ (see \cite{cosmolike_mask} and appendix C of \cite{y3-covariances} for more details). Redshift space distortions are included through the $C_{\ell}$'s in \autoref{eq:cov-c_ell}.

In the context of harmonic space analysis, we commence by employing the \cosmolike predictions for the angular power spectra.
Subsequently, we calculate analytical Gaussian covariance matrices that account for broadband binning and partial sky coverage within the context of the PCL estimator, as outlined in \cite{2004MNRAS.349..603E,2019JCAP...11..043G}.
The coupling terms are computed using the \namaster implementation \cite{2019JCAP...11..043G,2019MNRAS.484.4127A}.

Similarly to Y3, we have validated the \cosmolike covariance with estimates from the ICE-COLA, FLASK mocks \cite{Xavier:2016elr} and also with the covariance developed in \cite{Chan:2018gtc}, finding consistent results  (see \textcolor{blue}{Tables}~\ref{tab:ACF_mocks}, \ref{tab:APS_mocks} and \ref{tab:PCF_mocks}). 

\subsubsection*{Covariance for PCF}

For the projected correlation function, we also rely on a theoretical covariance. In this case the method follows \cite{PCF_method_Y3}, which builds up on the covariance for ACF developed in \cite{Chan:2018gtc}. That latter ACF covariance follows a similar approach as the \cosmolike\ one explained above and has been validated against that code during this study. Furthermore, in line with the \cosmolike\ covariance, we have included the mask correction as well \cite{cosmolike_mask}. 

Following from \autoref{eq:3D_model}, and using the same $f_{ijk}$ coefficients described there, we can simply construct the 3D clustering covariance, $C^{\xi_p}$ as a sum over the angular covariance, $C^w$: 
\begin{equation}
    C^{\xi_p}_{\{s,\mu\}\{s',\mu'\}}=\frac{\sum_{ijk} \sum_{lmn} f_{ijk} f_{lmn} C^w_{\{z_i, z_j,\theta_k\}\{z'_l, z'_m,\theta_n\}}}{\sum_{ijk} f_{ijk} \sum_{lmn} f_{lmn} } \, .
\end{equation}
We then get the covariance for $\xi_{\rm p }(s_\perp) $ by projecting the covariance to the transverse direction using the weight $W_{\rm G} $ in \autoref{eq:cutoffGaussian_stacking}. We do not apply the covariance  corrections introduced in \cite{PCF_method_Y3} as it has little effect on the final results. 

A visual representation of the Y3 covariance for ACF and APS is shown in \cite{y3-baokp}, whereas the PCF covariance is shown in \cite{Chan21}. The Y6 covariances are not shown here but follow a similar structure to those from Y3.

\subsection{Parameter inference}
\label{sec:inference}

Given our data vector $\bm{d}$ from the clustering measurements (ACF, APS or PCF in \autoref{sec:measurements}), the model $\bm{M}$ for a given set of parameters $\bm{p}$  (\autoref{eq:template_all_parameters} in \autoref{sec:template} )  and the covariance ${\bf C}$, (\autoref{sec:covariance}), the $\chi^2$ describes the goodness of fit between the data and the model and it is given by 
\begin{align}
\chi^2(\bm{p}  |  \bm{d}  ) =\sum_{ij} \big[ \bm{d} - \bm{M}( \bm{p} )\big]_i C^{-1}_{\, ij } \big[ \bm{d} - \bm{M}( \bm{p} )\big]_j.
\end{align}

Then, assuming a Gaussian likelihood $\mathcal{L}$, we have  

\begin{align}
\mathcal{L} ( \bm{p} |  \bm{d}  ) \propto e^{ - \frac{\chi^2}{2} } .
\end{align}

We then consider our best fit  as the model with the highest likelihood or, equivalently, lowest $\chi^2$. We follow a similar procedure to \cite{Chan:2018gtc} to minimize the $\chi^2$. This implies, first, to analytically fit the broad-band parameters $A_i$ (\textcolor{blue}{Equations} \ref{eq:broadband_terms_w}, \ref{eq:broadband_terms_Cl} \& \ref{eq:broadband_terms_xi}), profiting from their linear contribution to the model. After that, the $\chi^2$ is numerically minimized with respect to the amplitude $B$. Finally, we end up with a $\chi^2$ as a function of $\alpha$ which is our reported likelihood for each of the three methods (ACF, APS, PCF).

From this point, we consider our error $\sigma_\alpha$ as the half-width of the $\alpha$ region with $\Delta\chi^2=1$ around the minimum. 
If the 1-$\sigma$ region defined this way falls outside the $\alpha \in [0.8,1.2]$ region, then we consider this as a non-detection. 
We will see in \autoref{sec:mocktests} that the individual errors obtained from this method agree reasonably well with the scatter of the best fit $\alpha$.
At the stage of combination of ACF, APS and PCF (\autoref{sec:comb_mocks}), we estimate the covariance of the three best fits from the mocks and implicitly assume that they are Gaussianly distributed. As we will see in \autoref{tab:combination}, the resulting combined measurement (AVG) has an estimated error that captures very well the scatter of the best fit estimate. 

We also tested different ways to report the 1-$\sigma$ error as a summary of the likelihood. 
We could define $\sigma_{\bar \sigma^2}$ as the standard deviation of the likelihood from the second moment or $\sigma_{L68}$ as the half-width of the region that contains 68\% of the integral of the likelihood. These different definitions had some small impact on the error (somewhat larger for APS) but were found not to affect the conclusions drawn in the mocks tests or on the data. 

At this stage we remind the reader that $\alpha$ measures a shift of the BAO position with respect to the BAO position in the template, computed at our fiducial cosmology (defined in \autoref{sec:setup}). This relates to cosmology, through the comoving size of the sound horizon $r_{\rm d}$ and  angular distance $D_{\rm M}$:
\begin{equation}
\label{eq:alpha}
\alpha = \frac{D_M(z)}{r_{\rm d}} \frac{r^{\rm fid}_{\rm d}}{D^{\rm fid}_M(z)}. 
\end{equation}

This equation needs to be evaluated at an effective redshift that we define as
\begin{equation}
    \begin{aligned}
    z_{\rm eff} & = \frac{  \sum_i  w_{i,{\rm sys}}\cdot w_{i,{\rm FKP}} \cdot z_{\rm ph}  }
    {  \sum_i w_{i,{\rm sys}}\cdot w_{i,{\rm FKP}}  } \\
        &= 0.851 ,
    \end{aligned}
\end{equation}
where the $w_{\rm FKP}$ weights are inverse-variance weights that we compute following Eq. 16 of \cite{Ross:2017} and $w_{i,{\rm sys}}$ are the systematic weights described in \autoref{sec:systematics}.

We note that the definition of $\zeff$ is somewhat arbitrary. Different definitions we have tried lead to differences of up to $\Delta \zeff=0.035$. However, since $\alpha$ contains a ratio $D_M(z)$ to $D^{\rm fid}_M(z)$, as long as both functions evolve slowly with redshift, the uncertainty on $\zeff$ does not have much effect on cosmology. For example, comparing $D_M(z)$ from \mice cosmology to $D_M(z)$ from \planck cosmology (already a big change), only leads to a difference of $\Delta \alpha=0.001$ for $\Delta \zeff=0.035$, this is at the level of $1/20\ \sigma_\alpha$. Hence, we choose the definition above for consistency with Y1 and Y3 analyses. 
Finally, if we consider the different redshift calibrations discussed in \autoref{sec:redshifts} and \cite{Mena23}, there is an uncertainty on the mean redshift of the sample of about $\Delta z = 0.004$, this is one order of magnitude below the uncertainty associated to the $\zeff$ definition. 

\subsection{Combination of BAO from ACF, APS and PCF}
\label{sec:comb_method}

We follow the methodology described in \cite{avery1996combining,erler2015combination} to combine our three correlated statistics. We express the covariance matrix between ACF, APS and PCF as
\begin{equation}\label{eq:cov_matrix}
    {\rm COV}_{ij}=\langle\delta\alpha_i\delta\alpha_j\rangle,
\end{equation}
where $i,j\in\{\rm ACF,\rm APS,\rm PCF\}$ and $\delta\alpha_i=\alpha_i-\langle\alpha_i\rangle$, with $\langle\alpha_i\rangle$ the plain average of the three measurements.
We define the optimally weighted average of $\alpha$ as
\begin{equation}\label{eq:alpha_average}
    \alpha_{\rm AVG}=\sum_iw_i\alpha_i,
\end{equation}
where the optimal weights $w_i$ are to  be found. Writing $\delta\alpha_{\rm AVG}=\sum_iw_i\delta\alpha_i$ and using the definition of covariance matrix, \autoref{eq:cov_matrix}, we find
\begin{equation}\label{eq:sigma_alpha}
    \sigma^2_{\rm AVG}\equiv{\rm COV}_{\rm AVG,\rm AVG}=\sum_{ij}w_iw_j{\rm COV}_{ij}.
\end{equation}
To minimize $\sigma^2_{\rm AVG}$ subject to the condition $\sum_iw_i=1$, we use the Lagrange multiplier technique. Writing
\begin{equation}
     \sigma^2_{\rm AVG}=\sum_{ij}w_iw_j{\rm COV}_{ij}+\lambda\left(\sum_iw_i-1\right)
\end{equation}
and setting the derivative of $\sigma^2_{\rm AVG}$ with respect to the $w_i$ and $\lambda$ to 0, we find
\begin{equation}
    w_i=\frac{\sum_k{\rm COV}^{-1}_{ik}}{\sum_{jk}{\rm COV}^{-1}_{jk}}.
\end{equation}
We then calculate the error associated to $\alpha_{\rm AVG}$ via \autoref{eq:sigma_alpha}, but using the errors ($\sigma_{\alpha_i}$) measured on the $\alpha$ for the different estimators instead of the variance from the covariance matrix. Explicitly,
\begin{equation}\label{eq:sigma_average}
    \begin{split}
        \sigma_{\alpha_{\rm AVG}}^2 & = (w_{\rm ACF}^2\sigma_{\alpha_{\rm ACF}}^2 + w_{\rm APS}^2\sigma_{\alpha_{\rm APS}}^2 + w_{\rm PCF}^2\sigma_{\alpha_{\rm PCF}}^2 \\
        & \quad + 2w_{\rm ACF}w_{\rm APS}\sigma_{\alpha_{\rm ACF}}\sigma_{\alpha_{\rm APS}}\rho_{\rm ACF,APS} \\
        & \quad + 2w_{\rm APS}w_{\rm PCF}\sigma_{\alpha_{\rm APS}}\sigma_{\alpha_{\rm PCF}}\rho_{\rm APS,PCF} \\
        & \quad + 2w_{\rm ACF}w_{\rm PCF}\sigma_{\alpha_{\rm ACF}}\sigma_{\alpha_{\rm PCF}}\rho_{\rm ACF,PCF}),
    \end{split}
\end{equation}
where $\rho_{i,j}$ is the cross correlation coefficient measured from the mocks and will be detailed in \autoref{sec:comb_mocks}.

\section{Analysis validation}
\label{sec:validation}

Once we have set up the methodology, we validate it in this section. 
First, we will study the robustness of our method to the choice of redshift calibration in \autoref{sec:redshift_val}. 
Then, in \autoref{sec:mocktests} we will use the simulations presented in \autoref{sec:sims} to validate the accuracy of our methodology for our three estimators: Angular Correlation Function, Angular Power Spectrum and Projected Correlation Function. Finally, in \autoref{sec:comb_mocks}, we validate the method to combine the statistics. 
From these tests, we can derive a systematic error associated to each of the estimators. 

\subsection{Robustness against redshift calibration}
\label{sec:redshift_val}

\begin{table*}
  \caption{Impact of redshift calibration on the BAO estimation in each redshift bin (1-6) and the combination of the 6 (`All'). We generate a mock data vector assuming our fiducial $n(z)$ distribution (and the \datalike setup, \autoref{sec:setup}) and fit for the BAO shift $\alpha$ assuming a different $n(z)$, as marked in the first row. The entries at the body of the table show the best fit and error obtained in each case, following the methodology described in \autoref{sec:methods}. We compute this for each of the 6 tomographic bins (labeled in the first column), presented in 6 tiers, each of them with the results from the three estimators: Angular Correlation Function (ACF), Angular Power Spectrum (APS) and Projected Correlation Function (PCF). A seventh tier contains the results for the combination of all the bins together (`All') and a last entry considers the combination of ACF, APS and PCF into AVG. The different redshift calibrations are described in \autoref{sec:redshifts}.
  }
  \label{tab:redshift_effect}
    \begin{tabular}{llcccccZ}
    \toprule\toprule
     bin & method & fid. & DNF $\zmc$ & VIPERS & WZ & DNF PDF &  \gray{SOMPZ}  \\
     \midrule
      1 & ACF & $1.0001\pm0.0548$ & $0.9899\pm0.0550$ & $0.9931\pm0.0530$ & $1.0014\pm0.0548$ & $0.9849\pm0.0556$ & \gray{$0.9971\pm0.0556$} \\
1 & APS & $1.0000 \pm 0.0617$ & $0.9899 \pm 0.0612$ & $0.9927 \pm 0.0610$ & $1.0009 \pm 0.0623$ & $0.9852 \pm 0.0610$ & \gray{$0.9964 \pm 0.0619$} \\
     1 & PCF & $0.9998 \pm 0.0446 $ & $0.9922\pm 0.0458 $ & $0.9930\pm0.0426$ & $0.9994\pm0.0440$ & $0.9882\pm0.0460$ &  \gray{---} \\

    \midrule
      2 & ACF & $1.0001\pm0.0483$ & $0.9921\pm0.0481$ & $0.9950\pm0.0463$ & $0.9987\pm0.0486$ & $0.9924\pm0.0482$ & \gray{$0.9997\pm0.0529$} \\
2 & APS & $1.0000 \pm 0.0518$ & $0.9920 \pm 0.0514$ & $0.9945 \pm 0.0512$ & $0.9987 \pm 0.0518$ & $0.9924 \pm 0.0514$ & \gray{$0.9995 \pm 0.0537$} \\
  2 & PCF & $0.9998 \pm 0.0426 $ & $0.9938 \pm0.0432$ & $0.9954\pm 0.0408$ & $1.0002\pm0.0426$ & $0.9930\pm0.0436$ &    \gray{---} \\
 
    \midrule
      3 & ACF & $1.0001\pm0.0420$ & $0.9957\pm0.0422$ & $0.9918\pm0.0410$ & $0.9993\pm0.0417$ & $0.9953\pm0.0431$ & \gray{$0.9925\pm0.0421$} \\
    3 & APS & $1.0000 \pm 0.0438$ & $0.9957 \pm 0.0438$ & $0.9914 \pm 0.0435$ & $0.9991 \pm 0.0440$ & $0.9954 \pm 0.0439$ & \gray{$0.9923 \pm 0.0434$} \\   
    3 & PCF & $0.9998 \pm 0.0412$ & $0.9982 \pm 0.0418 $ & $0.9942\pm0.0392$ &  $0.9994\pm0.0406$ & $0.9970\pm0.0426$ &   \gray{---} \\
       
    \midrule
      4 & ACF & $1.0001\pm0.0410$ & $1.0019\pm0.0419$ & $1.0112\pm0.0398$ & $0.9983\pm0.0398$ & $1.0026\pm0.0427$ & \gray{$1.0184\pm0.0442$} \\
4 & APS & $1.0000 \pm 0.0402$ & $1.0017 \pm 0.0408$ & $1.0106 \pm 0.0405$ & $0.9981 \pm 0.0403$ & $1.0025 \pm 0.0408$ & \gray{$1.0182 \pm 0.0421$} \\
      4 & PCF & $0.9998\pm0.0404$ & $1.0026\pm 0.0422$ & $1.0082\pm0.0390$ & $1.0010\pm0.0388$ & $1.0026\pm0.0428$ &  \gray{---} \\
        
    \midrule
      5 & ACF & $1.0001\pm0.0472$ & $1.0030\pm0.0494$ & $0.9985\pm0.0452$ & --- & $0.9991\pm0.0518$ & \gray{$1.0063\pm0.0559$} \\
      5 & APS & $1.0000 \pm 0.0401$ & $1.0030 \pm 0.0409$ & $0.9971 \pm 0.0402$ & --- & $0.9995 \pm 0.0410$ & \gray{$1.0068 \pm 0.0427$} \\
      5 & PCF & $0.9994\pm0.0446$ & $1.0018\pm0.0509$ & $1.0026\pm 0.0434$ & --- & $0.9978\pm0.0507$ &   \gray{---} \\
      
    \midrule
      6 & ACF & $1.0001\pm0.0683$ & $1.0062\pm0.0741$ & $1.0048\pm0.0699$ & --- & $1.0012\pm0.0767$ & \gray{$1.0135\pm0.0874$} \\
      6 & APS & $1.0000 \pm 0.0458$ & $1.0067 \pm 0.0475$ & $1.0047 \pm 0.0466$ & --- & $1.0022 \pm 0.0469$ & \gray{$1.0151 \pm 0.0491$} \\
      6 & PCF & $0.9998\pm0.0831$ & $ 1.0130 \pm 0.0941$ & $1.0234\pm0.0773$ &  --- & $1.0078\pm0.0985$ &   \gray{---} \\

    \midrule
    \midrule
      {\bf All} & ACF & $1.0001\pm0.0201$ & $0.9972\pm0.0206$ & $0.9985\pm0.0195$ & --- & $0.9955\pm0.0210$ & \gray{$1.0027\pm0.0219$} \\
      {\bf All} & APS & $1.0000 \pm 0.0190$ & $0.9988 \pm 0.0194$ & $0.9989 \pm 0.0192$ & --- & $0.9971 \pm 0.0193$ & \gray{$1.0050 \pm 0.0200$} \\
     {\bf All} & PCF & $0.9998\pm0.0202$ & $ 0.9982\pm 0.0214$ & $1.0002\pm0.0196$ &  --- & $0.9962\pm0.0216$ &  \gray{---} \\
     \midrule
     {\bf All} & {\bf AVG} & $0.9998\pm0.0193$ & $0.9984\pm0.0204$ & $1.0001\pm0.0189$ & --- & $0.9965\pm0.0205$ & \gray{$\pm$} \\
    \bottomrule\bottomrule
    \end{tabular}
\end{table*}

As discussed in \autoref{sec:redshifts}, characterizing the redshift distribution of galaxy samples is one of the most important and challenging tasks in photometric surveys. A detailed comparison of different methods to characterize the redshift distribution ($n(z)$) of the 6 tomographic bins is presented in \cite{Mena23} and summarized in \autoref{sec:redshifts}. This results in a series of estimations of $n(z)$ for our tomographic bins, having three estimations largely independent (DNF, VIPERS and WZ). From a combination of those estimates, we obtain our fiducial $n(z)$. 

In this section, we estimate the offset we may obtain in the measured BAO if we assumed one $n(z)$ but the \textit{true} $n(z)$ were a different one. 
For that, we generate a data vector assuming the fiducial $n(z)$ and fit it with the methodology explained in \autoref{sec:methods} using a template generated with another $n(z)$. While we test the $n(z)$ assumption, the rest of the choices (cosmology and bias) follow the \datalike setup (\autoref{sec:setup}). 

The results are presented in \autoref{tab:redshift_effect}. The first column (fid.) represents the case in which the assumed and true redshift distributions are identical, naturally, giving unbiased results ($\langle \alpha\rangle=1.000$). The second column corresponds to the case where we use DNF $\zmc$ estimation, which corresponds to the redshift used to construct the mock catalogues described in \autoref{sec:sims}.
A different estimation from DNF, the PDF, is used in the fifth column. We also consider independent measurement from direct calibration with the spectroscopic survey VIPERS (third column) and clustering redshifts (WZ, fourth column). 
Given the great variety and independence of those methods, it is remarkable how small the observed shifts are in
the BAO parameter $\alpha$. Up to bin 5, the largest deviation is $\Delta \alpha =0.011$ (VIPERS, bin 4, ACF), corresponding to $<0.3\sigma$ (considering the error on each individual bins reported along with the measurement), but offsets are typically smaller. It is reassuring that these offsets  contribute in different directions for different bins and $n(z)$ calibrations, and no coherent offset is found (see also the discussion below when considering {\it All} the bins together). Remarkably, up to bin 5, the PCF method, which uses radial information, does not seem to be more sensitive to the $n(z)$ calibration than ACF or APS. 

For bin 6, the bias on the recovered $\alpha$ goes up to $\Delta \alpha=$ 0.013 ($\zmc$) and 0.023 (VIPERS) for the PCF method. However, given the large error bars on this last bin, this only represents $0.14\sigma$ and $0.30\sigma$, respectively. Since this bias is at the similar level in relative error as other redshift bins, its possible contribution to biasing the final result is expected  to be similar to other bins. Additionally, this relatively large bias only affects one of the three estimators. Hence, we do not expect this to be a relevant source of systematic error for the $\alpha$ derived from the  6 bins together and, especially, for the consensus measurement combining the 3 statistics. 

Finally, in the last part of \autoref{tab:redshift_effect} we show what we consider the main results of this subsection, where we show the results when considering all the redshift bins together (`All'), as done in our analysis. 
For this case, we do not only report these results on the individual methods ACF, APS, and PCF, but we also propagate our inferred values to the consensus measurement (AVG) using the method described in \autoref{sec:comb_method}.
Then, the largest bias found for AVG is taken to be the systematic error due to the redshift calibration: 
\begin{equation}
    \begin{aligned}
    & \sigma_{z \rm,  sys}^{\rm AVG}=0.0035 \,\, .
    \label{eq:sys_z}
    \end{aligned}
\end{equation}
In all 4 cases (ACF, APS, PCF, AVG), the maximum deviation from $\alpha=1$ comes from the DNF PDF, which is expected to have an over-estimation of the dispersion of the photo-$z$. 
Hence, this estimation can be considered as an upper limit on the systematic budget. The systematic errors found here are all below $0.22\sigma_{\rm stat}$, which if added in quadrature to the statistical error would only increase the total error by 2\%.

\subsection{Validation against simulations}
\label{sec:mocktests}

\begin{table*}
  \caption{BAO fits for the Angular Correlation Function (ACF, $w(\theta)$) on the 1952 \mocks\
  using by default the \mocklike setup (see \autoref{sec:setup})
  with different variations of the analysis in the different rows, 
  see discussion in \autoref{sec:mocktests}. The default analysis choice is shown in {\bf bold}. The last two rows also show results in log-normal mocks. We show: the mean ($\langle \alpha \rangle$) and standard deviation ($\sigma_{\rm std}$) of all best fits, 
   the semi-width of the inter-percentile region containing 68\% of the best fits, $\sigma_{68}$, the mean of all the individual error estimations ($\langle \sigma_\alpha\rangle$, from $\Delta \chi^2=1$, see \autoref{sec:inference}) 
   and, finally, the best fit and its associated error bar $\sigma_\alpha $ for the fit over the mean of the mocks.
  Note that for the \mice (default for this table) cosmology we expect $\bar{\alpha}=1$, while when using \planck cosmology templates, we expect $\bar{\alpha}=0.9616$.
}
  \label{tab:ACF_mocks}
    \begin{tabular}{lccccZZZZcZZZ}
    \toprule\toprule
     case & $\langle\alpha\rangle$ & $\sigma_{\rm std}$ & $\sigma_{68}$ & $\langle\sigma_\alpha\rangle$ & fraction encl.$\langle\alpha\rangle$ & $\langle d_{\rm norm}\rangle$ & $\sigma_{d_{\rm norm}}$ & $\langle\chi^2\rangle/$d.o.f. & mean of mocks & $\langle\sigma_\alpha\rangle/\langle\alpha\rangle$ & $N\sigma$ & frac detec \\ 
    \midrule
    $i=0$ & 1.0039 & 0.0187 & 0.0183 & 0.0180 & 67.1$\%$ & 0.0135 & 1.0281 & 85.9$/$119 & 1.0043$\pm$0.0178 & 1.86$\%$ & 0.21 & 100.00$\%$ \\
    $i=0,1$ & 1.0051 & 0.0202 & 0.0200 & 0.0190 & 65.9$\%$ & -0.0151 & 1.0540 & 80.8$/$113 & 1.0052$\pm$0.0188 & 2.01$\%$ & 0.25 & 100.00$\%$ \\
    $\boldsymbol{i=0,1,2}$ & \textbf{1.0057} & \textbf{0.0201} & \textbf{0.0202} & \textbf{0.0187} & \textbf{65.4\%} & \textbf{-0.0088} & \textbf{1.0718} & \textbf{76.3}$\boldsymbol{/}$\textbf{107} & \textbf{1.0059}$\boldsymbol{\pm}$\textbf{0.0185} & 2.00$\%$ & 0.28 & 99.95$\%$ \\\vspace{0.1cm}
    $i=-1,0,1,2$ & 1.0058 & 0.0202 & 0.0200 & 0.0188 & 65.1$\%$ & -0.0101 & 1.0747 & 70.6$/$101 & 1.0059$\pm$0.0185 & 2.01$\%$ & 0.29 & 100.00$\%$ \\
    Planck template $i=0,1$ & 0.9675 & 0.0197 & 0.0197 & 0.0205 & 69.8$\%$ & -0.0109 & 0.9601 & 86.7$/$113 & 0.9687$\pm$0.0202 & 2.04$\%$ & 0.30 & 99.95$\%$ \\
 Planck template    \boldsymbol{ $i=0,1,2$} & 0.9680 & 0.0193 & 0.0191 & 0.0182 & 65.4$\%$ & -0.0102 & 1.0651 & 77.3$/$107 & 0.9682$\pm$0.0180 & 1.99$\%$ & 0.34 & 100.00$\%$ \\\vspace{0.1cm}
    Planck template $i=-1,0,1,2$ & 0.9680 & 0.0195 & 0.0193 & 0.0182 & 65.6$\%$ & -0.0128 & 1.0710 & 71.4$/$101 & 0.9680$\pm$0.0180 & 2.01$\%$ & 0.33 & 100.00$\%$ \\
    $\Delta\theta=0.05$ deg & 1.0058 & 0.0202 & 0.0200 & 0.0188 & 64.9$\%$ & -0.0066 & 1.0629 & 190.0$/$515 & 1.0061$\pm$0.0186 & 2.01$\%$ & 0.29 & 99.95$\%$ \\
    $\Delta\theta=0.15$ deg & 1.0057 & 0.0202 & 0.0199 & 0.0188 & 64.9$\%$ & -0.0061 & 1.0665 & 95.8$/$155 & 1.0061$\pm$0.0186 & 2.01$\%$ & 0.28 & 99.95$\%$ \\
    \vspace{0.1cm}
    $\theta_{\rm min}=1$ deg & 1.0061 & 0.0203 & 0.0200 & 0.0189 & 65.8$\%$ & -0.0131 & 1.0714 & 67.9$/$95 & 1.0060$\pm$0.0186 & 2.02$\%$ & 0.30 & 100.00$\%$ \\
    Planck Cov. + Templ. & 0.9686 & 0.0194 & 0.0191 & 0.0209 & 72.2$\%$ & -0.0094 & 0.9250 & 60.3$/$107 & 0.9689$\pm$0.0206 & 2.00$\%$ & 0.37 & 100.00$\%$ \\
    COLA cov & 1.0063 & 0.0193 & 0.0187 & 0.0184 & 67.5$\%$ & 0.0011 & 1.0492 & 109.6$/$107 & 1.0066$\pm$0.0181 & 1.92$\%$ & 0.33 & 100.00$\%$ \\
    \hline
    Lognorm. Uncont. & 1.0117 & 0.0252 & 0.0230 & 0.0203 & 62.3$\%$ & -0.0032 & 1.2225 & 112.3$/$107 & 1.0116$\pm$0.0201 & 2.49$\%$ & 1.99 & 99.90$\%$ \\
    Lognorm. Cont. & 1.0119 & 0.0252 & 0.0235 & 0.0205 & 61.8$\%$ & -0.0056 & 1.2185 & 112.6$/$107 & 1.0117$\pm$0.0203 & 2.49$\%$ & 2.00 & 99.80$\%$ \\
    \bottomrule\bottomrule
    \end{tabular}
\end{table*}

\begin{table*}
  \caption{BAO fits for the Angular Power Spectrum (APS, $C_\ell$) 
on the 1952 \mocks\
  using by default the \mocklike setup 
  with different variations of the analysis in different rows, 
  see discussion in \autoref{sec:mocktests}. Similar structure  to \autoref{tab:ACF_mocks}.
  The default analysis choices, shown in bold use scale-cuts of $\ell_{\rm min}=10$ and $k_{\rm max}=0.211\, {\rm Mpc}^{-1}$, corresponding to $\ell_{\rm max}$ values for each redshift bin of 510, 570, 630, 710, 730 and 770.
  }
  \label{tab:APS_mocks}
  \begin{tabular}{lccccZZZZcZZZ}
  \hline
  \hline
case &  $\langle \alpha \rangle$     &   $\sigma_{\rm std}$   &   $\sigma_{68}$      & $\langle \sigma_{\alpha} \rangle$    &  ${\rm fraction\,encl.}   \langle \alpha \rangle$   &  $\langle d_{\rm norm} \rangle$   & $\sigma_{d_{\rm norm}}$  &   $\langle \chi^2 \rangle / {\rm d.o.f.}$  & ${\rm mean\,of\,mocks}$        \\ \hline   

$i=0$ & 1.0146 & 0.0165 & 0.0158 & 0.0146 & 64.5\% & -0.0081 & 1.1268 & 223.4/180(1.24) & 1.0146$\pm$0.0145 \\ 
$i=-2,-1,0$ & 1.0068 & 0.0197 & 0.0192 & 0.0180 & 64.9\% & 0.0286 & 1.0849 & 206.8/168(1.23) & 1.0071$\pm$0.0177 \\
$i=0,1,2$ & 1.0049 & 0.0191 & 0.0187 & 0.0170 & 63.4\% & 0.0237 & 1.1067 & 205.0/168(1.22) & 1.0052$\pm$0.0169 \\ 
$i=-2,-1,0,1$ & 1.0049 & 0.0207 & 0.0197 & 0.0171 & 60.8\% & 0.0094 & 1.2167 & 198.6/162(1.23) & 1.0051$\pm$0.0167 \\ 
$i=-1,0,1,2$ & 1.0064 & 0.0200 & 0.0194 & 0.0178 & 63.0\% & -0.0015 & 1.1152 & 198.2/162(1.22) & 1.0066$\pm$0.0176 \\ 
\vspace{0.1cm}
$\boldsymbol{i=-2,-1,0,1,2}$ & \textbf{1.0063} & \textbf{0.0216} & \textbf{0.0203} & \textbf{0.0178} & \textbf{60.8\%} & \textbf{-0.0168} & \textbf{1.2209} & \textbf{191.3}$\boldsymbol{/}$\textbf{156(1.23)} & \textbf{1.0061}$\boldsymbol{\pm}$\textbf{0.0174} \\ 

Planck temp, $i=0$ & 0.9166 & 0.0230 & 0.0214 & 0.0175 & 56.9\% & 0.1343 & 1.2797 & 246.8/180(1.37) & 0.9183$\pm$0.0170 \\ 
Planck temp, $i=-2,-1,0$ & 0.9555 & 0.0196 & 0.0187 & 0.0167 & 61.7\% & 0.0726 & 1.1533 & 215.0/168(1.28) & 0.9564$\pm$0.0164 \\ 
Planck temp, $i=0,1,2$ & 0.9576 & 0.0194 & 0.0188 & 0.0168 & 62.0\% & 0.0502 & 1.1374 & 214.4/168(1.28) & 0.9583$\pm$0.0165 \\ 
Planck temp, $i=-2,-1,0,1$ & 0.9577 & 0.0223 & 0.0204 & 0.0182 & 61.4\% & 0.0089 & 1.2108 & 207.2/162(1.28) & 0.9580$\pm$0.0177 \\ 
Planck temp, $i=-1,0,1,2$  & 0.9688 & 0.0197 & 0.0191 & 0.0184 & 64.9\% & 0.0031 & 1.0613 & 201.5/162(1.24) & 0.9690$\pm$0.0182 \\ 
\vspace{0.1cm}
Planck temp, \boldsymbol{$i=-2,-1,0,1,2$} & 0.9685 & 0.0225 & 0.0201 & 0.0187 & 61.7\% & -0.0364 & 1.1808 & 194.4/156(1.25) & 0.9678$\pm$0.0182 \\ 

$\Delta\ell=10$ & 1.0062 & 0.0209 & 0.0198 & 0.0175 & 59.9\% & -0.0238 & 1.1981 & 428.6/350(1.22) & 1.0060$\pm$0.0171 \\ 
$\Delta\ell=30$ & 1.0062 & 0.0239 & 0.0219 & 0.0186 & 58.3\% & -0.0247 & 1.3041 & 115.0/93(1.24) & 1.0059$\pm$0.0182 \\ 
$\ell_{\rm max}$=500 & 1.0068 & 0.0226 & 0.0218 & 0.0182 & 59.7\% & -0.0229 & 1.2349 & 130.6/113(1.16) & 1.0064$\pm$0.0178 \\ 
$\ell_{\rm max}$=550 & 1.0069 & 0.0224 & 0.0210 & 0.0181 & 59.0\% & -0.0278 & 1.2342 & 146.8/125(1.17) & 1.0063$\pm$0.0176 \\ 
$\ell_{\rm max}$=600 & 1.0066 & 0.0218 & 0.0204 & 0.0179 & 60.5\% & -0.0214 & 1.2226 & 171.2/143(1.20) & 1.0062$\pm$0.0174 \\ 
$k_{\rm max}$=0.167 & 1.0066 & 0.0225 & 0.0206 & 0.0181 & 59.7\% & -0.0250 & 1.2504 & 134.9/115(1.17) & 1.0062$\pm$0.0176 \\ 
\vspace{0.1cm}
$k_{\rm max}$=0.255 & 1.0058 & 0.0215 & 0.0199 & 0.0178 & 60.9\% & -0.0225 & 1.2221 & 253.2/198(1.28) & 1.0054$\pm$0.0172 \\ 

COLA cov & 1.0057 & 0.0215 & 0.0196 & 0.0195 & 66.2\% & 0.0009 & 1.2958 & 156.0/156(1.00) & 1.0061$\pm$0.0190 \\ 
Planck Cov. + Templ. & 0.9689 & 0.0222 & 0.0207 & 0.0215 & 68.3\% & -0.0214 & 1.0346 & 155.3/156(1.00) & 0.9687$\pm$0.0209 \\ 

\hline
  \end{tabular}
\end{table*}

\begin{table*}
 \caption{ 
 BAO fits for the Projected Correlation Function (PCF,  $\xi_p(s_\perp)$) 
on the 1952 \mocks\
  using by default the \mocklike setup 
  with different variations of the analysis in different rows, 
  see discussion in \autoref{sec:mocktests}. Similar structure  to \autoref{tab:ACF_mocks}.
 }
  \label{tab:PCF_mocks}
  \begin{tabular}{lccccZZZZcZZZ}
  \hline
  \hline
case &  $\langle \alpha \rangle$     &   $\sigma_{\rm std}$   &   $\sigma_{68}$      & $\langle \sigma_{\alpha} \rangle$    &  ${\rm fraction\,encl.}   \langle \alpha \rangle$   &  $\langle d_{\rm norm} \rangle$   & $\sigma_{d_{\rm norm}}$  &   $\langle \chi^2 \rangle / {\rm d.o.f.}$   & ${\rm mean\,of\,mocks}$        \\ \hline   
$i=0$                   &  1.0006    &   0.0176  & 0.0170  &   0.0185  & 72 \%  &  0.019  &  0.935  &  44.5/107 (0.42)   &   $ 1.0010 \pm 0.0184 $     \\  
$i=0,1$                 &  1.0007    &   0.0191  & 0.0182  &   0.0189  & 69 \%  & -0.004  &  0.993  &  41.3/101 (0.41)   &   $ 1.0014 \pm 0.0188 $      \\  
${\bf i=0,1,2}$         &  {\bf 1.0012}   &  {\bf 0.0187}  & {\bf 0.0180}  & {\bf 0.0189}  &  {\bf 69\%}  & {\bf -0.009}  & {\bf 0.98}   &  {\bf  37.1/95 (0.39)}   &  {\bf $1.0014\pm0.0192$ }       \\  
\vspace{0.1cm}
$i=-1,0,1,2$            &  1.0014    &  0.0191   & 0.0184  & 0.0193   &  69\%   &  -0.015  &  0.98    & 33.1/89 (0.37)   &  $1.0014\pm 0.0192$     \\

$ {\rm {\tt Planck} \,temp.} \, i=0 $   &   0.9597      &  0.0163  &  0.0158   &  0.0173  &  71\%  &  0.020  &  0.93   &  46.1/107 (0.43)   & $0.9610 \pm 0.0176$   \\
$ {\rm {\tt Planck} \,temp.} \, i=0, \, 1 $  &   0.9636  &  0.0180    &  0.0176  &  0.0189    & 69\%    & -0.013  & 0.944   & 40.9/101 (0.41)   & $0.9638 \pm 0.0188$   \\

$ {\rm {\tt Planck} \,temp.} \, \boldsymbol{  i=0, \, 1, \, 2 }$ &    0.9631   &  0.0180    &  0.0176  & 0.0182    & $69\%$    & -0.010  & 0.980  & 35.9/95 (0.39)   & $0.9622 \pm 0.0184$  \\
\vspace{0.1cm}
$ {\rm {\tt Planck} \,temp.} \, i=-1, \, 0, \, 1, \, 2 $  &  0.9632   &  0.0185    &  0.0180  &   0.0186   & $68\%$    &  -0.014 & 0.987  &  31.6/89 (0.36)   & $ 0.9626 \pm 0.0184$    \\

$\Delta s_\perp = 10$    & 1.0011     & 0.0191    & 0.0182   & 0.0189    & 69\%   & -0.005  & 1.002    & 22.1 /35 (0.63)   & $1.0010 \pm 0.0188$    \\          
$\Delta s_\perp = 8 $    & 1.0014     & 0.0191    & 0.0184   & 0.0190    & 68\%   & -0.006   & 0.998   & 27.6 /53 (0.52)   & $1.0014 \pm 0.0190$    \\          
$ \Delta s_\perp = 3 $   & 1.0015    & 0.0187     & 0.0186   & 0.0189    & 68\% & -0.008     & 0.981   & 50.0/179 (0.28)   &  $1.0014 \pm 0.0190 $    \\
$ \Delta s_\perp = 2 $   & 1.0016    & 0.0185     & 0.0182   & 0.0189    & 69\% & -0.008     & 0.977   & 61.0/275 (0.22)   &  $1.0018 \pm 0.0190 $    \\
$\text{Fit range } [70,130]$ & 0.9998   & 0.0204   & 0.0198   &  0.0232  &  75\%   & 0.045  & 0.864   & 18.25/47 (0.39)   &   $1.0014 \pm 0.0228$    \\
$N_z=1$   &  1.0031    &   0.0214  &  0.0208   &  0.0206    & \%    &    &    &    &$ 1.0026 \pm 0.0202 $    \\ 
\vspace{0.1cm}
$N_z=3$   &  1.0016    &   0.1929   &  0.0186   & 0.0190    & \%    &    &    &   & $1.0018\pm0.0190$ \\ 

$ {\rm {\tt Planck} }$ Cov. + Templ. & 0.9631 &  0.0177    &  0.0170  & 0.0208     & 76 \%    & -0.010  & 0.84   & 34.6/95 (0.36)   & $ 0.9622 \pm 0.0208 $    \\

$ {\rm COLA\,cov} $    &  1.0005    &   0.0192   &  0.0183   & 0.0175    & 67\%    &  0.0003  & 1.099    & 114.4/95 (1.2)   & $ 1.0010 \pm 0.0176$   \\

\hline
  \end{tabular} 
\end{table*}

One important part of validation of LSS analyses is to verify in cosmological simulations that we are able to recover the known input cosmology. Here, we use the ICE-COLA mocks described in \autoref{sec:sims} to validate the methodology explained in \autoref{sec:methods} and to guide different analysis choices. 

The tests are summarized in \autoref{tab:ACF_mocks}, \autoref{tab:APS_mocks} \& \autoref{tab:PCF_mocks} for ACF, APS and PCF, respectively. 

On the first part of the tables we vary the number of broad band terms $A_i$ from \autoref{eq:broadband_terms_w} / \autoref{eq:broadband_terms_Cl} / \autoref{eq:broadband_terms_xi} and we show in {\bf bold} the fiducial results. 
We find that for ACF results (namely, $\langle \alpha \rangle$ and $\langle \sigma_\alpha \rangle$) stabilize (see below) when using 3 broad band terms ($i=0,1,2$) to $\langle \alpha \rangle\approx1.0057$. For APS, we find the result only stabilizes when using as many as 5 parameters and that we need to include negative broad band terms. This implies that both negative and positive powers of $\ell$ are needed. Then, the results stabilize to $\langle \alpha \rangle \approx1.0063$. Finally, the results from the PCF do not change much with the number of broad band terms ($\langle \alpha \rangle \approx1.0012$). In order to judge {\it stabilization}, we run a larger number of $A_i$ configurations (not all of them shown here) and find that as we keep adding terms, the mean results converge to a given $\langle \alpha \rangle$, with some remaining small variations ($\lesssim0.1\sigma$). We choose the $A_i$ that has already approximately converged to that value, with the minimal number of terms. We also check that the recovered $\langle \sigma \rangle$ for the selected configuration is similar to other configurations of $A_i$ with similar or equal number of broad band terms.

To help guiding the decision on the number of broad band terms, in the second tier of the tables, we show these variations but now assuming {\it Planck} cosmology for the template. The importance of the broad band terms is expected to be larger for this case where the template cosmology does not agree with the cosmology of the mocks ({\tt MICE}) and these terms can absorb part of the differences, making the measurement of the BAO position more robust. For these tests here we use a hybrid setup with the \mocklike covariance and $n(z)$, but with the bias from \datalike and Planck cosmology. 
Given the differences in cosmology of the mocks, at $\zeff=0.85$, we expect to measure $\alpha=0.9616$. We see that for the three estimators, the results are already stable (at the level of $\Delta \langle \alpha \rangle < 0.0010$) for the number of broad band terms used as default. 

At this point, we note that we find a bias of $\alpha$ that we quantify with $\biasalpha\equiv (\langle \alpha \rangle - \bar \alpha)/\bar \alpha$ \footnote{We will use the alternative $\Delta \langle \alpha \rangle$ symbol for differences in $\langle \alpha \rangle$ without re-normalizing by $\bar \alpha$.}, where we define $\bar \alpha$ as the theoretical expected value: $1$ for {\tt MICE} (default) and $0.9616$ for {\tt Planck}. 
On \textcolor{blue}{Tables} \ref{tab:ACF_mocks}, \ref{tab:APS_mocks} \& \ref{tab:PCF_mocks} (bold values), which is later summarized in \autoref{tab:combination}, we find $\biasalpha= +0.57 \%$, $\biasalpha= +0.63 \%$ and $\biasalpha= +0.12 \%$ for ACF, APS and PCF, respectively in the mocks cosmology (MICE). These biases slightly rise to 
$\biasalpha=0.67\%, \, 0.75\% \, \& \, 0.16\%$
when assuming Planck cosmology. 
These biases stay at the level of $\Delta \langle \alpha \rangle/\sigma_{\rm std} \approx 0.3$ for both ACF and APS in \mice cosmology, rising up to $\Delta \langle \alpha \rangle/\sigma_{\rm std} \approx 0.36$ for \planck APS. 
The latter will be reported 
as the systematic error coming from the modeling. 
We now discuss the possible physical origin of these biases, and the fact that they are partially mitigated in our fiducial analysis that combines the three measurements. 

Non-linear evolution of the LSS predicts a shift in the BAO position of the order of $\biasalpha\sim +0.5\%$ (with respect to the linear case), with the exact value depending on the redshift range, linear bias and halo occupation distribution of the sample \cite{CrocceScoccimarro_2008,PadmanabhanWhite_2009}. Hence, most of the observed bias in ACF \& APS is expected to have a physical origin. Additionally, although not shown in the table, we also try for the ACF to use the alternative \textsc{Cosmoprimo}\footnote{\url{https://github.com/cosmodesi/cosmoprimo}} template with a different modeling of the BAO damping. {\sc Cosmoprimo} has several different ways to compute the no-wiggle power spectrum, but we use the one based on the method developed in \cite{wallisch2019cosmological}. We recover similar results for MICE ($\langle \alpha \rangle=1.0059$) and Planck cases ($\langle \alpha \rangle=0.9675$).
We will also see below (\autoref{sec:comb_mocks}) that when combining ACF, APS and PCF, the biases in $\langle \alpha\rangle$ get significantly mitigated. 
Taking into account all of this, we consider our default analysis to be robust, given the statistical uncertainty of our measurements (see discussion in  \autoref{sec:comb_mocks}).

On the third tier of the \autoref{tab:ACF_mocks}, \autoref{tab:APS_mocks} \& \autoref{tab:PCF_mocks} we test variations with respect to our fiducial scale choices: $\theta_{\rm min}=0.5$deg, $\Delta\theta=0.20$deg and $\theta_{\rm max}=0.5$deg for ACF; $\ell_{\rm min}=10$, $k_{\rm max}=0.211\,{\rm Mpc}^{-1}$ and $\Delta \ell=20$ for APS, and 
$s_{\perp,{\rm min}}=40  \, \mathrm{Mpc} \, h^{-1}$, $s_{\perp,{\rm max}}=140 \, \mathrm{Mpc} \, h^{-1}$, and $\Delta s_{\perp} = 5 \, \mathrm{Mpc} \, h^{-1} $ for PCF. 
We find all the changes of scale choices to have a negligible impact on the recovered statistics, with the largest deviation ($\Delta \langle \alpha \rangle \sim 0.10\% \sim 0.05 \sigma$) found when changing the APS maximum scales. 
For PCF, we also have the option to have $N_z$ tomographic bins, with $N_z=6$ being our fiducial option. We find that when using $N_z=1$ and $N_z=3$, the $\langle \alpha \rangle$ moves only by $+0.19\%\approx0.1\sigma$ and $0.04\% \approx 0.02\sigma$, respectively. The driving decision to choose $N_z=6$ was based on a better agreement between $\langle \sigma_\alpha \rangle$ and $\sigma_{68}$, and easier comparison to ACF \& APS when removing redshift bins and having a lower expected error (for all their estimations), even when considering the full combination with ACF and APS (AVG in \autoref{sec:comb_method}).   

On the fourth tier we test the change of choice for the covariance. First, we include the \textit{Planck cosmology} entry, which implies using \datalike (Planck) covariance and \datalike bias, together with the \planck template, but the $n(z)$ of the mocks.  This introduces a negligible shift in the $\langle \alpha \rangle$ ($\sim 0.06\%$). 
As further validation on the covariance, for the ACF (but not shown in \autoref{tab:ACF_mocks} in order to avoid overcrowding the table), we also tested  removing the non-Gaussian component of the \textsc{Cosmolike} covariance or switching to the covariance developed in \cite{Chan:2018gtc}, in both cases resulting in negligible changes on the results. 

Finally in this covariance discussion, we also tested the usage of the covariance estimated by the \mocks\ themselves, having a very small impact on $\alpha$ ($\Delta \langle \alpha \rangle < 0.07\% <0.05\sigma$).
We note again that due to replications in the construction of the mocks (\autoref{sec:sims}), this covariance is not realistic for data, as it introduces spurious correlations on parts of the data vector.
However, it will represent the true covariance of the mocks themselves. For this reason, although not shown here, the $\chi^2/d.o.f.$ of fits on the mocks gets close to unity for this covariance, but differs from our default \cosmolike covariance. 
As expressed in our previous paragraph and in \autoref{sec:covariance}, we remark that this covariance has been validated against other model covariances and mock estimates from FLASK lognormal simulations. 
Unfortunately, these differences between the ICE-COLA mocks inherent covariance and our fiducial covariance and their impact on the $\chi^2$ tell us that we can not consider the calibration of the absolute $\chi^2$ given by our pipeline as validated. Hence, we will not use absolute $\chi^2$ as a driving criteria on the data, although we may consider variations of $\chi^2$ when changing analysis choices. 
Regarding our usage of $\Delta\chi^2=1$ as our 1-$\sigma$ definition, we validate it below against the dispersion in the measurements of $\alpha$. 

Up to this point we have not commented much on the results for the different estimations of the error $\sigma$, which are somewhat heterogeneous. Nevertheless, all the estimators of $\sigma$, for the three BAO measurements (ACF, APS, PCF) give us errors of the order $\sigma \sim (2.0\pm 0.2 )\ \%$.
For the fiducial choice of ACF, we find that the estimated error $\sigma_\alpha$ (from $\Delta \chi^2=1$) is 7\% below the scatter observed in the distribution of best fit $\alpha$, when estimated with the standard deviation ($\sigma_{\rm std}$) or the inter-percentile region ($\sigma_{68}$). For the APS, this difference raises to 18\% or 12\%, respectively, whereas for the PCF, the differences in error estimation stay below $\sim 4\%$ (and switch sign). 
Those percentages stay similar when moving to \planck template. 
For the ACF (the most validated method), these differences reduce to below $5\%$ when using the COLA covariance. Finally, for the \datalike covariance (`\planck Cov. + Templ.') $\sigma_\alpha$ switches to an overestimation of the scatter of $7-9\%$.

As we will see in the next subsection, once we combine the three statistics, not only the biases in the mean are mitigated, but also the difference among different $\sigma$. 

Finally, only for ACF, we also did some tests on the lognormal mock catalogues 
used to study observational systematics. The main feature here is that we can include the imprint of the observational systematics on them (an earlier version of the weights summarized in \autoref{sec:systematics}, see also \cite{Mena23}). We show the results on these mocks in the last tier of \autoref{tab:ACF_mocks} and by comparing the results on the uncontaminated mocks to the contaminated ones, we find that the  results are unchanged for the BAO when we add these observational systematics. This shows the exceptional robustness of BAO to these effects.

\subsection{Validation of combination}
\label{sec:comb_mocks}

In this section, we study how the method described in \autoref{sec:comb_method} to combine 3 correlated statistics performs when combining our 3 analyses on the \mocks. 
For that, we start by  measuring covariance of the best fits of ACF, APS and ACF from the mocks. For that, we first eliminate the 11 mocks in which at least one of the three methods finds a non-detection. 
Then, this covariance is decomposed into the variance of the three measurements and correlation coefficient across measurements. 
The variance is simply the square of the $\sigma_{\rm std}$, which we now summarize in \autoref{tab:combination} for \planck and \mice cosmologies (there are some slight differences in the last digit with respect to \textcolor{blue}{Tables} \ref{tab:ACF_mocks}, \ref{tab:APS_mocks} \& \ref{tab:PCF_mocks}, due to removing the non-detections). The Pearson correlation between ACF and APS is $\rho_3=0.863$; between ACF and PCF, $\rho_2=0.905$; and between APS and PCF, $\rho_1=0.789$. 
These correlations are slightly lower than those found in spectroscopic surveys. For example, we have $\rho_{\rm ACF,APS}=0.863$, whereas the eBOSS LRG in BAO measurements in configuration \cite{Bautista21} and Fourier space \cite{GilMarin20} have a correlation of 90\% \cite{combination_eBOSS}. We note that, although we are using the same data, we expect part of the noise to be de-correlated. We believe that the de-correlation can increase when projecting $r$/$k$ onto $\theta$/$\ell$/$s_\perp$ and also when making different analysis choices such us the different number of broadband terms. 

One curiosity, is that in previous analyses (Y1, Y3) APS \& APS were found more correlated among themselves than to the PCF. In Y6, this pairing is broken and the ACF is found more correlated to the PCF than to the APS. The main driver for the increase of the correlation is the fact that in Y6 we are analysing the PCF in $N_z=6$, $\Delta \zph=0.1$ tomographic bins like the ACF, whereas in previous analyses the PCF was considering the entire redshift range altogether ($N_z=1$), making PCF and ACF less correlated. 
One of the reasons why we used $N_z=6$ is that the error in the PCF was smaller than using $N_z=1$. One could wonder if the information gained by using $N_z=6$ is somehow lost by the fact that it is more correlated with ACF. Following the same methodology presented here, we checked that the combined ACF+PCF error on $\alpha$ is still smaller for $N_z=6$ than for $N_z=1$.

Once we have the covariance between ACF, APS \& PCF, we combine them using \textcolor{blue}{Equations} \ref{eq:alpha_average} to \ref{eq:sigma_average}. In \autoref{eq:sigma_average}, the error that we propagate is the individual error $\sigma_\alpha$ estimated from $\Delta \chi^2=1$, as we plan to use the same method on the data, where we cannot use ensemble estimates. As a result, we obtain a new combined best fit $\alpha_{\rm AVG}$ and a new estimated error $\sigma_{\alpha_{\rm AVG}}$ for each mock. 
With this, we can again estimate the mean ($\langle \alpha \rangle$) and standard deviation ($\sigma_{\rm std}$) of the best fits $\alpha_{\rm AVG}$, the 68 inter-percentile region ($\sigma_{68}$) and the mean estimated error ($\langle \sigma_{\alpha_{\rm AVG}}\rangle$). 

All these summary statistics are shown in the fourth row (AVG) of each section of \autoref{tab:combination}.
We find that the AVG statistics have a small bias in $\langle \alpha \rangle$: $\biasalpha=0.19\%$ for \mice cosmology and $\biasalpha=0.23\%$ for the \planck cosmology. 
The larger one will be considered our systematic error from the modeling side in \autoref{sec:results} (\autoref{tab:robustness}): 
\begin{equation}
    \sigma_{\rm th, sys}^{\rm AVG} = 0.0023 \,\, .
    \label{eq:sys_th}
\end{equation}
This is at the level of $0.15\sigma_{\rm stat}$ (considering the error expected from the mocks) and if added in quadrature, would only increase the total error budget by 1\%. 

Concerning error bars, we find them to be very well behaved: our mean estimated uncertainty gives us $\langle \sigma_{\alpha_{\rm AVG}}\rangle = 0.0181$ (1.81\%), which agrees to better than $3\%$ with the scatter measured on the best fit $\alpha$'s on \mice cosmology. For Planck cosmology, we obtain an estimated uncertainty of $\langle  \sigma_{\alpha_{\rm AVG}}\rangle=0.0175$ (1.83\%), which agrees with the measured scatter to better than $2\%$.
When using the cosmology of the mocks (\mice) the pull distribution (see definition in \autoref{tab:combination}) also shows excellent agreement with Gaussianity to the 1\% level ($\langle d_n\rangle$=0.01) and the fraction of mocks enclosed in $[\langle \alpha \rangle - \langle  \sigma_\alpha \rangle, \langle \alpha \rangle + \langle  \sigma_\alpha \rangle]$ matches the Gaussian case exactly to the third significant figure (68.6\%). 
When assuming \planck cosmology, the degradation of  these two measures of Gaussianity is still very small.
Indeed, by combining different signals we do expect that the resulting estimates become more Gaussian. Additionally, we do also expect that different methods can be affected by small different theoretical errors and that the combination of them would give more robust results. 

\begin{table*}[]
\setlength{\tabcolsep}{4pt} 
\caption{Summary of fiducial analyses for individual estimators (ACF, APS, PCF) and their combination (AVG) for the 1941 out of 1952 mocks that show a detection in the three estimators. We show the results for the \mice cosmology (where $\alpha=1$ is expected as this is the cosmology of the mocks) and for the \planck template (where $\alpha=0.9616$ is expected from the theoretical perspective, \autoref{eq:alpha}). We show: the mean ($\langle \alpha \rangle$) and standard deviation ($\sigma_{\rm std}$) of all best fits, 
   the semi-width of the inter-percentile region containing 68\% of the best fits ($\sigma_{68}$), the mean of all the individual error estimations ($ \langle \sigma_\alpha \rangle $, from $\Delta \chi^2=1$, see \autoref{sec:inference}), we also include the fraction of mocks with the best fit $\alpha$ enclosed in $\langle \alpha \rangle\pm \langle  \sigma_\alpha \rangle $, and the mean ($\langle d_{\rm n}\rangle$) and standard deviation ($\sigma_{d_{\rm n}}$) of the pull statistics ($d_{\rm n}=(\alpha - \langle \alpha \rangle)/ \sigma_\alpha$).
 }
\label{tab:combination}
  \centering
    \begin{tabular}{ZlcrrrrZZZcrr}
    \toprule
     & case & meth. & $\langle\alpha\rangle$ & $\sigma_{\rm std}$ & $\sigma_{68}$ & $\langle\sigma_\alpha\rangle$ &  $\langle\sigma_{\Delta \chi^2}\rangle$ &  $\langle\sigma_{\bar{\alpha^2}}\rangle$  &  $\langle\sigma_{L68}\rangle$ & mocks $\in \langle \alpha \rangle \pm \langle  \sigma_\alpha \rangle$ & $\langle$$d_{\rm n}$$\rangle$ & $\sigma_{d_{\rm n}}$ \\
    \midrule
    0 &  MICE  & ACF & 1.0057 & 0.0202 & 0.0202 & 0.0187 & 0.0187 & 0.0192 & 0.0190 &  65.2$\%$ & -0.0086 & 1.0730 \\
    1 &   & APS & 1.0063 & 0.0216 & 0.0204 & 0.0178  & 0.0178 & 0.0197 & 0.0185 &  62.3$\%$ & -0.0168 & 1.2208 \\
    2 & & PCF & 1.0012 & 0.0187 & 0.0182 & 0.0189 & 0.0189 & 0.0196 & 0.0191 &  69.6$\%$ & -0.0084 & 0.9819 \\
    3 & & AVG & 1.0019 & 0.0185 & 0.0180 & 0.0181 & -- & -- & -- & 68.6$\%$ & -0.0100 & 1.0189 \\
    \midrule
     4 & Planck & ACF & 0.9680 & 0.0193 & 0.0191 & 0.0181 & 0.0181 & 0.0187 & 0.0185 &  65.3$\%$ & -0.0106 & 1.0665 \\
     4 &      & APS & 0.9685 & 0.0225 & 0.0203 & 0.0187 & 0.0187 & 0.0214 & 0.0197 & 64.5$\%$ & -0.0364 & 1.1805 \\
     4 &  & PCF & 0.9631 & 0.0180 & 0.0176 & 0.0182 & 0.0182 & 0.0189 & 0.0185 & 69.5$\%$ & -0.0095 & 0.9827 \\
    4 & & AVG & 0.9638 & 0.0180 & 0.0177 & 0.0175 & -- & -- & -- & 67.6$\%$ & -0.0137 & 1.0215 \\
    \bottomrule
    \end{tabular}
    
\end{table*}

\section{Pre-unblinding tests on data}
\label{sec:blindtest}

Before we start looking at the clustering results on the data, we have performed a thorough validation based on theory (with different $n(z)$ calibrations, \autoref{sec:redshift_val}) and on mock catalogues (\autoref{sec:mocktests}, \autoref{sec:comb_mocks}). 
Once we decide to move on to tests on the data, in order to avoid confirmation bias, the analysis is performed blinded to the cosmological information. 
In this case, this means that we are not allowed to see the value of the BAO shift $\alpha$ measured in the data. For that reason, most of the tests proposed here are carried out with scripts that only look at the differences in $\alpha$ between two analyses and not at $\alpha$ itself. When this is not possible,  we blind $\alpha$ by shifting each best fit by the same unknown amount ($\Delta \alpha \in [-0.2,0.2]$) with a common script and the same random seed for the three analyses. 
The error values $\sigma_\alpha$ are also blinded such that the only information accessible are relative changes in $\sigma_\alpha$ between two analysis setups (typically, the fiducial analysis and a variation of it). This is achieved by having the errors of each estimator (ACF, APS, PCF) rescaled by a factor such that they are equal to the mean error seen in the mocks for the fiducial case. 

We also blind all the clustering measurements, except for the $0.5<\theta<1$ deg scales of the ACF that were used to calibrate the mocks. At a later stage, when the sample, weights and redshift validation were finalized, we also 
allowed the fit to the galaxy bias to take a slightly larger
range, $0.5<\theta<2$ deg. These bias values were then used to build the final version of the \datalike covariance matrices. 

\subsection{Pre-unblinding tests on ACF, APS and PCF}
\label{sec:blindtest_ind}

Before finalizing our analysis pipeline, we perform a series of blinded tests, detailed below. 
The general guiding criterion is that, if something that happens on the data also occurs in $90-95\%$ or more of the mocks, we consider the test fully passed. Some mild revision is envisioned if some particularity on the data is found to happen in less than $5-10\%$ of the mocks. If it happens in less than $1\%$ of the mocks, we will regard the test as failed and consider a major revision of the methodology before continuing with our analysis and the unblinding of the results. 

Unless otherwise stated we will be using the \datalike setup for the data and the \mocklike\ setup for the mocks (see \autoref{sec:setup}).

\begin{table}
\centering
\caption{Pre-unblinding test 1: Detection rate of BAO. We show the BAO detection rate in the ICE-COLA mocks for the Angular Correlation Function (ACF), Angular Power Spectrum (APS) and the Projected Correlation Function (PCF). The first row represents the results for all the tomographic bins combined, whereas the following 6 rows show results for individual bins. In brackets, we show whether [Y] or not [N] there is a detection on the data. On the second part of the table, we show the percentage of mocks that have 0, 1, 2, 3 or 4 tomographic bins with non-detections. Here, we mark in {\bf bold} where the data fall.  }
\begin{tabular}{lccc} \\
\hline
\hline
	Bin & ACF & APS & PCF  \\
\hline
    All  &  99.95 \% [Y]        & 99.49 \% [Y]             & 100 \% [Y] \\
      1  &  90.32 \% [{N}]  & 74.49 \% [{N}]       & 95.39 \%  [{N}] \\
      2  &  94.98 \% [Y]        & 82.12 \% [Y]             & 97.34 \% [Y] \\
      3  &  97.39 \% [Y]        & 86.73 \% [Y]             & 97.69 \% [Y] \\
      4  &  97.59 \% [Y]        & 91.55 \% [Y]             & 97.84 \% [Y] \\
      5  &  96.67 \% [Y]        & 90.73 \% [Y]             & 95.39 \% [Y] \\
      6  &  91.19 \% [Y]        & 87.76 \% [Y]             & 86.22 \% [Y] \\
      \midrule
      Non-detections & & & \\
      \midrule
      0  &  72.90 \%        & 41.80 \%              &  73.77 \%  \\
      1  &  {\bf 22.85} \%  & {\bf 36.42} \%        & {\bf 22.69 }\%  \\
      2  &  3.84 \%         & 16.03 \%              &  3.23 \%  \\
      3  &  0.31 \%         & 4.82 \%               &  0.26 \%  \\
      4  &  0.10 \%         & 0.92 \%               &  0.05 \%  \\
\hline
\hline
\label{tab:detection}
\end{tabular}
\end{table}

\begin{enumerate}
	\item{ {\bf Is the BAO detected?} This test is summarized in \autoref{tab:detection}.
	
	In the ICE-COLA mocks, we have detections (i.e. $\alpha\pm\sigma_\alpha\in [0.8,1.2]$, see \autoref{sec:inference}) in $>99\%$ of the cases for the full dataset with any of the three estimators: ACF, PCF, APS. Therefore, we should strongly expect a detection in the data.
	
	Additionally, for most cases, we expect to have detections in most individual redshift tomographic bins. 
 Based on \autoref{tab:detection}, for ACF \& PCF  we impose as a pre-unblinding criterion that we would envision a major revision if there are 3 bins or more non-detections ($\lesssim0.5\%$), a mild revision for 2 non-detections ($\sim 4\%$), and we would consider the test passed for 0 or 1 non-detections ($\sim 95\%$). For APS, we would consider a major revision for 4 or more non-detections ($\lesssim1\%$), mild for 3 ($\sim 5\%$) and a pass for 0 to 2 non-detections. 
 
    {\bf Results:} 
    \begin{itemize}
        \item We find a detection in ACF, PCF, and APS when we use the full dataset (` All'), thus passing the first part of the test. 
        
    \item When looking at individual tomographic bins for ACF, APS \& PCF, we find 1 non-detection (in the first bin in all cases), hence passing this test. We notice that the non-detection in the first bin has been consistent across all DES BAO analyses, and it is considered a statistical fluke due to cosmic variance. 

    A natural question that arises here is whether it is worth removing the first bin from the data set once we know we do not find a detection (under our definition). We investigate this further in \appendixcite{app:non-detection}, without drawing strong conclusions in either direction. Since our method has been validated in \autoref{sec:validation} based on the full data set (6 bins), and for consistency with the adoption in Y1 and Y3 analyses, we proceed with the entire data set. 
      \end{itemize}
	
  }

	\item{ {\bf Is the measurement robust?}
	}

\begin{figure}
    \centering
    \includegraphics[trim={0 10 0 5}, clip=true,width=0.90\linewidth]{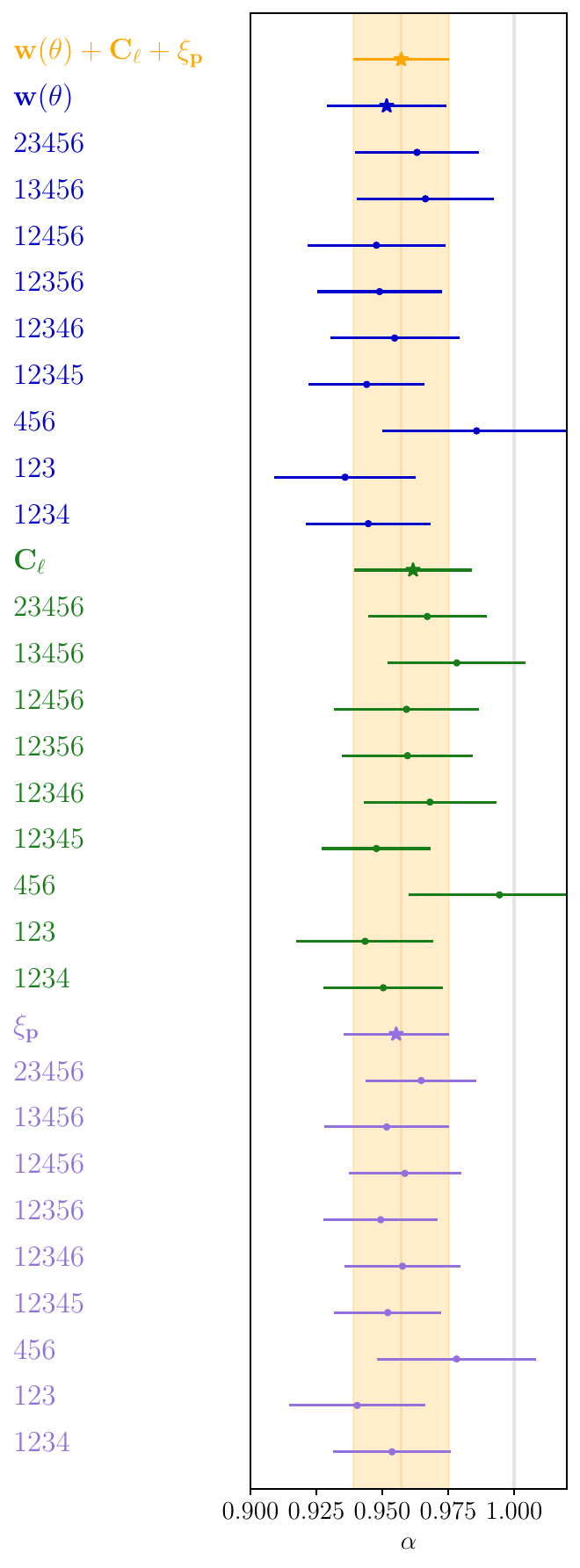}
    \caption{
    Unblinded representation of the pre-unblinding tests regarding partial data removal. 
    We show the fiducial AVG BAO measurement from \autoref{sec:results} with an orange star and a shaded area. For each of the individual estimators, ACF ($w(\theta)$), APS ($C_\ell$) and PCF ($\xi_p$), we show the fiducial result and how much it changes when we only keep some $z$-bins (indicated by the numbers). More details in \autoref{sec:blindtest}. 
    }
    \label{fig:preunblinding_tests}
\end{figure}

\begin{figure}
    \centering
    \includegraphics[width=0.9\linewidth]{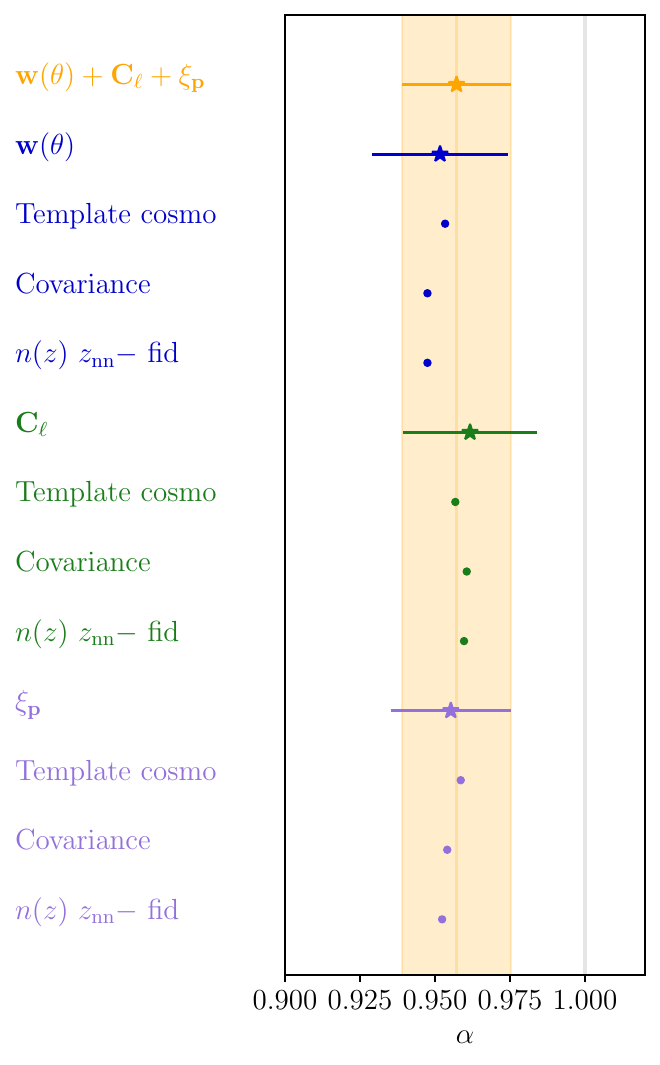}
    \caption{Unblinded representation of some pre-unblinding tests regarding robustness. 
    Main combined BAO measurement (AVG or $w(\theta)+C_\ell+\xi_p$) from \autoref{sec:results} shown with an orange star and a shaded area. For each of the individual estimators, ACF ($w(\theta)$), APS ($C_\ell$) and PCF ($\xi_p$), we show the fiducial result (with a star) and how much the best fit $\alpha$ changes when we change the assumed cosmology in the template, the covariance, or the $n(z)$ estimation. We also show a vertical gray line for the Planck BAO prediction ($\alpha=1$). The tests presented here are part of a series of pre-unblinding tests tabulated in \appendixcite{app:preunblind-tab} and discussed in \autoref{sec:blindtest}. 
    }
    \label{fig:preunblinding_tests_2}
\end{figure}

    These sets of tests are summarized in \textcolor{blue}{Figures} \ref{fig:preunblinding_tests} and \ref{fig:preunblinding_tests_2} (now shown unblinded) and tabulated in \appendixcite{app:preunblind-tab}. We test how much the best-fit $\alpha$ changes when we modify some choice in the analysis, and quantify it with $\Delta \alpha$. Similarly to the rest of the pre-unblinding tests, we assess the significance of the shifts in $\alpha$ by comparing to the distribution in the mocks. While \textcolor{blue}{Figures}~\ref{fig:preunblinding_tests} and \ref{fig:preunblinding_tests_2} show the results of this test on the data after unblinding (using the \textit{data-like} setup), the information we used for the pre-unblinding tests is shown is \textcolor{blue}{Tables}~\ref{tab:unblind_acf}, \ref{tab:unblind_aps} \& \ref{tab:unblind_pcf}. 
    There, we show for each test the limits of the $\Delta \alpha$ intervals containing 90\%, 95\%, 97\% and 99\% of the mocks. We consider an individual test failure if the $\Delta \alpha$ falls outside one of these intervals.  

    Given that we are performing a large number of tests, we expect that some of them could fail individually, without posing a global challenge. We quantify this  
    with the same guiding criteria we stated at the beginning of the section: mild revision if $5-10\%$ of the mocks show similar levels of failure, major revision if only $\sim < 1\%$ do.
    
	\begin{itemize}
		\item{ {\bf Impact of removing one tomographic bin.} 
		In \autoref{fig:preunblinding_tests} we show the change in best-fit $\alpha$ and $\sigma_{\alpha}$ when removing one tomographic bin at a time. These shifts are compared to the equivalent distribution in the COLA mocks in \appendixcite{app:preunblind-tab}. The quantity being measured on both the mocks and data is $\Delta \alpha = \alpha_{5\text{-}\rm bins} - \alpha_{6\text{-}{\rm bins}}$. 
		While we do not set strict pre-unblinding criteria on the $\sigma_{\alpha}$ values, we regard any significant changes as informative.
        
		}

        \item{ {\bf High-$z$ vs. low-$z$}. 
        In \autoref{fig:preunblinding_tests}, we also check the consistency of the results when only keeping the high-redshift half of the data (bins 456), only keeping the low-redshift half (bins 123) or removing the last two bins (bins 1234). The aim of this redshift split is to assess the consistency between different parts of the data, in particular, by checking if the high-$z$ data, for which the control of the observational systematics and the redshift validation is more challenging, could be dragging the results in one particular direction.  
        }

		\item{ {\bf Impact of template cosmology.} 
        Here we test whether the results vary as expected when changing the assumed cosmology in the template. 
        For the mocks, we compute a new $\alpha$ based on the \datalike Planck cosmology template 
        and compare it to the default \mocklike setup. 
        For the data, we change the template from \datalike\ to \datalikemice setup, while we keep the covariance unchanged. 
        Then, our test is given by the variable $\Delta \alpha=\alpha_{\rm Planck}-\alpha_{\rm MICE}+0.0384$, taking into account the $0.0384$ 
        difference expected by the change of cosmology. The best-fit $\alpha$ values on the data are shown in \autoref{fig:preunblinding_tests_2} for each estimator, while the results for the mocks are tabulated in \appendixcite{app:preunblind-tab}.

        We note that taking into account the biases ($\langle \alpha \rangle-1$) found in \textcolor{blue}{Tables} \ref{tab:ACF_mocks}, \ref{tab:APS_mocks} \& \ref{tab:PCF_mocks}, which differ from Planck and MICE cosmologies, we do not expect $\Delta \alpha$ to be centered at 0, but at $-0.0007$, $-0.0009$, $-0.0003$, respectively.
      
        }

		\item{ {\bf Impact of changing covariance.} 
        We check the difference when changing from our \datalike covariance (Planck cosmology and fiducial data setup) to the  \mocklike covariance (MICE cosmology and properties from the mocks, see \autoref{sec:setup}) or vice-versa. We define this test with $\Delta \alpha= \alpha_{\rm mock,cov}-\alpha_{\rm data,cov}$, noticing that, for the mocks, the fiducial choice is $\alpha_{\rm mock,cov}$, whereas for the data the fiducial choice is $\alpha_{\rm data,cov}$. We show the corresponding $\alpha$ values for the data in \autoref{fig:preunblinding_tests_2}, while the results for the mocks are tabulated in \appendixcite{app:preunblind-tab}.
        }
		
		\item{ {\bf Impact of n(z) estimation.} 
        Similarly, we now assess the impact of changing the assumed redshift distribution in the template from the data fiducial choice to $n(z)$ estimated from DNF $\zmc$: $\Delta \alpha=\alpha_{\rm znn}-\alpha_{\rm fid}$. Again, the fiducial choice of the mocks appears on the left ($\alpha_{\rm znn}$), whereas the fiducial choice on the data is on the right side of the difference. In this test, the covariance is left unchanged. We show the best-fit $\alpha$ values for the data in \autoref{fig:preunblinding_tests_2}, while the results for the mocks are tabulated in \appendixcite{app:preunblind-tab}.
        }

	\end{itemize}

   {\bf Results:} 
   \begin{itemize}
       \item 

   For ACF, the data does not fail any tests. This happens in 47\% of the mocks.
   Hence, we consider the robustness tests to be passed. 
   
   \item For APS, the data fails 1 test (removing bin 2) at the 90\% level (see \autoref{tab:unblind_aps}). 50\% of the mocks fail at least one of the tests at the 90\% level, and 21\% of the mocks fail exactly one test. Thus, we consider the robustness tests passed.
   
   \item For the PCF, the data does not fail any tests. On the mock catalogs, we find that 45\% of the mocks do not fail any of these tests. Thus, we consider the tests passed.  
   \end{itemize}

    We find another particular feature when looking at the impact of removing bin 6 on the error. 
    The error becomes smaller when removing this bin for the ACF (failing this test at the 97\% level, see \autoref{tab:unblind_acf}) and APS (failing at 90\%, see \autoref{tab:unblind_aps}), whereas for PCF the error does not become smaller. This led us to investigate this a bit further.
    First, we checked that 17\% of the mocks fail one or more $\Delta\sigma$ tests at the 97\% level for ACF.  
    Second, typically $\sim 10\%$ of the mocks show a smaller error when 1 particular redshift bin is removed. This is investigated further in \appendixcite{app:removing_bin}, where we check how the estimated error $\sigma_\alpha$ behaves in those particular cases. We find that the $\sigma_\alpha$ from the full set of 6 bins is a better representation of the best-fit $\alpha$ scatter compared to when using the $\sigma_\alpha$ estimated from the first 5 bins. Said otherwise, the $\sigma_\alpha$ from the first 5 bins becomes smaller, but just because it underestimates the underlying scatter, not because $\alpha$ is better determined. 

    In light of those results, we decided to continue with the full 6-bin case. Nevertheless, we will also report the results from bins $1-5$, bearing in mind that the last bin might be more prone to observational systematics. 

    Incidentally, although not listed in the tables from \appendixcite{app:preunblind-tab}, at some later stage but prior to unblinding, 
    we realized that the difference between the $\alpha$ values preferred by 123 and by 456 are somewhat large compared to the error bars (see \autoref{fig:preunblinding_tests}). This difference is highly correlated with the high- and low-redshift split tests discussed above (123 vs 456 in \autoref{fig:preunblinding_tests}, corresponding to entries 7 (456) and 8 (123) in \textcolor{blue}{Tables} \ref{tab:unblind_acf}, \ref{tab:unblind_aps} \& \ref{tab:unblind_pcf}). Nevertheless, we measured $\lvert \alpha_{123} - \alpha_{456}\lvert$ on the mocks, finding that 18\%, 19\% and 27\% of the mocks have a more extreme value than what is found in the data for ACF, APS and PCF, respectively. 
    
    At that stage, we also compared the (blinded) $\alpha$ preferred by individual bins (blinded version of \autoref{fig:bin_by_bin_bao} below), finding for ACF and APS a somewhat large difference between bin 6 and bin 2. However, we checked that the difference between the largest and lowest individual $\alpha$ found in mocks is compatible with what we see in the data for bin 2 and 6: for the case of ACF, 24\% of the mocks show a more extreme case, whereas for APS, this rises to 43\%.

\item{ {\bf Is it a likely draw?}
       Here we consider whether the ensemble of the 12 tests discussed above, each with a $\Delta \alpha$ (and shown in the top half of \textcolor{blue}{Tables} \ref{tab:unblind_acf}, \ref{tab:unblind_aps} \& \ref{tab:unblind_pcf}), is within expectations. For that, we measure the covariance of the 12 $\Delta \alpha$ on the mocks. We then compute the $\chi^2$ from this covariance and the $\Delta \alpha$ array in the data and compare it to the $\chi^2$ distribution seen in the 1952 mocks.
        
       {\bf Results:}
        \begin{itemize}
            \item ACF: The maximum values of the $\Delta \alpha$-based $\chi^2$ that contain 90\% and 95\% of the mocks are 26.16 and 37.78, respectively. For the \planck data, we get a $\chi^2$ value of 18.01, which is well below these limits. 

            \item APS: On mocks 90\% and 95\% have $\chi^2<28.42$, $\chi^2<42.22$. We find on the data that the $\Delta \alpha$-based $\chi^2$ is $11.22$, well within those limits.      
    
            \item PCF: The maximum $\Delta \alpha$-based $\chi^2$ that contains 90\% and 95\% of the mocks are 22.99 and 31.31, respectively. For the data, we get a $\chi^2$ value of 6.50, 
            which is well within the interval.

        \end{itemize}
}

\end{enumerate}

Finally, we also check the goodness of fit for the clustering statistics, although we do not put any specific criterion on it. The reason is that the $\chi^2$ could not be validated against the \mocks, due to their spurious covariance, as discussed in \autoref{sec:mocktests}. We still expect the $\chi^2/$d.o.f. to be of order unity. 

For the case of ACF, we find $\chi^2/\text{d.o.f.}=84.5/107$ 
(\datalike setup), similar to what we find in the mocks ($76.3/107$ for \mocklike setup). For reference, 22.64\% of the mocks have a larger $\chi^2$ than that of the data. 
For APS, we find $\chi^2/{\rm d.o.f.} = 163.3/156 = 1.05$, 
well within the $\chi^2<229.16$ limit found for 95\% of the mocks. 
For PCF, the $\chi^2/{\rm d.o.f.} =39.8/95 = 0.42 $,  
similar to the mean values found in the mocks (37.1/95 for \mice, 35.9/95 for \planck). 
As explained before, the $\chi^2$ is not considered an unblinding criterion as it could not be validated on the \mocks. 
Additionally, for PCF we have the added difficulty that the covariance matrix needs some ad-hoc treatment discussed in \cite{PCF_method_Y3}, where the $\chi^2$/d.o.f. does not reach unity. 

At this point, we also remark that even though, in the mocks, the $\chi^2$/d.o.f.~does not approach unity, the errors derived from it are very consistent with the scatter found in the best fit $\alpha$ (see, e.g., \autoref{tab:combination}). Hence, we find that the $\sigma_\alpha$ reported are robust. We also note that if we considered the $\chi^2$ correct for ACF and PCF, this would be hinting at an overestimation of the uncertainties, hence, if anything, lying on the conservative side. On the other hand APS has $\chi^2/{\rm d.o.f.}$ very close to unity. 

\subsection{Pre-unblinding tests for combination}
\label{sec:blindtest_avg}

Once we have validated the individual measurements by ACF, APS and PCF, we need to check the compatibility among those measurements before proceeding to their combination. For that, first, we check the difference between different estimators and compare it with the mocks. This is performed in the first part of \autoref{tab:unblind_avg}. For example, the first entry shows the difference in the best fit between the ACF and APS, $\alpha_{\rm ACF} - \alpha_{\rm APS}$, together with the limits expected from the mocks 90\% inter-quantile regions.  
In this case, we would look a bit more carefully at the combinations if the data falls outside the 90\% bulk of the mocks (and would have pursued a strong scrutiny if they fall outside the 99\% range).

Once that test is concluded, we can look at the difference from one individual estimator and the combination of the other two (part two of \autoref{tab:unblind_avg}) or the difference between the combination of the three (AVG) and an individual measurement (third part of \autoref{tab:unblind_avg}). The details on how these combinations are performed are in \autoref{sec:comb_method} and \autoref{sec:comb_mocks}.
Again, by running such a large number of tests we are likely to statistically fail some of the tests. In that case we would consider the ensemble of the tests. 

One feature of the intervals reported in \autoref{tab:unblind_avg} is that they are not always symmetric around zero. This is already expected, since the different estimators give slightly different $\langle \alpha \rangle$ (\autoref{tab:combination}).

{\bf Results:}

We find compatibility among APS, ACF and APS, and among all the combinations tested: all the data points shown in \autoref{tab:unblind_avg} fall well within the 90\% intervals measured in the \mocks. Hence, not only the individual (ACF, APS, PCF) measurements are ready for unblinding, but also our consensus combined BAO measurement (AVG).

\begin{table}[ht!]
\setlength{\tabcolsep}{5pt} 
\caption{Pre-unblinding tests for combination of the 3 estimators: ACF, APS, PCF. We take two different estimators (labeled in the first column) of the BAO shift, $\alpha$, and measure their difference ($\Delta \alpha$) on the data (second column). We then compare with the symmetric inter-quantile region that contains 90\% 
of the mocks (third 
column
). 
}

\label{tab:unblind_avg}
\begin{tabular}{lccZ}
\hline
$\Delta \alpha \times 100$ & Data & 90\%-mocks & 99\%-mocks \\
\hline
ACF-APS & -1.00 & [-1.36, 1.12] & [-3.29, 2.42] \\
ACF-PCF & -0.36  & [-0.58, 1.51] & [-1.73, 2.72] \\
APS-PCF & 0.64  & [-1.04, 2.15] & [-2.53, 4.48] \\
\hline
ACF-\{APS+PCF\} & -0.48  & [-0.52, 1.24] & [-1.47, 2.18] \\
APS-\{ACF+PCF\} & 0.68  & [-1.02, 2.02] & [-2.43, 4.27] \\
PCF-\{ACF+APS\} & 0.10  & [-1.58, 0.61] & [-2.90, 1.73] \\
\hline
AVG-ACF & 0.54  & [-1.34, 0.58] & [-2.37, 1.63] \\
AVG-APS & -0.45  & [-1.78, 0.81] & [-3.70, 2.08] \\
AVG-PCF & 0.19  & [-0.23, 0.39] & [-0.53, 0.86] \\
\hline

\hline
\end{tabular}
\end{table}

\section{Results}
\label{sec:results}


Once the pre-unblinded tests presented in the previous section were passed, we entered a gradual unblinding phase. 
The check-list to unblind is described in \appendixcite{app:unblinding} and highlights the level of scrutiny that we put into this analysis before we allowed ourselves to know the consequences for cosmology.
This is likely the strongest blinding policy to date imposed on a BAO analysis.

At the end of this phase, we obtained our unblinded fiducial results, described below (\autoref{sec:mainresults}), and also their corresponding different variations when some choices in our analysis are changed. These are referred to as robustness tests and are described in \autoref{sec:robustness}.

\subsection{Main results}
\label{sec:mainresults}

\begin{figure*}
    \centering
    \includegraphics[width=1\linewidth]{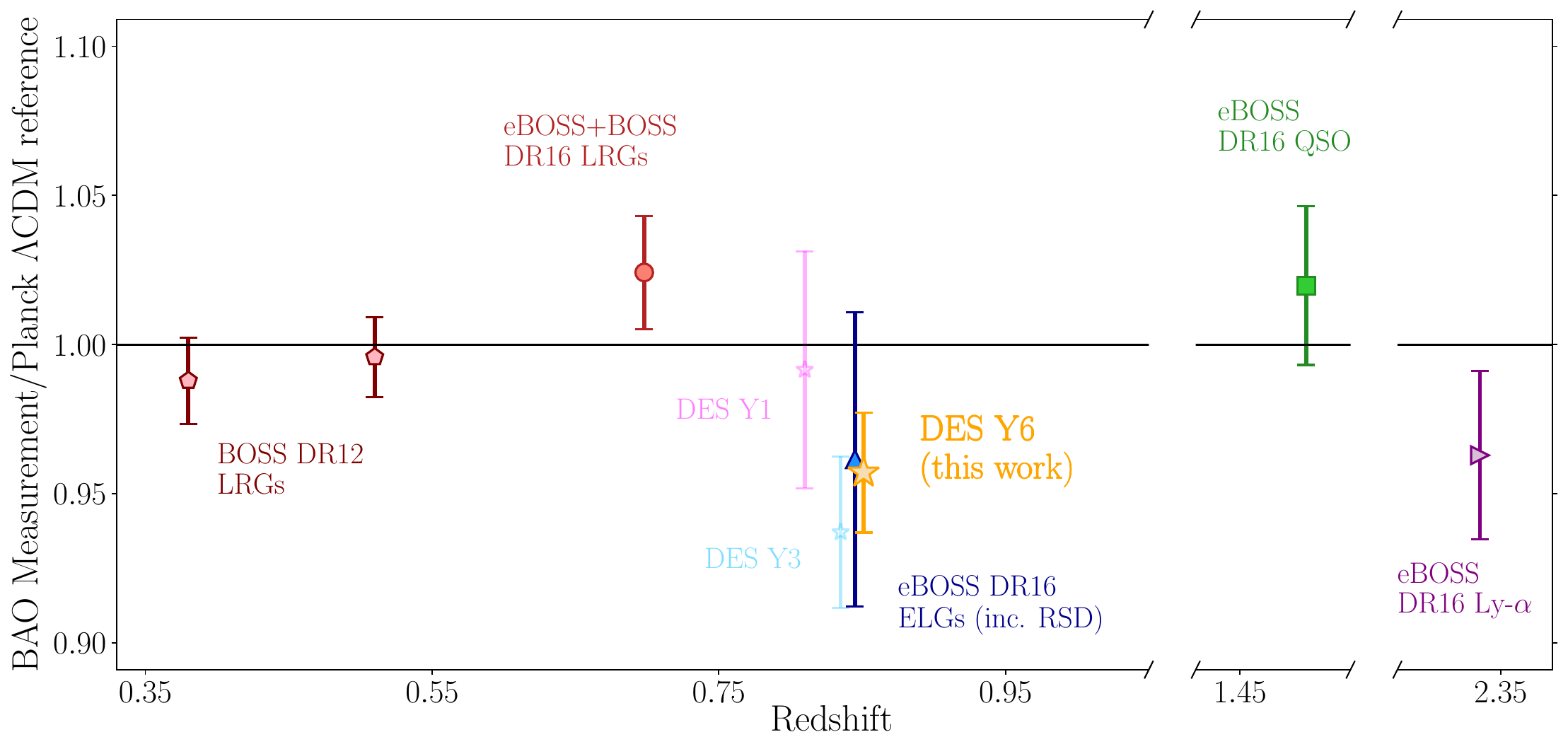}
    \caption{Ratio between the $D_M(z)/r_d$ measured using the BAO feature at different redshifts for several galaxy surveys and the prediction from the cosmological parameters determined by Planck, assuming $\Lambda$CDM. We include a series of measurements by SDSS, and also the DES Y1 and Y3 results. The DES Y6 measurement is shown with an orange star. This represents the most updated angular BAO distance ladder at the closure of stage III.
    }
    \label{fig:ladder}
\end{figure*}

\begin{figure}
    \centering
\includegraphics[width=1\linewidth]{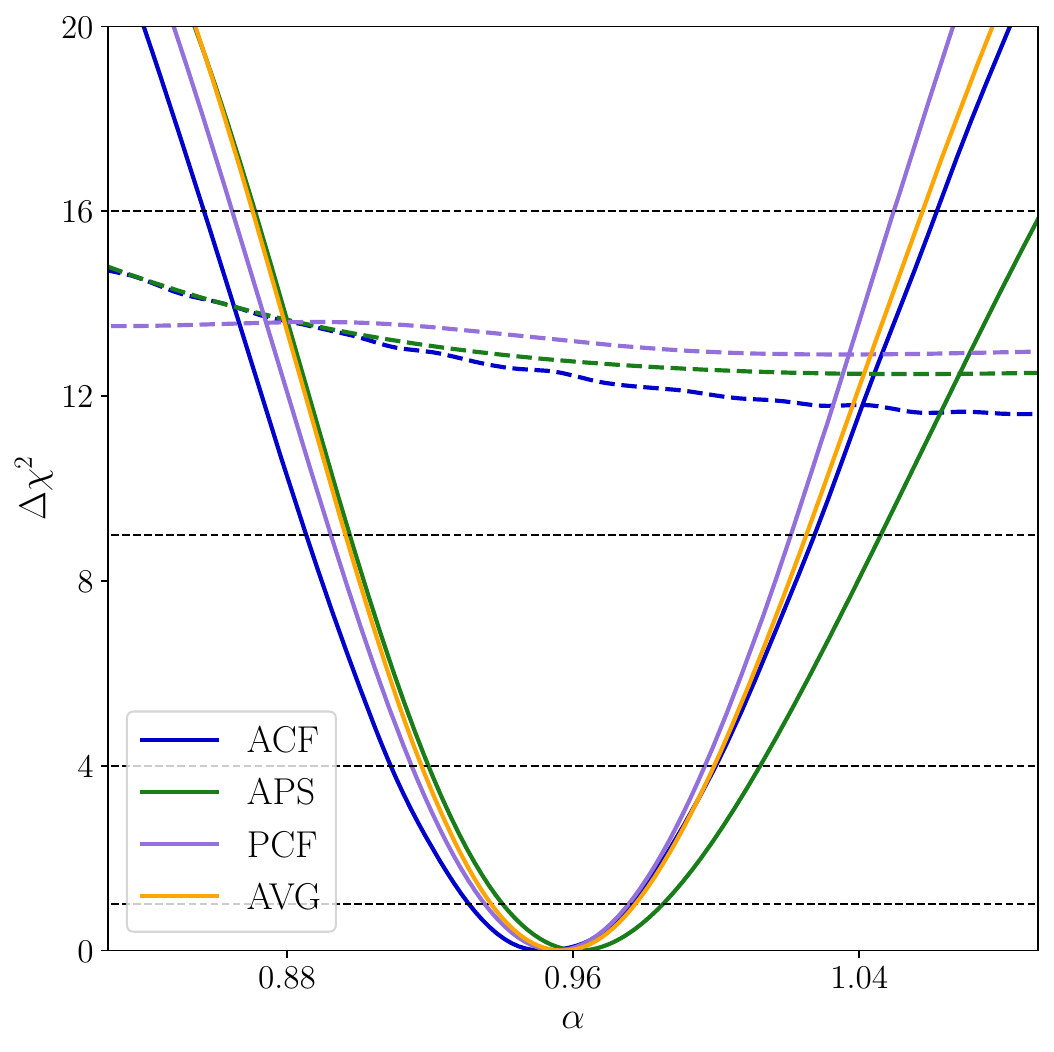}
    \caption{$\Delta\chi^2$ profile for the different estimators (ACF in blue, APS in green and PCF in purple). In colored dashed lines we show the $\Delta\chi^2$ obtained when trying to fit the data with a template without BAO. The combined $\Delta \chi^2$ profile (AVG, in orange) is the mean of the three $\Delta \chi^2(\alpha)$ curves, but  shifted and tightened so that its best fit and its width match our consensus measurement reported in \autoref{eq:results_alpha} (for the total error). The 1, 2, 3 and 4$\sigma$ limits are shown as horizontal black dashed black lines. 
    }
    \label{fig:likelihood}
\end{figure}

\begin{figure*}
    \centering
    \begin{minipage}{0.48\linewidth}
        \includegraphics[width=0.99\linewidth]{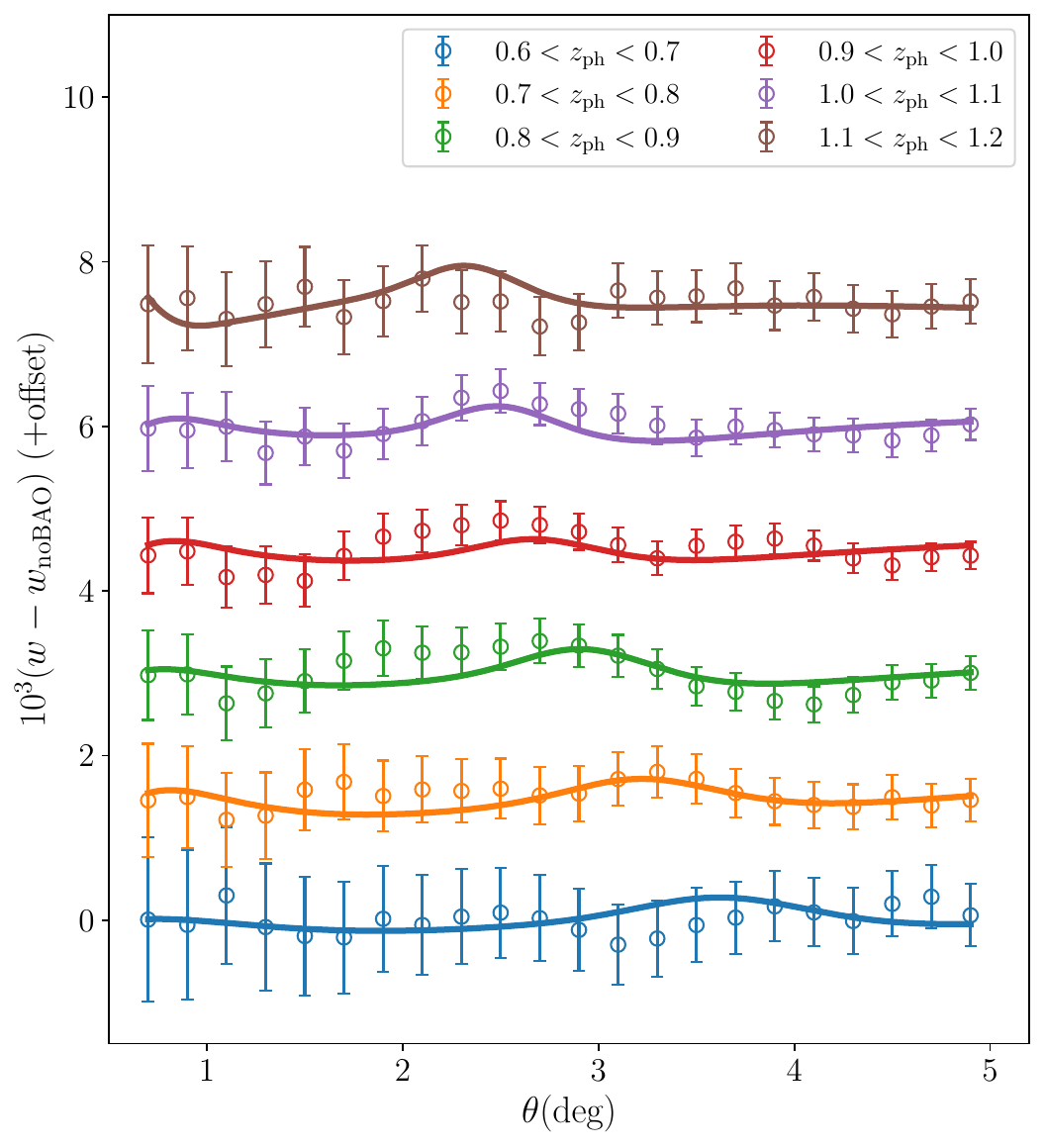}
        \caption{The isolated BAO feature, measured in configuration space using the angular correlation function, or $w(\theta)$. The curves have been re-scaled by a factor of $10^3$ and vertical offsets of +1.5 have been sequentially added to each tomographic bin, having our bin 1 (lowest redshift) at the bottom, and bin 6 at the top. Measurements are shown as markers with error bars (derived following \autoref{sec:covariance} and \planck cosmology), while the best fit model (with a single BAO shift $\alpha$ for the 6 $z$-bins) is shown in solid lines. The BAO feature moves to lower angular scales as the redshift increases, reflecting its constant comoving size. 
        Raw clustering measurements (without BAO template subtraction) of ACF, APS and PCF can be found in the companion paper \cite{Mena23}. 
        }
        \label{fig:ACF_data}
    \end{minipage}
    \begin{minipage}{0.01\linewidth}
       \hspace{0.01\linewidth}
    \end{minipage}
    \begin{minipage}{0.48\linewidth}
        \includegraphics[width=1\linewidth]{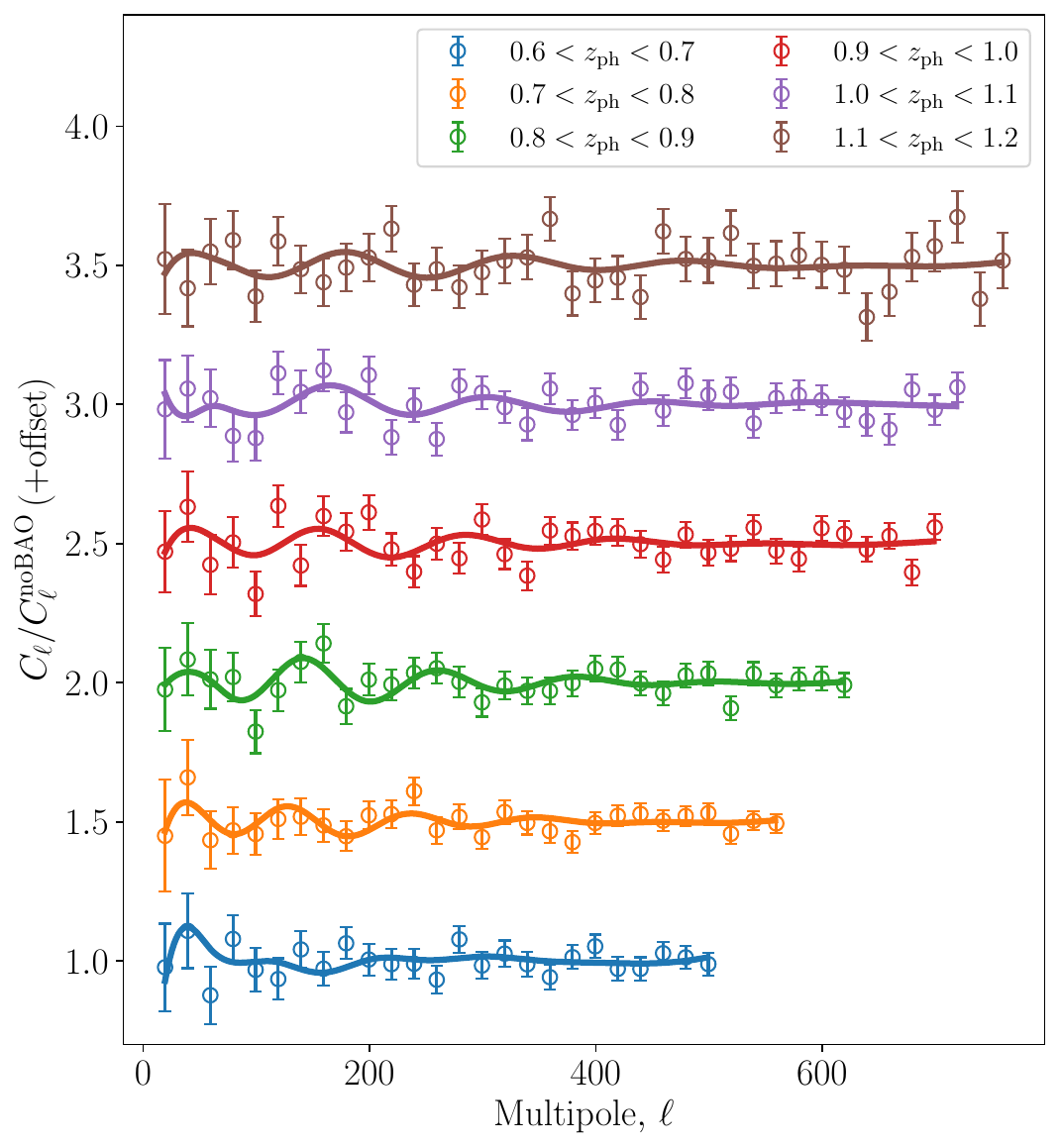}
        \caption{
        The isolated BAO feature in harmonic space. Same as \textcolor{blue}{Figures} \ref{fig:ACF_data} and \ref{fig:PCF_data} but using the Angular Power Spectrum (APS). Each tomographic bin has been sequentially offset vertically by +0.5. The BAO feature, with its constant comoving scale, expands to larger $\ell$ values (smaller scales) as redshift increases. The error bars are derived from the Planck fiducial covariance, and the solid lines represent the best fit. 
        }
        \vspace{63pt}
        \label{fig:APS_data}
    \end{minipage}
\end{figure*}

\begin{figure*}
    \centering  \includegraphics[width=0.49\linewidth]{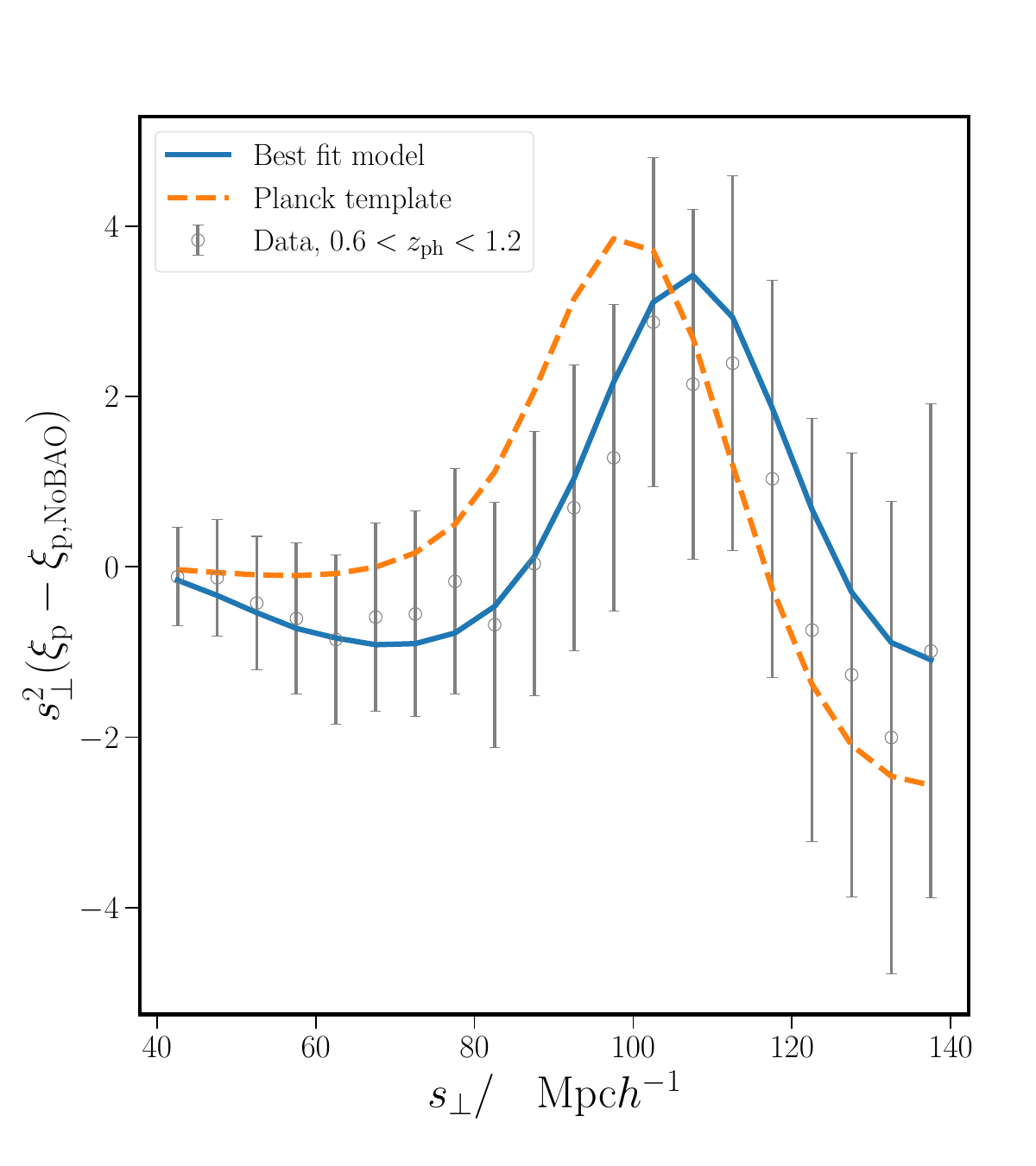}
    \includegraphics[width=0.49\linewidth]{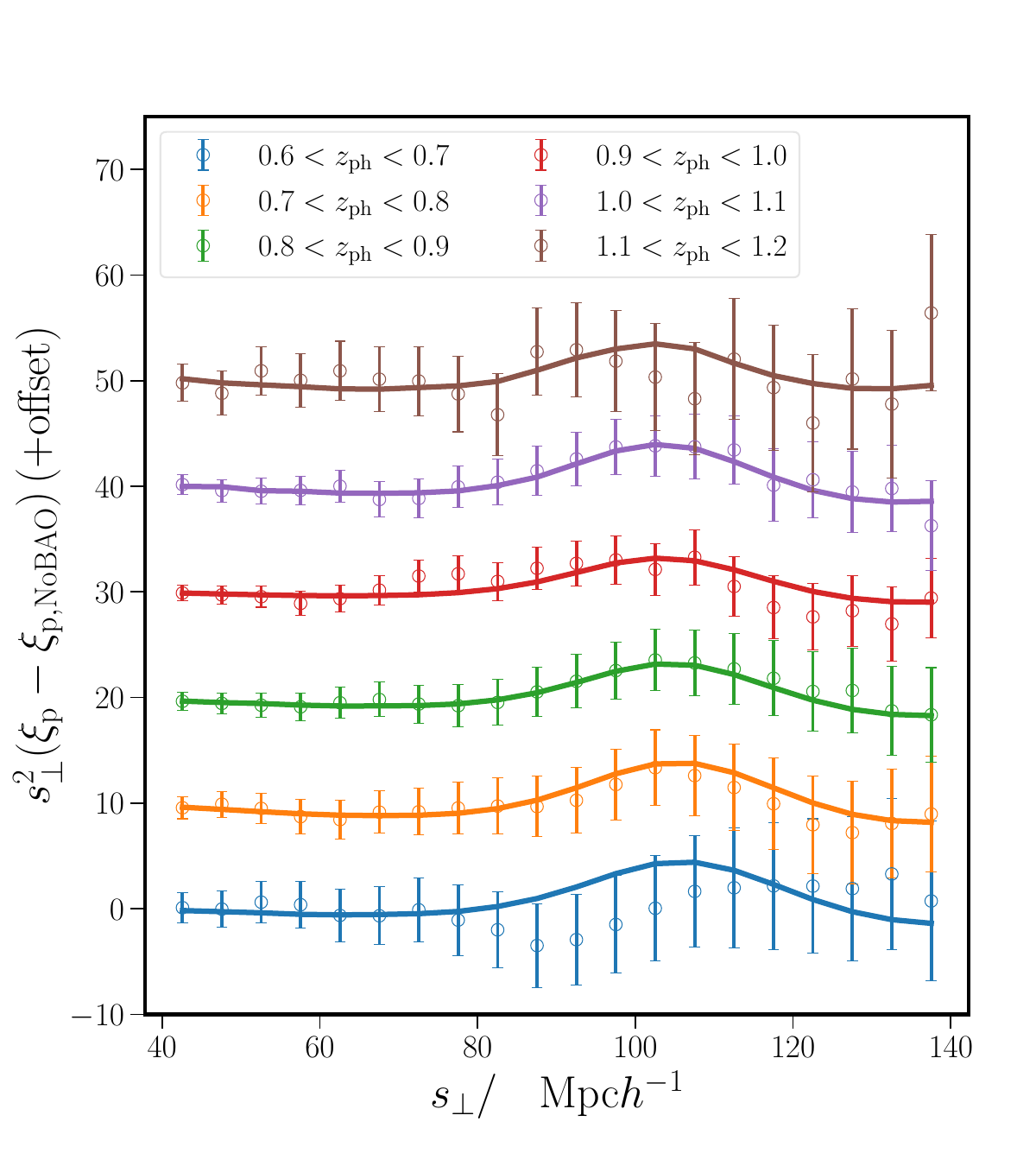}
    \caption{ The BAO feature measured using the Projected Correlation Function (PCF) in configuration space.  The markers are the data measurements and their error bars are derived from the fiducial Planck covariance.  The solid lines show the best fit model.  {\bf Left:} The PCF clustering is measured in a single bin ($N_z=1$) in order to concentrate all the BAO signal, for visualization purposes. Note that this is not the fiducial setup for the analysis. In addition to the best fit (solid, blue), the original Planck template (dashed, orange) is also overplotted.  {\bf Right:} The PCF measured in $N_z=6$ bins used for the fiducial analysis.  Each tomographic bin has been sequentially offset vertically by +10. Unlike the angular statistics,  the BAO feature in $\xi_{\rm p} $ does not change with redshift.  
  }
    \label{fig:PCF_data}
\end{figure*}

\begin{figure}
    \centering
    \includegraphics[width=1\linewidth]{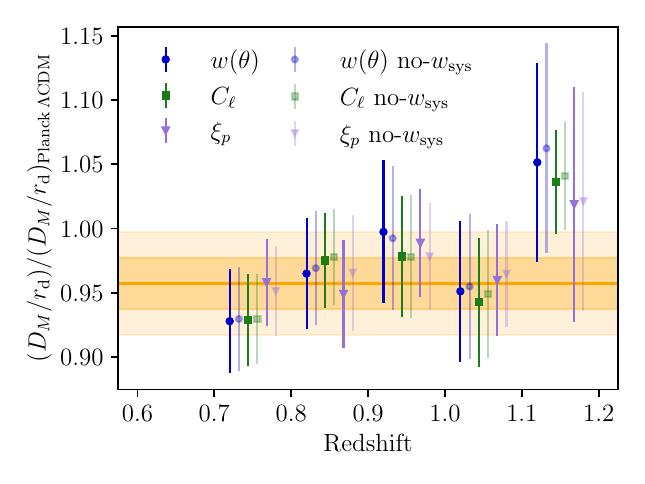}
    \caption{Constraints on $D_M/r_d$ from BAO measurements of individual tomographic bins by ACF, APS and PCF, in blue circles, green squares, and purple triangles, respectively. The $D_M/r_d$ values are normalized by the prediction from Planck, assuming $\Lambda$CDM. The orange bands depict the 1 and 2–$\sigma$ regions from the consensus measurement (AVG, fitting 6 $\zph$-bins simultaneously).
    We remind the reader that we do not find a detection on the first bin (lowest $z$-bin) and this is attributed to sample variance (see \autoref{tab:detection}). The vivid symbols show the measurements accounting for observational systematics (by default in our analysis), whereas the faded symbols show the measurements without those corrections.
     }
    \label{fig:bin_by_bin_bao}
\end{figure}

After unblinding, we find $\alpha= 0.9517 \pm 0.0227$, $\alpha=0.9617  \pm 0.0224$ and $\alpha=0.9553 \pm 0.0201$, for ACF, APS and PCF, respectively. Then, applying \autoref{eq:alpha_average} \& \autoref{eq:sigma_average}, our consensus combined measurement (AVG) is 
\begin{equation}
    \begin{aligned}
    \alpha =  0.9571 &\pm 0.0196\, \,{\rm [ stat. ]},  \\
                    &\pm 0.0041 \, \, {\rm [ sys. ]}, \\
        \alpha =  0.9571 &\pm 0.0201\, \,{\rm [ tot. ]}. 
    \label{eq:results_alpha} 
    \end{aligned}
\end{equation}
We report first the purely statistical error (by computing the $\Delta \chi^2 = 1$ criterion in ACF, APS and PCF and then combining them to AVG with \autoref{eq:sigma_average}), then the systematic error (adding in quadrature the AVG systematics from \autoref{eq:sys_z} \& \autoref{eq:sys_th}), and finally the total error by adding in quadrature the former two. When reporting only two significant figures on the error (as done in the abstract), the total uncertainty is indistinguishable from the statistical one.  

We remind the reader that, for our default analysis, we assume the {\it Planck} cosmology as fiducial (see \datalike in \autoref{sec:setup}). This implies that 
%
\begin{equation}
    D_M(z=0.85)/r_d = 19.51 \pm 0.41 \, \,{\rm [ tot. ]}. 
    \label{eq:results_DM}
\end{equation}

This result represents a $2.1\%$ precision measurement at $\zeff\sim0.85$ of the angular BAO. 
In \autoref{fig:ladder}, we show the value of $D_M/r_d$ divided by Planck's prediction for our Y6 measurement compared to the series of SDSS BAO measurements, and also including the ones from the DES Y1 \cite{Y1_BAO} and Y3 BAO \cite{y3-baokp} analyses. For SDSS, we include the combined BOSS LOWz+CMASS galaxy samples (at $0.2<z<0.5$ and $0.4<z<0.6$) \cite{2017MNRAS.470.2617A}, the eBOSS Luminous Red galaxies (LRG, $0.6<z<1.0$) \cite{2020MNRAS.500..736B,2020MNRAS.498.2492G} and Emission Line galaxies (ELG, $0.6<z<1.1$) \cite{2020MNRAS.499.5527T,2021MNRAS.501.5616D}, as well as the eBOSS Quasars ($0.8<z<2.2$) \cite{Tamone,deMattia} and the Lyman-$\alpha$ combination of auto-correlation and cross-correlation with quasars ($z>2.1$) \cite{2020ApJ...901..153D}. 

\autoref{fig:ladder} represents the state-of-the-art at the closure of stage-III galaxy surveys for angular BAO measurements. It shows that our measurement is competitive with spectroscopic surveys that were designed for BAO science, with the caveat that those surveys also report competitive results from radial BAO and redshift space distortions from anisotropic galaxy clustering. 
In terms of relative uncertainty, our measurement is the most precise angular BAO measurement from a photometric survey at any redshift and also the most precise one from any type of galaxy survey at $\zeff>0.75$. 
Our 2.1\% measurement of $D_M/r_d$ at $\zeff=0.85$ more than doubles the precision of the constraint from eBOSS ELGs at a similar redshift (5.1\% at $\zeff=0.85$). It also exceeds the relative precision of higher redshift measurement from quasar clustering (2.6\% at $\zeff=1.48$) and Lyman-$\alpha$ forests (2.9\% at $\zeff=2.33$). The eBOSS LRG measurement gives a more precise measurement, $\alpha=1.024\pm0.019$ ($\sim1.9\%$), at a lower redshift, $\zeff=0.70$, whereas the BOSS measurements are the most precise ones: 1.5\% at $\zeff=0.38$ and 1.3\% at $\zeff=0.51$. Next generation spectroscopic surveys such as DESI and Euclid are expected to improve upon these constraints. 

All of those measurements report angular distance constraints from post-reconstruction BAO-only fits except for eBOSS ELGs. This case only reports the isotropic BAO ($D_V$) from post-reconstruction, which combines information from the angular ($D_M$) and Hubble distances ($D_H=c/H(z)$) together: $D_V=(z D_M^2 D_H)^{1/3}$.  
In order to compare the purely angular constraints, we chose to show in \autoref{fig:ladder} the constraints on $D_M$ coming from a combination of BAO and redshift space distortions \cite{Tamone,deMattia}, $\alpha=0.962\pm0.049$.
Alternatively, one could pick the isotropic measurement, $\alpha_{\rm iso}=0.986\pm0.032$ at $\zeff=0.85$ (3.3\% precision) \cite{deMattia,raichoor21}, and increase the error-bar by $\times 1.5$ (4.9\%), taking into account that 2/3 of the isotropic information comes from the angular BAO. 
The ELG measurement including RSD agrees in central value with our measurements, as shown in \autoref{fig:ladder}. The isotropic constraint prefers a slightly higher value, but that shift is below half of the sigma reported by eBOSS ELGs.
Other measurements of the BAO in $0.70<\zeff<1.0$ tend to agree with the Planck predictions, but with larger uncertainties, a summary of these can be found in Fig. 17 of \cite{raichoor21}.

\subsection{The BAO signal}

In \autoref{fig:likelihood}, we report the $\Delta \chi^2$ as a function of the BAO shift $\alpha$ for each of the three individual measurements (ACF, APS and PCF). Although not shown, we compared these $\chi^2$ distribution to the assumption of a Gaussian likelihood, finding good agreement to $\sim2-3\sigma$. As hinted in the mock tests, the APS likelihood is found to be less Gaussian than the ones for ACF or PCF. 
For our consensus AVG error, we need to assume a Gaussian likelihood (implicitly assumed throughout \autoref{sec:comb_method}). 
As an alternative, 
we compute the mean of the three $\Delta\chi^2$.
This curve is then shifted and tightened to match the best-fit $\alpha$ value and 1-$\sigma$ error reported in \autoref{eq:results_alpha}, as shown by the orange curve. The four versions of the likelihood will be publicly released, see URL in \autoref{sec:conclusions}.
In colored dashed lines, we also show the $\Delta \chi^2$ obtained when trying to fit the data with a template without BAO. By comparing the curves with and without BAO, we can see a difference in $\chi^2$ of $\sim$12, which implies a detection of the BAO signal at the $\sim3.5\sigma$ level.

The best-fit models are compared to the clustering measurements in \autoref{fig:ACF_data}, \autoref{fig:APS_data} \& \autoref{fig:PCF_data} for ACF, APS and PCF, respectively. In order to highlight the BAO feature we subtract the no-BAO template. 
In the case of PCF, we also show the $N_z=1$ case in which all the BAO signal is concentrated into a single redshift bin, in order to visualize better the BAO feature. This is only possible for this statistic, where the BAO signal is expected to align in the $s_\perp$ coordinate. Nevertheless, $N_z=1$ is not used for our fiducial results of PCF, which rely on using six redshift bins ($N_z=6$) like the ACF and APS cases. 
The raw clustering statistics can be found in the companion paper (\cite{Mena23}, figure 7).

We note again that the fiducial fit is performed over all six redshift bins simultaneously with one single BAO shift $\alpha$. Hence, not all the tomographic bins are necessarily fitted equally well. 
In order to understand better the contribution from each tomographic bin, in \autoref{fig:bin_by_bin_bao} we show the results from fitting each bin individually. As previously noted, we do not have a detection in bin 1, but this is compatible with the results in the mock catalogues (\autoref{sec:blindtest}). The consensus orange band representing the AVG fit from the 6 bins altogether tends to agree more with bins 3, 4 \& 5, whereas bin 2 lies on the lower end (except for PCF), and bin 6 sits at the higher end. Overall, bearing in mind the error bars, we find an agreement between the consensus measurement (orange band, showing the combination of the six bins and the three methods) and the individual (bin and method) measurements (see more quantitative discussions in \autoref{sec:blindtest}). 
We note that the preferred value by each individual bin is somewhat different to the individual-bin information reported in Y3 \citep[Figure 8]{y3-baokp}, whereas the global measurement is very consistent. Nevertheless, only about $\sim$30\% of the galaxies in each Y6 bin were present in the same bin in Y3. Therefore, we expect substantial scatter in the clustering and best-fit BAO per individual bin, where the signal is not so strong. 
The fainter symbols of \autoref{fig:bin_by_bin_bao} show the results when not applying the weights that account for observational systematics, and we find negligible impact in all redshift bins and for all three estimators.

\subsection{Robustness tests}
\label{sec:robustness}

\begin{figure}[hbt!]
    \centering
    \includegraphics[width=\linewidth]{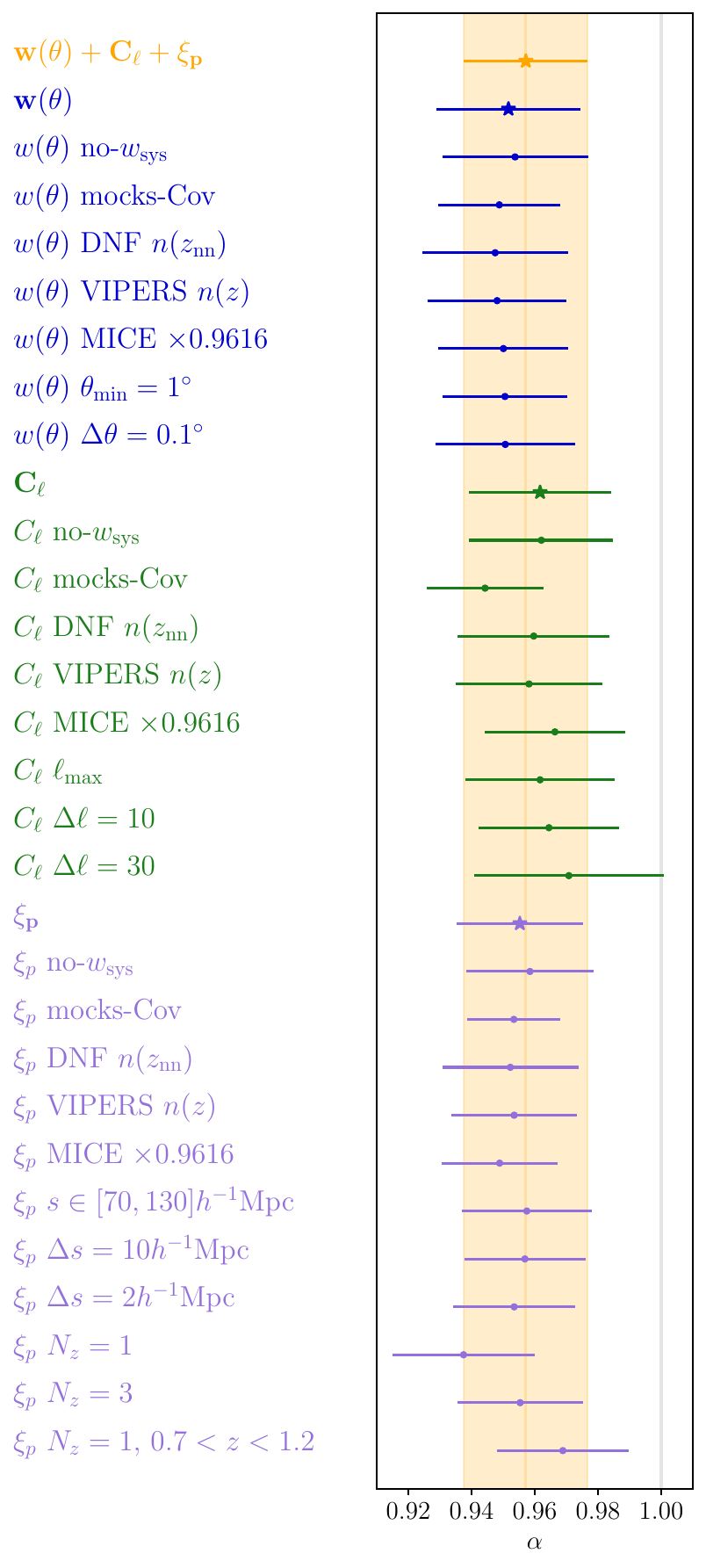}
    \caption{Main BAO measurement shown with an orange star and a shaded area together with several variations of the analysis. Variations of the ACF, APS and PCF analyses are presented in blue, green and purple, respectively. These results are also shown in \autoref{tab:robustness} and discussed in \autoref{sec:robustness}.}
    \label{fig:robustness}
\end{figure}

In this subsection, we evaluate how our main results vary when we change assumptions or choices made during our analysis. 
The variations considered are shown in \autoref{tab:robustness}, where the main $D_M/r_d$ constraints are at the top with the total error included. 
The rest of the constraints are given in terms of $\alpha=(D_M/r_d)/(D_M/r_d)_{\rm Planck}$, reporting their best-fit values and statistical errors. We first show it for our main result (AVG or `$w(\theta)+C_\ell+\xi_p$') and report below the systematic error contribution from the redshift calibration (\autoref{sec:redshift_val}) and from the modeling (\autoref{sec:comb_mocks}). The $\alpha$ values presented in this table with their statistical errors are also shown in \autoref{fig:robustness}. We note again that all these tests were studied first blinded and unblinded a posteriori.

The remainder of the table reports variations from the individual estimators considering only statistical errors. We split this into three parts, one per method: ACF or $w(\theta)$, APS or  $C_\ell$ and PCF or $\xi_p$. We start by reporting the individual fiducial measurement for each of those methods (in {\bf bold}). 
First, we remark on the good agreement between the three methods, with the largest $\alpha$ value preferred by $C_\ell$. This tendency is somewhat different to what we find in the mocks, where the pairing between ACF and APS is more common, and the PCF tends to be lower. However, we already checked in \autoref{sec:blindtest} that these results are statistically compatible with the mocks. The combination AVG indeed represents a good consensus measurement, being closer to ACF and PCF than to APS.

The first robustness test consists in removing the systematic weights (`no-$w_{\rm sys}$'), where we see this has a very small effect (0.09$\sigma$, 0.02$\sigma$ and 0.16$\sigma$ for ACF, APS, PCF, respectively). This highlights again the robustness of BAO against observational systematics. 
We then look at changing the $n(z)$ assumed in the template to that calibrated from DNF $\zmc$ and VIPERS (see discussion in \autoref{sec:redshifts}), finding differences below $0.2\sigma$. We note that these differences are similar to those found in \autoref{sec:redshift_val} and are also accounted for by the systematic error.

We also test changes in some analysis choices such as assuming a \mice template (in this case we multiply the resulting $\alpha$ by 0.9616 so that we can do a direct comparison with the rest of $\alpha$ values reported) or changing the scale cuts or binning. For the PCF, we also include a test changing the number of redshift bins $N_z$ in which we split the sample, which is six for the fiducial case.

All these tests were performed while blinded. At that stage, we paid more attention to the cases in which the shift in $\alpha$ was larger than $\Delta \alpha=0.06$, which is approximately $1/3$ of the forecasted error. The corresponding results are marked in {\it italic} in \autoref{tab:robustness} and discussed in the following: 

\begin{itemize}
    
    \item APS $\Delta\ell=30$.  14.8\% of the mocks present such an extreme change in $\alpha$. We also understand that increasing the binning of $\ell$ can lead to an increment in the errors and a possible shift in the mean value since the wider binning makes it more challenging to resolve the wiggles.
    
    \item PCF MICE. We find a shift of $\alpha_{\rm fid}-\alpha_{\mice}\times0.9616=0.0064$ (raw $\Delta \alpha=0.0332$), but 16\% of the mocks have an equivalent or larger negative shift. We also note that this shift is just below the $\sigma/3$ limit.
    
    \item PCF $N_z=1$. Although only 3.5\% of the mocks present such a big negative shift between $N_z=6$ and $N_z=1$, this fraction rises to $10\%$ if we consider changes in absolute value, and to 12\% if we only consider those mocks with one non-detection.

    To understand the origin of this difference, we look at the results from individually fitting each redshift bin (\autoref{fig:bin_by_bin_bao}). We identify the particularities of bin 6, which prefers a high value of $\alpha$, and of bin 1, which does not have a detection but whose likelihood peaks at low $\alpha$. We already argued in \cite{PCF_Y3_BAO} that combining data at the clustering statistic level (i.e. a single $\xi_p(s_{\perp})$ measurement, $N_z=1$), can give unstable results when the preferred value of $\alpha$ varies significantly from bin to bin.
    
    Hence, we checked the results for $N_z=1$ when removing bin 1,  bin 6, and both. We see that, in these cases, $\alpha$ moves to $1.0178\pm0.0203$ (bins 2-6, last entry in \autoref{tab:robustness} and \autoref{fig:robustness}), $0.9984\pm0.00226$ (bins 1-5), and $0.09983\pm0.0226$ (bins 2-5), respectively. Thus, we conclude that the results for $N_z=1$ are unstable and less reliable than the fiducial 
    analysis (based on $N_z=6$), for which we already studied in \autoref{sec:blindtest} the stability of the results.
    
\end{itemize}

\begin{table}[hbt!]
\centering
\caption{Main results and robustness tests, discussed in detail in \autoref{sec:results} and also represented in \autoref{fig:robustness}. {\it Italic} fonts are used for tests that are found to imprint substantial deviation in either the central value or the uncertainty, accordingly, and are further discussed in the text. Overall, our measurement is very robust.}

\begin{tabular}{lcZ}
  \hline
  \hline
  {\bf Y6 Measurement} & $D_M/r_{\rm d}$  \T  \\   $z_{\rm eff}=0.85$ & $19.51 \pm 0.41$ \B \\
  \hline
  \T case \B & $\alpha$  &  $\chi^2$/d.o.f. \\
  \hline
  \T $\boldsymbol{ w(\theta)+C_\ell+\xi_p}\,$  [Fid.]  & $0.9571 \pm {0.0196}$   & - \\
  \T  Redshift Sys. Err.  &   \hspace{1cm}$ \pm\, 0.0035$   & - \\
  \T  Modelling Sys. Err.  & \hspace{1cm}$ \pm\, 0.0023$   & - \\
  \hline
  \boldsymbol{ $w(\theta)$ }  &    $0.9517 \pm 0.0227$  & 84.9/107 \B \\
  $w(\theta)$ no-$w_{\rm sys}$ &    $0.9538 \pm 0.0231$   & - \B \\
  $w(\theta)$ DNF $n(\zmc)$ &    $0.9475 \pm 0.0230$   & - \B \\
  $w(\theta)$ VIPERS $n(z)$ &    $0.9481 \pm 0.0219$   & - \B \\
  $w(\theta)$  \mice\ $\times 0.9616$  &    $0.9501 \pm 0.0197$   & - \B \\
  $w(\theta)$ $\theta_{\rm min} = 1^{\circ}$  & $0.9506 \pm 0.0226$ & - \\
  $w(\theta)$ $\Delta \theta =0.1^{\circ}$ & $0.9507 \pm 0.0220$ & -  \\
  \hline
  \boldsymbol{ $C_\ell$ }          & $0.9617 \pm {0.0224}$  & - (-) \B \\
    $C_\ell$ no-$w_{\rm sys}$       & $0.9621 \pm 0.0228$          & - (-) \B \\
    $C_\ell$ DNF $n(\zmc)$             & $0.9597 \pm 0.0239$          & -  \B \\
    $C_\ell$ VIPERS $n(z)$          & $0.9582 \pm 0.0232$          & -  \B \\
    $C_\ell$ \mice\ $\times 0.9616$ & $0.9664 \pm 0.0220$          & - (-) \B \\
    $C_\ell$ $\ell_{\rm max}=500$   & $0.9617 \pm 0.0235$          & - -) \\
    $C_\ell$ $\Delta\ell=10$        & $0.9645 \pm 0.0221$          & - (-) \\
    $C_\ell$ $\Delta\ell=30$        & $\mathit{0.9708} \pm \mathit{0.0300}$ &    (-) \\
 \hline
  \boldsymbol{ $\xi_p$ }   &    $0.9553 \pm {0.0201}$     & 44.4 / 95 (=0.47)      \B \\
  $\xi_p$ no-$w_{\rm sys}$   &    $0.9585 \pm 0.0201 $   & 40.0 / 95 (=0.42)      \B \\
  $\xi_p$ DNF $n(\zmc)$       &    $0.9523 \pm 0.0215  $   & 45.0/95 (=0.47)        \B \\
  $\xi_p$ VIPERS $n(z)$    &    $0.9535 \pm 0.0199 $   & 46.6 / 95 (=0.49 )     \B \\
  $\xi_p$  \mice\ $\times  0.9616$ &    $\mathit{0.9489} \pm 0.0184 $   &   46.7 /101 (=0.46)    \B \\
  $\xi_p$ $s\in [70,130] \, h^{-1}$ Mpc &    $  0.9575 \pm 0.0205  $   &  -  (=0.31)    \B \\
  $\xi_p$ $\Delta s_\perp= 10 \, h^{-1}$ Mpc &   $  0.9569 \pm 0.0191 $   & - (=0.56)   \B \\
  $\xi_p$ $\Delta s_\perp= 2 \, h^{-1}$ Mpc  &   $  0.9535 \pm 0.0193  $   & - (-)   \B \\
   $\xi_p$ $N_z=3$          &        $  0.9554  \pm 0.0199 $            &
  - \B \\
  $\xi_p$ $N_z=1$          &         $\mathit{0.9375} \pm 0.0225  $            & 15.6/15 (=1.04)     \B \\
  $\xi_p$ $N_z=1$,  $0.7< z< 1.2$         &     $\mathit{0.9689} \pm 0.0203$            &  ()     \B \\
  \hline
  \hline
  \label{tab:robustness}
\end{tabular}
\end{table}

Regarding the errors, we investigate the cases in which the error changes by more than 0.003, which is approximately 15\% of our fiducial uncertainty. Such a difference only happens for APS $\Delta \ell =30$. We again checked that it is compatible with the shifts in the mocks. 

Additionally, we confirmed that, when assuming the mock covariance, the results do not change in a qualitatively big way. We find some differences in the best fit and error, but these changes are compatible with what we see in the mocks. Since we do not trust the covariance from the mocks due to the spurious correlations discussed in \autoref{sec:sims}, we do not include this test in \autoref{tab:robustness} but it is shown in \autoref{fig:robustness} to understand that the changes are not dramatically different. 

When assuming the \mice cosmology for the template, the results shown in \autoref{tab:robustness} can be combined (AVG) to $\alpha \times 0.9616=0.9529\pm 0.0184$ [stat.], which translates to $D_M/r_s=19.43\pm 0.38$ [stat.]. This shows that, even though the $\alpha$ value depends on the assumed cosmology, the recovered physical constraints remain practically unchanged (in this case within $0.2\sigma$).

Finally, we report the results when considering only the first five bins ($0.6<\zph<1.1$), as discussed 
in \autoref{sec:blindtest} and \appendixcite{app:removing_bin} we discussed the possibility of removing bin 6 as this seemed to reduced the uncertainty. We concluded that it would not be removed for our fiducial analysis, but that it would also be reported. When considering only the first 5 bins ($0.6<\zph<1.1$)
In this case, we obtain $\alpha_{\rm ACF} = 0.9441\pm 0.0220$, $\alpha_{\rm APS} = 0.9478\pm 0.0206$ and $\alpha_{\rm PCF} = 0.9521\pm 0.0203$, leading to $\alpha_{\rm AVG} = 0.9519\pm 0.0195$ [stat.] and
\begin{equation}
    \big(D_M(z=0.85)/r_d \big)_{\zph<1.1}=  19.41\pm  0.40\, \,{\rm [ stat. ]}, 
    \label{eq:results_nobin6}
\end{equation}
which is compatible with our fiducial result, $D_M(z=0.85)/r_d = 19.51 \pm 0.41$. 

With all the tests performed in this section, we conclude that our fiducial result is robust and that it represents well the consensus of the different variations in the analysis and data calibration.

\section{Conclusions}
\label{sec:conclusions}

\subsection{Summary}

We have measured the BAO angular position using galaxy clustering from the final data (Year 6 or Y6) of the Dark Energy Survey with a significance of $3.5\sigma$. 
This measurement translates to a constraint on the ratio of the angular diameter distance to the acoustic scale of $D_M/r_d = 19.51 \pm 0.41 $ at an effective redshift of $\zeff\sim0.85$.
When comparing to the prediction from Planck $\Lambda$CDM cosmology ($D_M/r_d=20.39$), we obtain  $\alpha=(D_M/r_d)/(D_M/r_d)_{\rm Planck}=0.957 \pm 0.020$.

The DES Y6 BAO measurement
\begin{itemize}
    \item represents a 2.1\% precision measurement and it is 2.1$\sigma$ below Planck's prediction. 
    \item is the tightest BAO measurement from a photometric survey. 
    \item is the most precise angular BAO measurement at $\zeff>0.75$ from any survey to date. 
    \item represents a competitive constraint on $D_M/r_s$ even when compared with current results from spectroscopic surveys with BAO as their main science driver. This is clearly well depicted by \autoref{fig:ladder}, which represents the state-of-the-art for the angular BAO distance ladder and its snapshot at the closure of the stage-III dark energy experiments. The second tightest angular BAO constraint at similar redshift comes from eBOSS ELG \cite{deMattia,Tamone} with $\alpha=0.962\pm0.049$ at $\zeff=0.85$, in agreement in central value but with a $\sim \times 2.5$ larger uncertainty.
    \item agrees with previous DES analyses, improving the uncertainties by  $\sim 25\%$ with respect to Y3 \cite{y3-baokp} and by a factor of 2 compared to Y1 \cite{Y1_BAO}. A comparison between Y6 and Y3 data and analysis is detailed in \appendixcite{app:Y3}.
    
    \end{itemize}


For this work, we made use of the final data set from DES, consisting of 6 years (Y6) of observations of the southern galactic sky over $\sim5,000$ deg$^2$ in the optical bands $g$, $r$, $i$, $z$ and $Y$. 
From that data, we have constructed a galaxy sample optimized for BAO science: the Y6 BAO sample, described in our companion paper, \cite{Mena23}. 
To select this sample, we impose a color selection targeting red galaxies at $z>0.6$ (\autoref{eq:redselection}) and a redshift-dependent magnitude cut (\autoref{eq:maglim}) that is tuned to maximize the BAO precision based on Fisher forecasts. This sample is then corrected from observational systematics using the Iterative Systematic Decontamination (ISD) method \cite{y3-galaxyclustering}. 

We split the sample in six tomographic redshift bins (using the $\zph$ estimates from DNF), and we calibrate the redshift distributions ($n(z)$) using three independent methods. These include the Directional Neighboring Fitting (DNF) machine learning photo-$z$ code \cite{DNF}, clustering redshifts by angular cross-correlating our sample with spectroscopic surveys (WZ, following the method from \cite{cawthon22}), and direct calibration with the VIPERS survey \cite{vipers_pdr2}, which is complete for our sample and overlaps in 16 deg$^2$. By studying the impact of the different redshift calibrations on the BAO analysis in \autoref{sec:redshift_val}, we estimate a systematic error on $\alpha$ of $\sigma_{z \rm,  sys}=0.0035$.

We use a template-fitting method to constrain the BAO position using three different clustering estimators: 
the angular correlation function (ACF, $w(\theta)$), the angular power spectrum (APS, $C_\ell$) and the projected correlation function (PCF, $\xi_p$). We then combine the three BAO measurements into our consensus (AVG) constraints by taking into account their correlation.
In \autoref{sec:mocktests} and \autoref{sec:comb_mocks}, the model is optimized and validated against 1952 ICE-COLA mock catalogues described in \autoref{sec:sims} and following the method from \cite{y3-baomocks}. As a result of this validation, we estimate a systematic error from modeling of $\sigma_{\rm th,  sys}=0.0023$.

After validating the method, we run a large set of robustness tests on the data while keeping the results blinded (\autoref{sec:blindtest} and \autoref{tab:robustness}). Eventually, the results were unveiled, obtaining $\alpha_{\rm ACF}= 0.952 \pm 0.023$, $\alpha_{\rm APS}=0.963 \pm 0.022$ and $\alpha_{\rm PCF}=0.955\pm 0.020$ for each of the three estimators, finding them very consistent with each other. The consensus result is $\alpha_{\rm AVG}=0.957 \pm 0.020$, which translates to $D_M(z=0.85)/r_d = 19.51 \pm 0.41$, already including systematic error contributions. We find these results robust to removing parts of the data (individual redshift bins, high-$z$ data, and low-$z$ data) and variations in scale cuts, analysis choices, redshift calibration and treatment of observational systematics.

All the cosmological information obtained in this work is contained in the reported data point, $D_M(z=0.85)/r_d = 19.51 \pm 0.41$ or, more precisely, in the consensus likelihood that will be released in \textsc{Cosmosis}\footnote{\url{https://cosmosis.readthedocs.io/}} once the paper is accepted. A study of the consequences from this measurement on different cosmological parameters and models will follow up in a separate paper.

\subsection{Outlook}

This work does not only report a measurement on the angular BAO position that is among the most precise measurements at {\it high} redshift but also shows the success of the Dark Energy Survey Collaboration to use Galaxy Clustering (GC) from photometric surveys as a robust and competitive probe. Some of the techniques and ideas developed and lessons learned within the DES BAO project were or have the potential to be transferred to other GC analyses and vice-versa. For example, the construction of an optimal sample based on forecasts was done first for BAO \cite{DESY1baosample} and served as an inspiration to later create the MagLim sample \cite{porredon21} for the combination of GC and WL in the so-called 3$\times$2pt analysis \cite{porredon22,y3-3x2ptkp}. The use of APS ($C_\ell$) in DES was first developed for the BAO analysis \cite{Camacho:2018mel} and is now being applied for the combination of GC with WL (3$\times$2pt)\cite{2021MNRAS.505.5714A,2022MNRAS.516.5799C,2022MNRAS.515.1942D,Faga23}. 
Other ideas, such as the PCF method, constructing the order of 2000 realistic simulations to better understand the significance of features in the data, and how to combine different statistics, could potentially also be extrapolated to 3$\times$2pt analyses.

Certainly, some of the lessons learnt from the DES BAO analyses can also be transferred to other surveys, including spectroscopic ones. In particular, DES has clearly pioneered with regard to the blinding policies for BAO, with this work likely being the analysis with the most stringent blinding criteria to this date. 
For upcoming photometric surveys such as Vera Rubin's LSST \cite{2009arXiv0912.0201L} or Euclid \cite{Euclid}, the transfer of the techniques used here is more immediate, as they will need to deal with similar challenges (e.g. the calibration of the redshift distribution and how it affects the inferred cosmology).

With increasingly precise and accurate galaxy clustering measurements from photometric surveys, one can envision other ways to extract cosmological information. One example is the study of Primordial Non-Gaussianities, a probe forecasted to beat CMB and spectroscopic constraints \cite{deputter} if different sources of systematic errors are kept under control. Preparations from DES in this direction are presented in \cite{Riquelme23}.

The other main promising avenue for photometric galaxy clustering is the combination with other probes in order to break parameter degeneracies, check consistency across probes, and mitigate the impact of systematics. In this direction, DES is preparing its final flagship 3$\times$2pt analysis combining three 2-point functions: galaxy position auto-correlation, cosmic shear auto-correlation, and the cross-correlation between galaxy positions and shear. 
DES is also prepared to combine galaxy clustering with many probes, including CMB(-lensing) and galaxy cluster number counts. Additionally, the completed DES supernovae (SN) cosmology results were recently released \cite{SNkey}, constraining the expansion history of the Universe in a complementary way to the BAO. 
In a follow-up work we will study implications of the combined constraints on the expansion history (DES BAO + SN) for cosmological parameters sensitive to it, such us those characterizing dark energy (e.g. $\Omega_\Lambda$, $w$), curvature ($\Omega_k$) or the current  rate of expansion of the Universe ($H_0$). 
Once all other probes finalize their analysis, DES will combine them together, completing its mission of pioneering the field of multi-probe cosmology.

\
\begin{acknowledgments}

{\bf Author Contributions:}
All authors contributed to this paper and/or carried out infrastructure work that made this analysis possible. Some highlighted contributions from the authors of this paper include:
\textit{Scientific management and coordination}: S. Avila and A. Porredon (Large-Scale Structure working group conveners).
\textit{Significant contributions to project development, including paper writing and figures}: S. Avila, H. Camacho, K. C. Chan, J. Mena-Fern\'andez, A. Porredon, M. Rodriguez-Monroy, and E. Sanchez.
\textit{Data analysis and methods validation}: H. Camacho, K. C. Chan, and J. Mena-Fern\'andez.
\textit{Fitting and running simulations}: I. Ferrero.
\textit{Redshift characterisation}: R. Cawthon, G. Giannini, J. De Vicente, J. Mena-Fern\'andez, and L. Toribio San Cipriano.
\textit{Correction for observational systematics}: J. Elvin-Poole, M. Rodriguez-Monroy, I. Sevilla-Noarbe, and N. Weaverdyck.
\textit{Internal reviewing of the paper}: T. M. Davis, W. J. Percival, and A. J. Ross. 
\textit{Advising}: M. Crocce and C. Sánchez.
\textit{Construction and validation of the DES Gold catalog}: M. Adamow, K. Bechtol, A. Carnero Rosell, T. Diehl, A. Drlica-Wagner, R. A. Gruendl, W. G. Hartley, A. Pieres, E. S. Rykoff, I. Sevilla-Noarbe, E. Sheldon, and B. Yanny.
The remaining authors have made contributions to this paper that include, but are not limited to, the construction of DECam and other aspects of collecting the data; data processing and calibration; developing broadly used methods, codes, and simulations; running the pipelines and validation tests; and promoting the science analysis. 

Funding for the DES Projects has been provided by the U.S. Department of Energy, the U.S. National Science Foundation, the Ministry of Science and Education of Spain, 
the Science and Technology Facilities Council of the United Kingdom, the Higher Education Funding Council for England, the National Center for Supercomputing 
Applications at the University of Illinois at Urbana-Champaign, the Kavli Institute of Cosmological Physics at the University of Chicago, 
the Center for Cosmology and Astro-Particle Physics at the Ohio State University,
the Mitchell Institute for Fundamental Physics and Astronomy at Texas A\&M University, Financiadora de Estudos e Projetos, 
Funda{\c c}{\~a}o Carlos Chagas Filho de Amparo {\`a} Pesquisa do Estado do Rio de Janeiro, Conselho Nacional de Desenvolvimento Cient{\'i}fico e Tecnol{\'o}gico and 
the Minist{\'e}rio da Ci{\^e}ncia, Tecnologia e Inova{\c c}{\~a}o, the Deutsche Forschungsgemeinschaft and the Collaborating Institutions in the Dark Energy Survey. 

The Collaborating Institutions are Argonne National Laboratory, the University of California at Santa Cruz, the University of Cambridge, Centro de Investigaciones Energ{\'e}ticas, 
Medioambientales y Tecnol{\'o}gicas-Madrid, the University of Chicago, University College London, the DES-Brazil Consortium, the University of Edinburgh, 
the Eidgen{\"o}ssische Technische Hochschule (ETH) Z{\"u}rich, 
Fermi National Accelerator Laboratory, the University of Illinois at Urbana-Champaign, the Institut de Ci{\`e}ncies de l'Espai (IEEC/CSIC), 
the Institut de F{\'i}sica d'Altes Energies, Lawrence Berkeley National Laboratory, the Ludwig-Maximilians Universit{\"a}t M{\"u}nchen and the associated Excellence Cluster Universe, 
the University of Michigan, NSF's NOIRLab, the University of Nottingham, The Ohio State University, the University of Pennsylvania, the University of Portsmouth, 
SLAC National Accelerator Laboratory, Stanford University, the University of Sussex, Texas A\&M University, and the OzDES Membership Consortium.

Based in part on observations at Cerro Tololo Inter-American Observatory at NSF's NOIRLab (NOIRLab Prop. ID 2012B-0001; PI: J. Frieman), which is managed by the Association of Universities for Research in Astronomy (AURA) under a cooperative agreement with the National Science Foundation.

The DES data management system is supported by the National Science Foundation under Grant Numbers AST-1138766 and AST-1536171.
The DES participants from Spanish institutions are partially supported by MICINN under grants ESP2017-89838, PGC2018-094773, PGC2018-102021, SEV-2016-0588, SEV-2016-0597, and MDM-2015-0509, some of which include ERDF funds from the European Union. IFAE is partially funded by the CERCA program of the Generalitat de Catalunya.
Research leading to these results has received funding from the European Research
Council under the European Union's Seventh Framework Program (FP7/2007-2013) including ERC grant agreements 240672, 291329, and 306478.
We  acknowledge support from the Brazilian Instituto Nacional de Ci\^encia
e Tecnologia (INCT) do e-Universo (CNPq grant 465376/2014-2).

This manuscript has been authored by Fermi Research Alliance, LLC under Contract No. DE-AC02-07CH11359 with the U.S. Department of Energy, Office of Science, Office of High Energy Physics.

KCC is supported by the National Science Foundation of China under the grant number 12273121 and the science research grants from the China Manned Space Project. APo acknowledges support from the European Union’s Horizon Europe program under the Marie Skłodowska-Curie grant agreement 101068581. 

\end{acknowledgments}

\appendix

\section{Unblinding checklist}
\label{app:unblinding}

The checklist to unblind the results goes in the following order:

\begin{enumerate}
    \item Finalize the sample, mask and systematic weights. Also, finalize the decision of fiducial and alternative redshift distributions. These are presented in \autoref{sec:data}. 
    
    \item Finish the validation of the analysis pipeline. Check the robustness of the redshift calibration and of the modeling against the mock catalogues. This validation leads to an estimation of the systematic errors  
    (see \autoref{sec:validation}).

    \item Perform the pre-unblinding tests described in \autoref{sec:blindtest_ind} and \autoref{sec:blindtest_avg}, following the order described there. 

    \item Circulate an advanced draft of this paper with all the previous tests carefully explained to the DES collaboration and request feedback and unblinding approval from the internal reviewers.

    \item Use the blinded data (with a coherent random shift on $\alpha$ and a factor applied to the errors, as described in \autoref{sec:blindtest}) to fill up the robustness tests shown in \autoref{tab:robustness} and \autoref{fig:robustness} with the different obtained $\alpha$. 
    We discuss these tests in \autoref{sec:robustness}.

    \item Check and compare the blinded measurements of $\alpha$ in individual bins to our fiducial measurement (with all 6 bins together). The unblinded version of this figure is \autoref{fig:bin_by_bin_bao}.

   \item Unblind the errors $\sigma_{\alpha}$. At this advanced stage, this allowed us to check whether the errors met our expectations and to understand better the significance of the relative changes in $\alpha$ we saw when performing the robustness tests. For example, we found that in general our errors are larger than the mean error from the mocks. However (for ACF), we checked that 12\% of the mocks have an error larger than what we measure on the data. This rises to 26\% if we only look at the mocks with a non-detection on bin 1. 

    \item Last, present a new draft of this and the companion paper \cite{Mena23} to the collaboration\footnote{  At this stage, we received a comment on the possible relevance of the $Y$-band on the systematic weights. After some investigation, this led us to an update of the systematic weights described in \cite{Mena23}. The pre-unblinding tests on this and previous sections were updated without any qualitative major difference.}. We also show our results in a video conference and, provided no further tests are required, we proceed to unblind.

\end{enumerate}

This stage-by-stage unblinding aims to test our analysis and data without knowing the implications for cosmology. For that reason, we start with the parts that are further away from this information, and eventually get closer and closer to the measurement of $\alpha$ once we take the corresponding decisions based on the previous step. 
At the final unbinding phase (last point above), we hope all the tests we would like to run on the data are already run. Nevertheless, we would allow further investigation if new tests are considered necessary once the data is unblinded. The guiding philosophy for the decision-making in that case would be similar to the one presented in \autoref{sec:blindtest}.
For example, one case discussed before unblinding the errors was the possibility of AVG showing a larger error than one of the individual estimators (ACF, APS or PCF) or the combination of two of them. We did not find an easy implementation of these tests prior to unblinding them, so we decided that this would be tested once the errors were unblinded. If we were to find that possibility, we would then use the statistics from the mocks, similarly to what we did in 
\textcolor{blue}{Appendices} \ref{app:removing_bin}  \& \ref{app:non-detection} 
for the cases of the $z$-bin that enlarges the error and a single $z$-bin with no detection. We would thus complete a table with the mocks that have that property (e.g. AVG having a larger error than ACF) and check for this case which error estimation is better behaved (compared to the scatter $\sigma_{68}$ or $\sigma_{\rm std}$) and if the bias ($\langle \alpha \rangle -1$) is worse for one of the two options.  

At the point of unblinding the paper draft was nearly final from \autoref{sec:data} to \autoref{sec:blindtest} and parts of  \autoref{sec:results} (Results) were completed with blinded data/figures. The discussion of robustness tests (\autoref{sec:robustness}) was also completed.

\section{Comparison with previous DES BAO analyses}
\label{app:Y3}

We have reduced the uncertainty by $\sim 25 \%$ with respect to our previous analysis based on the first 3 years of observations (Y3) \cite{y3-baokp} and by a factor of 2 with respect to the results from the first year (Y1) \cite{Y1_BAO}. Here we summarize some key differences: 

\begin{itemize}
    \item The most obvious change is that we include the data from the completed survey (6 years), approximately doubling the cumulative exposure time with respect to Y1 and Y3, impacting mostly the maximum depth in our galaxy sample.  
    In Y6, the depth in r, i and z bands increase by 0.7, 0.6, 0.5 magnitudes respectively.
    The improvement in depth resulted in more precise measurement of band fluxes and, hence, better estimates of redshifts. The Y6 data also has a more uniform depth coverage, and so observational systematics affecting galaxy counts across the sky are less pronounced. 
    On the footprint, the area considered for each analysis is 1,336 deg${}^2$, 4,108.47 deg${}^2$, and 4,273.42 deg${}^2$ for the Y1, Y3, and Y6 samples, respectively.\footnote{We note that we have used slightly different masking criteria across data batches.}
    
    \item While in Y3 we used the sample selection criteria optimized in Y1, in Y6, we have re-optimized the selection in the $i$ band using a Fisher forecast. This results in a sample containing nearly twice the number of galaxies: 15,937,556 galaxies in the Y6 sample vs 7,031,993 in the Y3 one. Given the increased depth, we also pushed the sample to a higher redshift, $\zph=1.2$, compared to $\zph=1.1$ in Y3 and $\zph=1.0$ in Y1.

    \item On the mitigation of observational systematics, we have required a more restrictive threshold ($T_{1D}=2$, see \autoref{sec:systematics} for details) on the correlation between survey property maps and our galaxy density maps. We have also imposed the masking of outliers using two additional maps that trace artefacts and galactic cirrus in the footprint (see \autoref{sec:mask}). In addition, we have accounted for residual additive stellar contamination, although we argue its impact in the BAO analysis would be negligible (see \autoref{sec:stellar_contamination}). 

    \item In Y6, on top of the DNF redshift characterisation (used in Y1 and Y3) and the validation with VIPERS (also used in Y3), we have also incorporated a further validation with clustering redshifts \autoref{sec:redshifts}. Unlike in our previous analyses, in Y6, we have propagated the impact from different $n(z)$ calibrations down to the BAO shift parameter $\alpha$, being able to quantify the systematic contribution from \autoref{sec:redshift_val}.

    \item For the first time in DES, we report our main results from the statistical combination (AVG) of three clustering estimators: ACF, APS, and PCF. In Y1, the main BAO results were reported only from the ACF method, although results from APS and PCF were shown, they were not considered sufficiently matured or robust at that stage. The Y3 main BAO results were reported by taking the log-mean likelihood of ACF and APS, whereas the PCF method was improved and reported results using Y3 data at a later stage \cite{PCF_method_Y3,PCF_Y3_BAO}.  
  
    \item Although the pre-unblinding tests of Y6 are mostly based on those defined in Y3, we have extended some of these tests by having blinded versions of \autoref{fig:preunblinding_tests}, \autoref{fig:bin_by_bin_bao} and \autoref{fig:robustness} (see also the discussion in \autoref{sec:blindtest} and \appendixcite{app:unblinding}).    

    \item When comparing the results, Y1, Y3 and Y6 are compatible. The results from Y3 and Y6 are both below Planck's prediction by just over $2\sigma$, having slightly different best-fit values and, the latter, with a 25\% tighter error.  
\end{itemize}

\section{Error reduction upon removal of one bin}
\label{app:removing_bin}

\begin{table}[th]
    \centering
    \setlength{\tabcolsep}{4pt} 
    \caption{ $\Delta\sigma<0$ test. 
     We select the \mocks\ for which the error on the ACF $\alpha$ decreases when we remove one redshift bin ($(\sigma-\sigma_{\rm All \ Bins})/\sigma_{\rm All \ Bins}<0$). This table has 6 sections, one corresponding to each of the 6 redshift bins meeting the condition above. Each section contains 2 entries: one where we have removed the bin with $(\sigma-\sigma_{\rm All \ Bins})/\sigma_{\rm All \ Bins}<0$ and one where we consider all the dataset. See \autoref{tab:ACF_mocks} for the definition of the summary statistics. On the last column, we report the fraction of mocks selected in each case over the entire 1952 mocks. 
    }
    \label{tab:sigma_neg}
    \begin{tabular}{ZlccccZZZZZZZcZ}
    \toprule
     & Bins & $\langle\alpha\rangle$ & $\sigma_{\rm std}$ & $\sigma_{68}$ & $\langle\sigma_\alpha\rangle$ & fraction encl.$\langle\alpha\rangle$ & $\langle d_{\rm norm}\rangle$ & $\sigma_{d_{\rm norm}}$ & $\langle\chi^2\rangle/$d.o.f. & mean of mocks & $\sigma_{\rm std}/\langle\alpha\rangle$ & $N\sigma$ & fraction of cases & p-value \\
    \midrule
    \midrule
    0 & 23456 & 1.0048 & 0.0237 & 0.0236 & 0.0203 & 61.4$\%$ & -0.0156 & 1.1631 & 59.3$/$89 & 0.0000$\pm$0.0000 & 2.36$\%$ & 0.20 & 12.35$\%$ & 0.99 \\
    1 & All & 1.0058 & 0.0217 & 0.0211 & 0.0210 & 67.6$\%$ & 0.0010 & 1.0487 & 79.2$/$107 & 0.0000$\pm$0.0000 & 2.16$\%$ & 0.27 & 12.35$\%$ & 0.98 \\
    \midrule
    2 & 13456 & 1.0045 & 0.0277 & 0.0252 & 0.0209 & 56.7$\%$ & -0.0035 & 1.2918 & 62.9$/$89 & 0.0000$\pm$0.0000 & 2.76$\%$ & 0.16 & 9.12$\%$ & 0.98 \\
    3 & All & 1.0050 & 0.0237 & 0.0239 & 0.0216 & 61.8$\%$ & -0.0110 & 1.1000 & 78.4$/$107 & 0.0000$\pm$0.0000 & 2.36$\%$ & 0.21 & 9.12$\%$ & 0.98 \\
    \midrule
    4 & 12456 & 1.0047 & 0.0283 & 0.0267 & 0.0205 & 54.0$\%$ & -0.0216 & 1.3369 & 64.7$/$89 & 0.0000$\pm$0.0000 & 2.82$\%$ & 0.17 & 9.68$\%$ & 0.98 \\
    5 & All & 1.0059 & 0.0235 & 0.0237 & 0.0212 & 63.5$\%$ & -0.0121 & 1.0870 & 79.5$/$107 & 0.0000$\pm$0.0000 & 2.34$\%$ & 0.25 & 9.68$\%$ & 0.98 \\
    \midrule
    6 & 12356 & 1.0011 & 0.0238 & 0.0253 & 0.0205 & 55.8$\%$ & -0.0134 & 1.1685 & 63.0$/$89 & 0.0000$\pm$0.0000 & 2.38$\%$ & 0.05 & 8.91$\%$ & 0.98 \\
    7 & All & 1.0041 & 0.0230 & 0.0240 & 0.0213 & 60.9$\%$ & -0.0030 & 1.0745 & 78.5$/$107 & 0.0000$\pm$0.0000 & 2.29$\%$ & 0.18 & 8.91$\%$ & 0.98 \\
    \midrule
    8 & 12346 & 1.0020 & 0.0254 & 0.0254 & 0.0206 & 56.5$\%$ & -0.0035 & 1.2200 & 64.0$/$89 & 0.0000$\pm$0.0000 & 2.53$\%$ & 0.08 & 9.53$\%$ & 0.98 \\
    9 & All & 1.0034 & 0.0234 & 0.0229 & 0.0211 & 59.7$\%$ & -0.0026 & 1.0949 & 77.4$/$107 & 0.0000$\pm$0.0000 & 2.33$\%$ & 0.15 & 9.53$\%$ & 0.99 \\
    \midrule
    10 & 12345 & 1.0057 & 0.0228 & 0.0237 & 0.0197 & 62.2$\%$ & 0.0017 & 1.1741 & 67.1$/$89 & 0.0000$\pm$0.0000 & 2.27$\%$ & 0.25 & 10.04$\%$ & 0.96 \\
    11 & All & 1.0060 & 0.0219 & 0.0211 & 0.0200 & 65.3$\%$ & 0.0053 & 1.1041 & 77.9$/$107 & 0.0000$\pm$0.0000 & 2.18$\%$ & 0.27 & 10.04$\%$ & 0.98 \\
    \bottomrule
    \bottomrule
    \end{tabular}
\end{table}

\begin{table}[t]
    \centering
    \setlength{\tabcolsep}{4pt} 
    \caption{
    Same as Table \ref{tab:sigma_neg} except for $\Delta\sigma<-0.03$ test. 
    }
        \label{tab:sigma_m3}
    \begin{tabular}{ZlccccZZZZZZZcZ}
    \toprule
     & Bins & $\langle\alpha\rangle$ & $\sigma_{\rm std}$ & $\sigma_{68}$ & $\langle\sigma_\alpha\rangle$ & fraction encl.$\langle\alpha\rangle$ & $\langle d_{\rm norm}\rangle$ & $\sigma_{d_{\rm norm}}$ & $\langle\chi^2\rangle/$d.o.f. & mean of mocks & $\sigma_{\rm std}/\langle\alpha\rangle$ & $N\sigma$ & fraction of cases & p-value \\
    \midrule
    0 & 23456 & 1.0050 & 0.0248 & 0.0244 & 0.0218 & 61.9$\%$ & -0.0318 & 1.1408 & 59.9$/$89 & 0.0000$\pm$0.0000 & 2.47$\%$ & 0.20 & 4.30$\%$ & 0.99 \\
    1 & All & 1.0046 & 0.0220 & 0.0185 & 0.0233 & 75.0$\%$ & -0.0007 & 0.9666 & 82.5$/$107 & 0.0000$\pm$0.0000 & 2.19$\%$ & 0.21 & 4.30$\%$ & 0.96 \\
    \midrule
    2 & 13456 & 1.0004 & 0.0314 & 0.0297 & 0.0225 & 58.2$\%$ & 0.0534 & 1.3834 & 63.9$/$89 & 0.0000$\pm$0.0000 & 3.14$\%$ & 0.01 & 2.82$\%$ & 0.98 \\
    3 & All & 1.0033 & 0.0264 & 0.0244 & 0.0242 & 67.3$\%$ & 0.0212 & 1.1171 & 81.6$/$107 & 0.0000$\pm$0.0000 & 2.63$\%$ & 0.13 & 2.82$\%$ & 0.97 \\
    \midrule
    4 & 12456 & 1.0057 & 0.0320 & 0.0280 & 0.0215 & 56.0$\%$ & -0.0478 & 1.4691 & 66.0$/$89 & 0.0000$\pm$0.0000 & 3.18$\%$ & 0.18 & 3.84$\%$ & 0.97 \\
    5 & All & 1.0061 & 0.0260 & 0.0243 & 0.0230 & 62.7$\%$ & -0.0203 & 1.1197 & 82.2$/$107 & 0.0000$\pm$0.0000 & 2.58$\%$ & 0.23 & 3.84$\%$ & 0.96 \\
    \midrule
    6 & 12356 & 0.9946 & 0.0229 & 0.0249 & 0.0217 & 60.5$\%$ & -0.0236 & 1.0765 & 63.3$/$89 & 0.0000$\pm$0.0000 & 2.30$\%$ & 0.24 & 3.89$\%$ & 0.98 \\
    7 & All & 0.9983 & 0.0233 & 0.0242 & 0.0232 & 64.5$\%$ & -0.0007 & 1.0039 & 79.9$/$107 & 0.0000$\pm$0.0000 & 2.33$\%$ & 0.07 & 3.89$\%$ & 0.98 \\
    \midrule
    8 & 12346 & 0.9986 & 0.0312 & 0.0299 & 0.0230 & 53.3$\%$ & 0.0239 & 1.4042 & 64.5$/$89 & 0.0000$\pm$0.0000 & 3.12$\%$ & 0.04 & 2.31$\%$ & 0.98 \\
    9 & All & 1.0007 & 0.0259 & 0.0224 & 0.0246 & 66.7$\%$ & 0.0361 & 1.1162 & 79.6$/$107 & 0.0000$\pm$0.0000 & 2.59$\%$ & 0.03 & 2.31$\%$ & 0.98 \\
    \midrule
    10 & 12345 & 1.0037 & 0.0315 & 0.0312 & 0.0238 & 46.7$\%$ & 0.0769 & 1.3453 & 71.2$/$89 & 0.0000$\pm$0.0000 & 3.14$\%$ & 0.12 & 0.77$\%$ & 0.92 \\
    11 & All & 1.0040 & 0.0267 & 0.0272 & 0.0252 & 66.7$\%$ & 0.0715 & 1.1388 & 84.3$/$107 & 0.0000$\pm$0.0000 & 2.66$\%$ & 0.15 & 0.77$\%$ & 0.95 \\
    \midrule
    \bottomrule
    \end{tabular}
\end{table}

Here, we investigate the scenario when removing one tomographic bin reduces the estimated error $\sigma_\alpha$. In this case, we want to find out what is the best course of action to take. Potentially, we picture two solutions: taking the error and best fit from the full data set (the 6 bins) or the reduced data set with one bin eliminated (5 bins). 

In \autoref{tab:sigma_neg}, we select the mocks for which $(\sigma-\sigma_{\rm All \ Bins})/\sigma_{\rm All \ Bins}<0$ when removing one redshift bin. For these mocks, we perform the ACF fit with and without the bin causing $\Delta \sigma<0$ and 
compare the different summary statistics. 
We find that even though the estimated error $\langle \sigma_\sigma \rangle$ reduces when removing those bins, the actual scatter of the best fit  ($\sigma_{68}$ or $\sigma_{\rm std}$) increases. Hence, we conclude that this reduction is due to an underestimation of the error on the 5-bin cases, not to an actual gain in information, and that we should use the results from the combined 6-bins (`All'). We also see in the last column that, for each of the 6 bins, typically $\sim10\%$ of the mocks have a $\Delta \sigma <0$.

If we try to pin down the cases that are more similar to our result on the data ACF (\autoref{tab:ACF_mocks}), we can select the mocks for which $100(\sigma-\sigma_{\rm All \ Bins})/\sigma_{\rm All \ Bins}\leq-3$. This is the case shown in \autoref{tab:sigma_m3}, where we find the exact same effect as in the previous paragraph, but now with augmented differences.

\section{Non-detections: including them in the fit}
\label{app:non-detection}

\begin{table}[]
    \centering
    \setlength{\tabcolsep}{4pt} 
    \caption{Non-detection test. We select the \mocks\ for which there is a non-detection in bins 1 to 6 and analyse them in the 6 sections of this table. Each section contains 2 entries: one where we have removed the bin with non-detection and one where we consider all the dataset. See \autoref{tab:ACF_mocks} for the definition of the summary statistics. On the last column, we report the fraction of mocks selected in each case over the entire 1952 mocks.  }
    \label{tab:non-detections}
    \begin{tabular}{ZlccccZZZZZZZcZ}
    \toprule
     & Bins & $\langle\alpha\rangle$ & $\sigma_{\rm std}$ & $\sigma_{68}$ & $\langle\sigma_\alpha\rangle$ & fraction encl.$\langle\alpha\rangle$ & $\langle d_{\rm norm}\rangle$ & $\sigma_{d_{\rm norm}}$ & $\langle\chi^2\rangle/$d.o.f. & mean of mocks & $\sigma_{\rm std}/\langle\alpha\rangle$ & $N\sigma$ & fraction of cases & p-value \\
    \midrule
    \midrule
    0 & 23456 & 1.0084 & 0.0213 & 0.0196 & 0.0208 & 68.6$\%$ & 0.0020 & 1.0154 & 59.2$/$89 & 0.0000$\pm$0.0000 & 2.11$\%$ & 0.39 & 9.63$\%$ & 0.99 \\
    1 & All & 1.0078 & 0.0222 & 0.0222 & 0.0207 & 66.0$\%$ & -0.0013 & 1.0623 & 79.4$/$107 & 0.0000$\pm$0.0000 & 2.20$\%$ & 0.35 & 9.63$\%$ & 0.98 \\
    \midrule
    2 & 13456 & 1.0054 & 0.0229 & 0.0222 & 0.0217 & 66.0$\%$ & 0.0179 & 1.0442 & 62.5$/$89 & 0.0000$\pm$0.0000 & 2.28$\%$ & 0.24 & 4.97$\%$ & 0.99 \\
    3 & All & 1.0053 & 0.0230 & 0.0233 & 0.0217 & 62.9$\%$ & 0.0240 & 1.0435 & 78.4$/$107 & 0.0000$\pm$0.0000 & 2.29$\%$ & 0.23 & 4.97$\%$ & 0.98 \\
    \midrule
    4 & 12456 & 1.0062 & 0.0258 & 0.0277 & 0.0217 & 56.9$\%$ & 0.0510 & 1.1646 & 66.5$/$89 & 0.0000$\pm$0.0000 & 2.56$\%$ & 0.24 & 2.61$\%$ & 0.96 \\
    5 & All & 1.0060 & 0.0245 & 0.0259 & 0.0216 & 51.0$\%$ & 0.0381 & 1.1279 & 80.2$/$107 & 0.0000$\pm$0.0000 & 2.44$\%$ & 0.24 & 2.61$\%$ & 0.98 \\
    \midrule
    6 & 12356 & 1.0085 & 0.0248 & 0.0236 & 0.0225 & 63.8$\%$ & -0.0106 & 1.0801 & 65.2$/$89 & 0.0000$\pm$0.0000 & 2.46$\%$ & 0.34 & 2.41$\%$ & 0.97 \\
    7 & All & 1.0092 & 0.0232 & 0.0232 & 0.0221 & 66.0$\%$ & -0.0038 & 1.0526 & 79.6$/$107 & 0.0000$\pm$0.0000 & 2.30$\%$ & 0.40 & 2.41$\%$ & 0.98 \\
    \midrule
    8 & 12346 & 1.0031 & 0.0241 & 0.0245 & 0.0221 & 60.0$\%$ & -0.0136 & 1.1005 & 64.7$/$89 & 0.0000$\pm$0.0000 & 2.40$\%$ & 0.13 & 3.33$\%$ & 0.98 \\
    9 & All & 1.0044 & 0.0242 & 0.0248 & 0.0220 & 58.5$\%$ & -0.0096 & 1.1054 & 78.1$/$107 & 0.0000$\pm$0.0000 & 2.41$\%$ & 0.18 & 3.33$\%$ & 0.98 \\
    \midrule
    10 & 12345 & 1.0040 & 0.0202 & 0.0196 & 0.0202 & 67.2$\%$ & -0.0214 & 1.0154 & 66.2$/$89 & 0.0000$\pm$0.0000 & 2.01$\%$ & 0.20 & 8.76$\%$ & 0.97 \\
    11 & All & 1.0064 & 0.0206 & 0.0206 & 0.0201 & 65.5$\%$ & -0.0305 & 1.0355 & 77.1$/$107 & 0.0000$\pm$0.0000 & 2.05$\%$ & 0.31 & 8.76$\%$ & 0.99 \\
    \bottomrule
    \end{tabular}

\end{table}

A case that requires our attention is when one of the bins does not show a detection. In this case we wonder if it is better to estimate $\alpha$ and its error from the whole data set (6 bins) or to eliminate the non-detection bin (5 bins).

The results are presented in \autoref{tab:non-detections} for ACF, where we compare the summary statistics of the best fits for the results without the non-detection bin (e.g.  "23456") and with it ("All"). In this case the results for the comparison of the estimated error ($\langle \sigma_\alpha \rangle$) and the scatter measures ($\sigma_{68}$ or $\sigma_{\rm std}$) are somewhat heterogeneous and it is hard to draw strong conclusions. 
Regarding the mean $\langle \alpha \rangle$ for the 23456 case, we seem to find a larger bias ($\langle \alpha \rangle -1$ ) than when considering the whole data set, although this situation changes when the bin under consideration is another one. 
In the absence of strong preference shown by this test, we move forward with our standard analysis, which includes the 6 bins altogether.

\section{Additional tables for pre-unblinding tests on data}
\label{app:preunblind-tab}

This appendix includes the tables used for the set of pre-unblinding tests performed to assess the robustness of our measurements, discussed in \autoref{sec:blindtest}. These sets of tests are summarized in \textcolor{blue}{Tables} \ref{tab:unblind_acf}, \ref{tab:unblind_aps} \& \ref{tab:unblind_pcf} for ACF, APS and PCF, respectively. 
We test how much the best fit $\alpha$ changes ($\Delta \alpha$, in \%) when we modify some choice in the analysis and compare it with the mocks. For each test, we print the minimum and maximum value $\Delta \alpha$ that contains 90\%, 95\%, 97\% and 99\% of the mocks, with equal number of mocks left outside each of the two extremes. We analyse the mocks with the \textit{mock-like} setup, and the data by default with the \textit{data-like} setup (labeled as {\planck}). For the tests that consist in removing part of the data, we also repeat them on the data assuming \mice cosmology (\textit{data-like-mice}), but this is considered a secondary test.
In the main text (\autoref{sec:blindtest}), we only show the results for the (\planck) data in \autoref{fig:preunblinding_tests} \& \autoref{fig:preunblinding_tests_2}, but the quantitative decision for the fail/pass criteria comes from the tables shown in this appendix. 

Specifically, we assess the impact of removing one tomographic bin at a time, removing the high- or low- redshift parts of the data, of changing the template cosmology, the covariance and the $n(z)$ estimation. For each test, we report variations in the best-fit $\alpha$ with respect to our fiducial analysis. For all cases except for the template cosmology, the covariance, and the $n(z)$ estimation, we also test the impact on the estimated uncertainty $\sigma_{\alpha}$, which is displayed in the bottom part of each table. While we do not impose strict pre-unblinding criteria for the changes in $\sigma_{\alpha}$, we regard them as informative.

\begin{table*}[h!]
\caption{
Table of pre-unblinding tests for the Angular Correlation Function from \autoref{sec:blindtest}, showing the impact of removing individual/several tomographic bins, of changing the assumed cosmology for the BAO template, changing the covariance and of considering an different estimate of the true redshift distributions. 
We report variations in $\alpha$ with respect to our fiducial analysis, to keep results blind. 
The middle  four (double) columns show the range of $\Delta \alpha$ values measured on the \mocks\ that enclose the fraction of mocks shown at the top of each column. The mocks are analysed with the \mocklike setup (\mice cosmology, $\sim n(\zmc)$, $b_{\rm mocks}$). The last column shows the $\Delta \alpha$ value measured on the data, by default for the \mocklike setup (\planck, $n(z_{\rm fid})$, $b_{\rm pl,data}$, main results), but for some tests also with the \datalikemice setup (\mice, $n(z_{\rm fid})$, $b_{\rm pl,data}$, secondary results). 
We mark in {\bf bold} the tests that fail on the data column and on the boundary that has been surpassed.
The bottom rows show the impact on the error in $\alpha$ of removing one/several tomographic bins of the data (although we do not impose specific criterion in these tests).
}
\label{tab:unblind_acf}
\begin{tabular}{l|ll|ll|ll|ll||c|c}
\hline
\hline
Threshold & \multicolumn{2}{l|}{90 \%} & \multicolumn{2}{l|}{95 \%} & \multicolumn{2}{l|}{97 \%} & \multicolumn{2}{l||}{99 \%} & \multicolumn{2}{l}{data} \\ \cline{2-11}
(Fraction of mocks) & min & max & min & max & min & max & min & max & {\tt \gray{MICE}} & {\bf Planck} \\ \hline
           \T        & \multicolumn{8}{c||}{$10^2 (\alpha-\alpha_{\rm fiducial})$}                      \B                                                                                 \\ \cline{2-11}

   Bins     23456 & -1.33 & 1.43 & -1.79 & 1.86 & -2.10 &2.17 & -2.44 & 2.76 & \gray{0.75} &  1.15 \\
   Bins     13456 & -1.39 & 1.63 & -1.83 & 1.99 & -2.03 &2.30 & -2.80 & 3.13 & \gray{1.03} &  1.47 \\
   Bins     12456 & -1.37 & 1.51 & -1.71 & 2.00 & -2.03 &2.35 & -2.52 & 3.23 & \gray{-0.21} & -0.39 \\
   Bins     12356 & -1.45 & 1.27 & -1.81 & 1.57 & -2.19 &1.88 & -2.80 & 2.76 & \gray{-0.66} & -0.27 \\
   Bins     12346 & -1.21 & 1.11 & -1.51 & 1.41 & -1.79 &1.72 & -2.48 & 2.02 & \gray{0.37} &  0.30 \\
   Bins     12345 & -0.86 & 0.76 & -1.07 & 0.96 & -1.30 &1.15 & -1.63 & 1.65 & \gray{-0.68} & -0.76 \\
   Bins       456 & -2.85 & 3.73 & -3.42 & 4.85 & -3.86 & 5.54 & -5.00 & 7.90 &  \gray{3.26} &  3.41 \\
   Bins       123 & -3.30 & 2.65 & -4.27 & 3.45 & -5.04 & 4.26 & -6.80 & 5.56 & \gray{-1.55} & -1.58 \\
   Bins      1234 & -1.83 & 1.67 & -2.25 & 2.13 & -2.55 & 2.35 & -3.67 & 3.22 & \gray{-0.39} & -0.70 \\
   \midrule
     Template Cosmo & -0.33 & 0.48 & -0.40 & 0.60 & -0.44 & 0.68 & -0.55 & 0.89 & \gray{x} &  0.17 \\
  Covariance &    -0.46 & 0.42 & -0.58 & 0.54 & -0.68 & 0.64 & -0.83 & 0.82 & \gray{x} & -0.42  \\
  $n(z)\ \zmc-$ fid & -0.56 & 0.08 & -0.60 & 0.14 & -0.64 & 0.20 & -0.72 & 0.31 & \gray{x} & -0.42 \\

\hline  \T      & \multicolumn{8}{c||}{100 $(\sigma-\sigma_{\rm All \ Bins})/\sigma_{\rm All \ Bins}$}            \B                                                                                           \\ \cline{2-11}

   Bins   23456 & -2.47 & 25.15 & -4.33 & 30.34 & -6.09 & 35.42 & -9.08 & 41.50  & \gray{5.37}  & 3.96  \\
   Bins   13456 & -1.60 & 26.16 & -3.55 & 31.21 & -5.18 & 35.18 & -8.95 & 45.61  & \gray{18.05} & 14.54 \\
   Bins   12456 & -2.00 & 26.22 & -4.53 & 31.44 & -5.84 & 36.80 & -8.93 & 45.86  & \gray{18.05} & 14.98 \\
   Bins   12356 & -2.29 & 25.17 & -4.09 & 30.79 & -5.51 & 35.11 & -9.35 & 41.35  & \gray{8.29} & 4.41  \\
   Bins   12346 & -1.39 & 19.89 & -2.84 & 24.51 & -4.07 & 27.92 & -6.22 & 34.80  & \gray{7.32}  & 7.93  \\
   Bins   12345 & -0.66 & 11.94 & -1.45 & 14.87 & \textbf{-1.97} & 17.79 & -3.56 & 22.50  & \gray{0.49}  & \textbf{-3.08} \\
   Bins     456 & 12.08 & 94.25 & 8.20 & 114.76 & 5.13 & 128.50 & -1.76 & 166.46 & \gray{66.34} & 57.71 \\
   Bins     123 & 10.14 & 80.86 & 5.92 & 95.62  & 3.02 & 109.42 & -2.36 & 144.43 & \gray{21.95} & 18.50 \\
   Bins    1234 & 1.37  & 35.50 & -0.99 & 42.58 & -1.84 & 45.70 & -4.23 & 55.74  & \gray{7.80}  & 3.96\\

\hline
\end{tabular}
\end{table*}

\begin{table*}[h!]
\caption{
Table of pre-unblinding tests for the Angular Power Spectrum (APS) from \autoref{sec:blindtest}. See description in \autoref{tab:unblind_acf} and text.  
}
\label{tab:unblind_aps}
\begin{tabular}{l|ll|ll|ll|ll||l|l}
\hline
\hline
Threshold       & \multicolumn{2}{l|}{0.9} & \multicolumn{2}{l|}{0.95} & \multicolumn{2}{l|}{0.97} & \multicolumn{2}{l||}{0.99} & \multicolumn{2}{l}{data} \\ \cline{2-11}
(Fraction of mocks)                &          min &          max &          min &          max &          min&          max &          min&          max & {\tt MICE}       & {\tt Planck}       \\ \hline
       \T        & \multicolumn{8}{c||}{$10^2 (\alpha-\alpha_{\rm fiducial})$}                      \B                                                                                 \\ \cline{2-11}            
       
Bins 23456 & -1.18 & 1.45 & -1.59 & 1.85 & -2.01 & 2.12 & -2.99 & 3.19 & \gray{0.24} & 0.54 \\
Bins 13456 & -1.46 & {\bf 1.55} & -1.88 & 2.26 & -2.24 & 2.76 & -3.54 & 3.60 & \gray{1.37} & {\bf 1.66} \\
Bins 12456 & -1.32 & 1.48 & -1.82 & 2.07 & -2.14 & 2.71 & -2.75 & 4.22 & \gray{-0.18} & -0.25 \\
Bins 12356 & -1.55 & 1.28 & -2.05 & 1.79 & -2.63 & 2.10 & -4.36 & 3.08 & \gray{-0.20} & -0.21 \\
Bins 12346 & -1.48 & 1.43 & -2.01 & 1.91 & -2.67 & 2.49 & -3.96 & 3.31 & \gray{1.22} & 0.64 \\
Bins 12345 & -1.59 & 1.45 & -2.10 & 2.00 & -2.67 & 2.42 & -3.95 & 3.60 & \gray{-1.52} & -1.39 \\
Bins 456 & -2.68 & 3.75 & -3.25 & 4.91 & -4.08 & 5.56 & -5.96 & 8.06 & \gray{2.35} & 3.28 \\
Bins 123 & -4.58 & 3.44 & -6.16 & 4.46 & -7.80 & 5.49 & -14.48 & 7.00 & \gray{-1.33} & -1.82 \\
Bins 1234 & -2.78 & 2.47 & -3.87 & 3.37 & -4.58 & 4.29 & -6.38 & 6.17 & \gray{-0.81} & -1.13 \\

\midrule

Template Cosmo & -0.59 & 0.62 & -0.72 & 0.83 & -0.89 & 0.99 & -1.20 & 1.60 & \gray{x} & -0.49 \\
Covariance & -0.57 & 0.62 & -0.75 & 0.79 & -0.91 & 0.91 & -1.30 & 1.38 & \gray{x} & -0.11 \\
$n(z)\ \zmc-$ fid & -0.35 & 0.61 & -0.47 & 0.69 & -0.55 & 0.75 & -0.78 & 0.89 & \gray{x} & -0.20 \\

\hline  \T      & \multicolumn{8}{c||}{100 $(\sigma-\sigma_{\rm All \ Bins})/\sigma_{\rm All \ Bins}$}            \B                                                                                           \\ \cline{2-11}            
\hline

Bins 23456 & -5.53 & 28.15 & -7.91 & 34.95 & -10.77 & 44.11 & -16.41 & 61.14 & \gray{-1.46} & 0.69 \\
Bins 13456 & -5.06 & 33.65 & -8.46 & 40.63 & -11.21 & 47.60 & -18.61 & 78.34 & \gray{12.64} & 16.85 \\
Bins 12456 & -5.09 & 29.11 & -8.63 & 37.93 & -10.67 & 45.63 & -15.98 & 54.38 & \gray{26.19} & 23.16 \\
Bins 12356 & -6.17 & 33.22 & -9.39 & 45.05 & -12.03 & 49.74 & -22.23 & 62.51 & \gray{8.43} & 10.56 \\
Bins 12346 & -5.67 & 31.74 & -9.73 & 41.92 & -12.71 & 47.79 & -19.20 & 73.87 & \gray{13.69} & 12.18 \\
Bins 12345 & {\bf -4.89} & 30.92 & -7.90 & 42.62 & -10.77 & 52.16 & -18.77 & 75.55 & \gray{\bf -6.37} & {\bf -7.72} \\
Bins 456 & -0.98 & 97.42 & -7.16 & 130.80 & -12.08 & 155.71 & -18.40 & 203.90 & \gray{46.07} & 53.65 \\
Bins 123 & 1.81 & 126.95 & -3.37 & 160.80 & -7.36 & 189.98 & -17.54 & 257.44 & \gray{24.08} & 16.14 \\
Bins 1234 & -3.89 & 70.89 & -7.83 & 86.84 & -12.27 & 104.47 & -22.21 & 156.87 & \gray{7.57} & 1.34 \\

\hline
\end{tabular}
\end{table*}

\begin{table*}[h!]
\caption{
Table of pre-unblinding tests for the Projected Correlation Function 
from \autoref{sec:blindtest}. See description in \autoref{tab:unblind_acf} and text.  
}

\label{tab:unblind_pcf}
\begin{tabular}{l|ll|ll|ll|ll||c|c}
\hline
\hline
Threshold & \multicolumn{2}{l|}{0.9} & \multicolumn{2}{l|}{0.95} & \multicolumn{2}{l|}{0.97} & \multicolumn{2}{l||}{0.99} & \multicolumn{2}{l}{data} \\ \cline{2-11}
(Fraction of mocks) & min & max & min & max & min & max & min & max & {\tt \gray{MICE}} & {\bf Planck} \\ \hline
\T        & \multicolumn{8}{c||}{$10^2 (\alpha-\alpha_{\rm fiducial})$}                      \B                                                                                 \\ \cline{2-11}

   Bins     23456 & -1.72 & 1.48 & -2.12 & 1.92 & -2.32 & 2.12 & -2.96 & 2.77 &  \gray{0.95} &  0.95  \\
   Bins     13456 & -1.48 & 1.32 & -1.86 & 1.72 & -2.12 & 1.92 & -2.65 & 2.63 &  \gray{-0.33} & -0.36  \\
   Bins     12456 & -1.28 & 1.28 & -1.64 & 1.70 & -2.00 & 2.04 & -2.53 & 2.89 &  \gray{0.12}  &  0.33  \\
   Bins     12356 & -1.16 & 1.16 & -1.46 & 1.60 & -1.68 & 1.92 & -2.57 & 2.37 &  \gray{-0.48}  & -0.59 \\
   Bins     12346 & -0.76 & 0.96 & -1.00 & 1.20 & -1.16 & 1.52 & -1.60 & 2.00 &  \gray{0.17}  &  0.24  \\
   Bins     12345 & -0.44 & 0.52 & -0.60 & 0.64 & -0.68 & 0.72 & -0.85 & 0.88 &  \gray{-0.21}  & -0.32  \\
   Bins       456 & -3.76 & 3.24 & -4.85 & 4.28 & -5.40 & 4.88 & -6.59 & 6.47 &  \gray{2.37}  &  2.29  \\
   Bins       123 & -1.92 & 2.44 & -2.48 & 3.12 & -2.96 & 3.52 & -4.35 & 4.36 &  \gray{-1.00} &  -1.48  \\
   Bins      1234 & -1.00 & 1.32 & -1.28 & 1.66 & -1.52 & 1.88 & -1.94 & 2.48 &  \gray{-0.08}  &  -0.16  \\
 \midrule
Template Cosmo    & -0.80 & 0.92 & -1.01 & 1.12 & -1.13 & 1.20 & -1.49 & 1.49 &  \gray{x}      &  0.33    \\
   
Covariance        & -0.40 & 0.48 & -0.48 & 0.60 & -0.56 & 0.68 & -0.73 & 0.84 &  \gray{x}      &  -0.12   \\

$n(z)\ \zmc-$ fid & -0.40 & -0.08 & -0.44 & -0.04 & -0.44 & 0.00 & -0.49 & 0.04 & \gray{x} & -0.29    \\

\hline  \T      & \multicolumn{8}{c||}{100 $(\sigma-\sigma_{\rm All \ Bins})/\sigma_{\rm All \ Bins}$}            \B                                                                                         
\\ \cline{2-11}

   Bins   23456 & -1.12 & 32.11 & -3.30 & 40.01 & -4.62 & 45.33 & -6.56 & 55.20 & \gray{4.55}  & 4.65 \\ 
   Bins   13456 & -1.12 & 27.14 & -2.80 & 32.06 & -3.99 & 37.51 & -6.23 & 47.73 & \gray{22.38} & 18.27 \\
   Bins   12456 & -1.97 & 23.91 & -3.41 & 28.78 & -4.54 & 33.69 & -6.91 & 40.64 & \gray{8.39} & 5.98 \\
   Bins   12356 & -1.02 & 22.99 & -2.61 & 27.89 & -3.21 & 33.89 & -5.85 & 40.96 & \gray{6.29 } & 7.97 \\
   Bins   12346 & 0.00 & 15.35  & -1.49 & 18.40 & -2.20 & 20.89 & -3.80 & 26.56 & \gray{7.34}  & 9.30 \\
   Bins   12345 & 0.00 & 7.00   & -1.14 & 8.93  & -1.34 & 10.48 & -2.12 & 12.19 & \gray{1.75}  & 1.00 \\
   Bins     456 & 18.11 & 124.49 & 14.14& 150.21& 10.89 & 169.93& 5.18 & 222.60 & \gray{69.23} & 50.83 \\
   Bins     123 & 6.09 & 58.36  & 4.35  & 69.50 & 2.92  & 75.66 & -1.03 & 94.05 & \gray{21.33} & 27.91 \\
   Bins    1234 & 8.46 & 19.08  & 0.00  & 26.12 & -0.97 & 29.96 & -2.36 & 37.06 & \gray{9.44}  & 10.96 \\

\hline
\end{tabular}
\end{table*}

\bibliography{y6kp_bao_des}

\end{document}